\documentclass[twocolumn,trackchanges]{aastex63}

\newcounter{magicrownumbersCP}

\newcounter{magicrownumbersCPLM}
\newcommand\rownumber{\stepcounter{magicrownumbersCP}\arabic{magicrownumbersCP}}

\newcommand\rownumberLowMES{\stepcounter{magicrownumbersCPLM}\arabic{magicrownumbersCPLM}}

\usepackage{tabularx}

\usepackage{bbding}
\usepackage{longtable}
\usepackage{microtype}
\usepackage{graphicx,subfigure}
\usepackage{booktabs} 
\usepackage{enumitem}
\usepackage{amsmath}
\usepackage{multirow}
\usepackage{makecell}
\usepackage{slashbox}
\usepackage[linesnumbered, ruled, vlined]{algorithm2e}
\usepackage{algorithmic}
\newcolumntype{Y}{>{\centering\arraybackslash}X}

\usepackage{array,etoolbox}
\setitemize{leftmargin=10pt,noitemsep,topsep=0pt,parsep=0pt,partopsep=0pt}
\usepackage{hyperref}

\usepackage[flushleft, para]{threeparttable}
\usepackage{booktabs}
\usepackage{array}

\usepackage{scrextend}

\usepackage{comment}

\SetKwInput{KwInput}{Input}
\SetKwInput{KwOutput}{Output}

\newcommand{\kepler}{Kepler}
\newcommand{\KeplerMission}{Kepler Mission}
\newcommand{\TESSMission}{TESS Mission}
\newcommand{\vespa}{\texttt{vespa}} 
\newcommand{\ExoMiner}{\texttt{ExoMiner}}
\newcommand{\AstroNet}{\texttt{AstroNet}}  
\newcommand{\ExoNet}{\texttt{ExoNet}}
\newcommand{\GPC}{\texttt{GPC}}
\newcommand{\RFC}{\texttt{RFC}}

\newcommand{\Robovetter}{\texttt{Robovetter}}
\newcommand{\Autovetter}{\texttt{Autovetter}}
\newcommand{\TEC}{\texttt{TEC}}

\newcommand{\MD}{\mathcal{D}}
\newcommand{\ME}{\mathbb{E}}

\shorttitle{ExoMiner: A Highly Accurate and Explainable Deep Learning Classifier to Mine Exoplanets}
\shortauthors{Valizadegan et. al.}

\begin{document}

\title{ExoMiner: \\A Highly Accurate and Explainable Deep Learning Classifier that Validates 301 New Exoplanets}

\correspondingauthor{Hamed Valizadegan}
\email{hamed.valizadegan@nasa.gov}

\author[0000-0001-6732-0840]{Hamed Valizadegan}
\affiliation{Universities Space Research Association (USRA), Mountain View, CA 94043, USA}
\affiliation{NASA Ames Research Center (NASA ARC), Moffett Field, CA 94035, USA}

\author[0000-0002-2188-0807]{Miguel J. S. Martinho}
\affiliation{Universities Space Research Association (USRA), Mountain View, CA 94043, USA}
\affiliation{NASA Ames Research Center (NASA ARC), Moffett Field, CA 94035, USA}

\author[0000-0001-9508-2739]{Laurent S. Wilkens}
\affiliation{Delft University of Technology, Delft, Netherlands}
\thanks{Author contributed while interning at USRA/NASA ARC}

\author[0000-0002-4715-9460]{Jon M. Jenkins}
\affiliation{NASA Ames Research Center (NASA ARC), Moffett Field, CA 94035, USA}

\author[0000-0002-6148-7903]{Jeffrey C. Smith}
\affiliation{The SETI Institute, Mountain View, CA  94043, USA}
\affiliation{NASA Ames Research Center (NASA ARC), Moffett Field, CA 94035, USA}

\author[0000-0003-1963-9616]{Douglas A. Caldwell}
\affiliation{The SETI Institute, Mountain View, CA  94043, USA}
\affiliation{NASA Ames Research Center (NASA ARC), Moffett Field, CA 94035, USA}

\author[0000-0002-6778-7552]{Joseph D. Twicken}
\affiliation{The SETI Institute, Mountain View, CA  94043, USA}
\affiliation{NASA Ames Research Center (NASA ARC), Moffett Field, CA 94035, USA}

\author[0000-0001-5019-7508]{Pedro C.L. Gerum}
\affiliation{Cleveland State University, Cleveland, OH 44115, USA}
\thanks{Author contributed while interning at NASA through $I\hat{\;}2$ program}

\author[0000-0002-2715-374X]{Nikash Walia}
\affiliation{University of Illinois at Urbana-Champaign, Urbana, IL, 61801, USA}
\thanks{Author contributed while interning at USRA/NASA ARC}

\author[0000-0001-5064-4283]{Kaylie Hausknecht}
\affiliation{Harvard University, Cambridge, MA 02138, USA}
\thanks{Author contributed while interning at USRA/NASA ARC}

\author[0000-0001-5428-262X]{Noa Y. Lubin}
\affiliation{Bar-Ilan University, Ramat Gan, Israel}
\thanks{Author contributed while interning at NASA through $I\hat{\;}2$ program}

\author[0000-0003-0081-1797]{Stephen T. Bryson}
\affiliation{NASA Ames Research Center (NASA ARC), Moffett Field, CA 94035, USA}

\author[0000-0002-5987-1033]{Nikunj C. Oza}
\affiliation{NASA Ames Research Center (NASA ARC), Moffett Field, CA 94035, USA}


\begin{abstract}
The \kepler\ and TESS missions have generated over 100,000 potential transit signals that must be processed in order to create a catalog of planet candidates. During the last few years, there has been a growing interest in using machine learning to analyze these data in search of new exoplanets. Different from the existing machine learning works, \ExoMiner, the proposed deep learning classifier in this work, mimics how domain experts examine diagnostic tests to vet a transit signal. \ExoMiner\ is a highly accurate, explainable, and robust classifier that 1) allows us to validate 301 new exoplanets from the MAST Kepler Archive and 2) is general enough to be applied across missions such as the on-going TESS mission. We perform an extensive experimental study to verify that \ExoMiner\ is more reliable and accurate than the existing transit signal classifiers in terms of different classification and ranking metrics. For example, for a fixed precision value of 99\%, \ExoMiner\ retrieves 93.6\% of all exoplanets in the test set (i.e., recall=0.936) while this rate is 76.3\% for the best existing classifier. Furthermore, the modular design of \ExoMiner\ favors its explainability. We introduce a simple explainability framework that provides experts with feedback on why \ExoMiner\ classifies a transit signal into a specific class label (e.g., planet candidate or not planet candidate).
\end{abstract}

\section{Introduction}
\label{introduction}

Recent space missions have revolutionized the study of exoplanets in astronomy by observing and comprehensively analyzing hundreds of thousands of stars in search of transiting planets. The CoRoT Mission \citep{baglin2006corot} observed 26 stellar fields of view between 2007 and 2012, detecting more than 30 exoplanets and more than 550 candidates \citep{deleuil2018corot}. \kepler~\citep{borucki2010kepler} continuously observed 112,046 stars\footnote{\kepler\ observed an additional 86,663 stars for some portion of its four-year mission \citep{Twicken_2016-autovetterlabels}.} in a 115 square degree region of the sky for four years and identified over 4000 planet candidates \citep{Thompson_2018} among which there are more than 2300 confirmed or statistically validated exoplanets\footnote{\url{https://exoplanetarchive.ipac.caltech.edu}}. K2, using the re-purposed \kepler\ spacecraft \citep{Howell_2014_K2}, detected more than 2300 candidates with over 400 confirmed or validated exoplanets. More recently, the Transiting Exoplanet Survey Satellite (TESS) Mission \citep{ricker2015transiting} surveyed $\sim$75\% of the entire sky in its first two years of flight starting in June 2018, an area 300 times larger than that monitored by \kepler, and detected 2241 candidates and $\sim$130 confirmed exoplanets \citep{guerrero2021TOI}. All of these large survey missions have produced many more candidate exoplanets than can easily be confirmed or identified as false positives using conventional approaches.

\begin{figure*}[htb!]
\begin{center}
\centerline{\includegraphics[width=\textwidth]{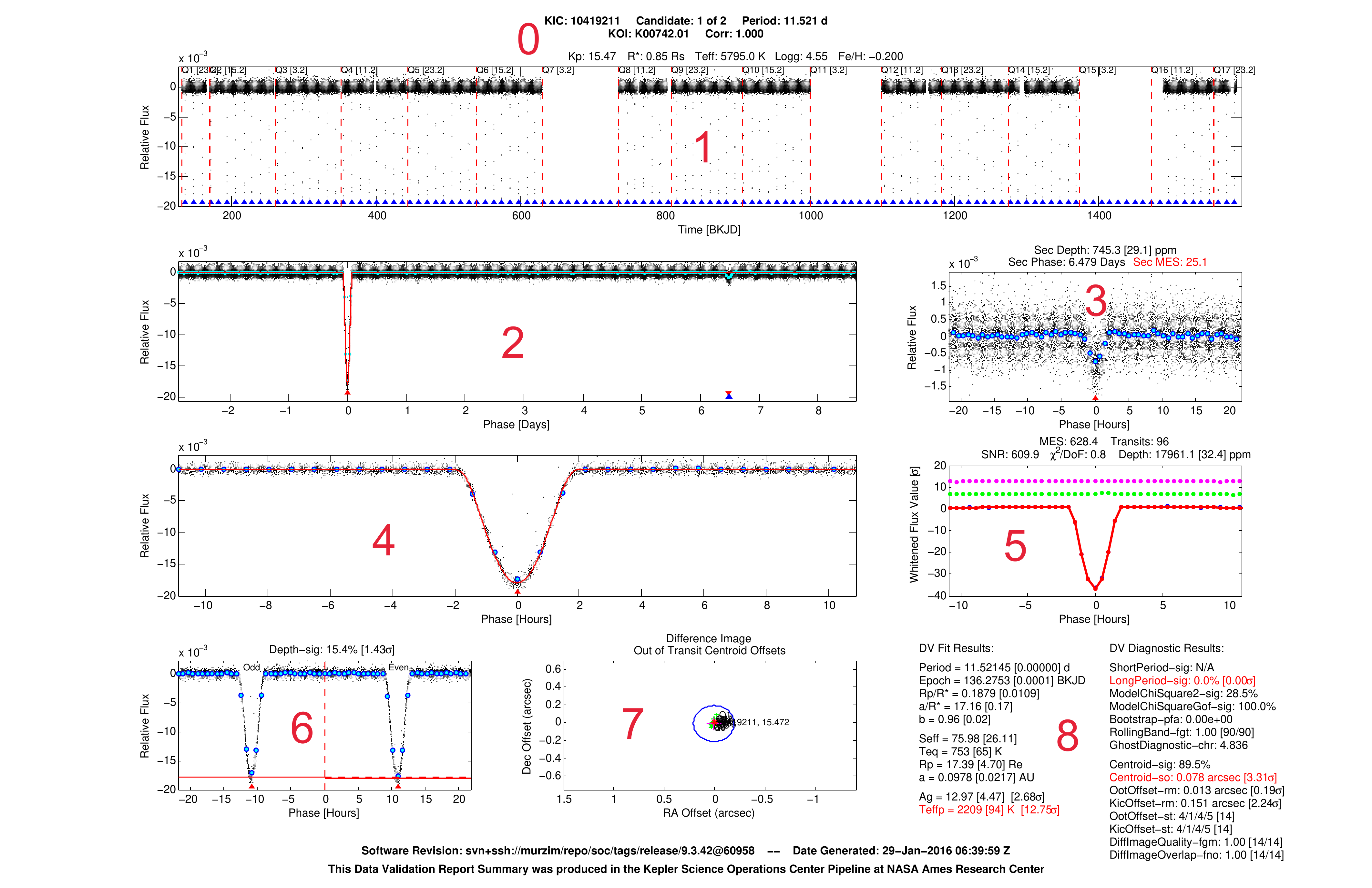}}
\caption{Example DV 1-page summary report. It includes multiple diagnostic plots and variables:  (0) Stellar parameters, (1) Full time series flux, (2)  Full-orbit phase-folded flux, (3) Transit-view phase-folded secondary eclipsing flux, (4) Transit-view phase-folded flux, (5) Transit-view phase-folded whitened flux, (6) Transit-view phase-folded odd \& even flux, (7) Difference image (out-of-transit) centroid offsets, and (8) DV analysis table of variables.}
\label{DVsummaryreport}
\end{center}
\end{figure*}

The most common contemporary\footnote{Early vetting efforts by the Kepler and TESS teams involved a two-step process wherein a team of vetters reviewed the Data Validation reports and other diagnostics of each TCE and then voted to elevate potential planet candidates to \kepler/TESS Object of Interest (KOI/TOI) status for prioritization for follow-up observations.} approach to detecting and vetting exoplanet candidates involves a three-step process: 1) The imaging data from a transit survey are processed on complex data processing and transit search pipelines,\footnote{Citizen scientist projects to identify transit signatures in large transit photometry datasets were initiated by the Planet Hunters Zooniverse project for the Kepler Mission \citep[see, e.g., ][]{PlanetHunters_Kepler_I_2013} and continue in the TESS era \citep{PlanetHunters_TESS_I_2020}.} \citep[see, e.g.,][]{Jenkins_2010_keplerpipeline,Jenkins2016SPIE}, that identify transit-like signals (called threshold-crossing events -- TCEs), perform an initial limb-darkened transit model fit for each signal, and conduct a suite of diagnostic tests to help distinguish between exoplanet signatures and non-exoplanet phenomena \citep[such as background eclipsing binaries --][]{Twicken_2018_DV, Li2019KeplerDataValidation2}. For the Kepler and TESS science pipelines, the results of the transit search, model fitting and diagnostic tests are presented in Data Validation (DV) reports~\citep{Twicken_2018_DV,Li2019KeplerDataValidation2} that include 1-page summary reports (Figure~\ref{DVsummaryreport}) along with more comprehensive reports. 2) The TCEs are filtered by either pre-defined if-then vetting rules \citep{Thompson_2018,Coughlin2017robovetter} or Machine Learning (ML) classifiers~\citep[\AstroNet][]{Yu-2019-TESS,shallue_2018} to identify those most likely to be exoplanets, and 3) The DV reports for top-tier TCEs that survive the filtering process in step 2 are then typically reviewed by vetting teams and released as \kepler\ or TESS Objects of Interest for follow-up observations \citep{Thompson2016dr25,guerrero2021TOI}. 

ML methods are ideally suited for probing these massive datasets, relieving experts from the time-consuming task of sifting through the data and interpreting each DV report, or comparable diagnostic material, manually. When utilized properly, ML methods also allow us to train models that potentially reduce the inevitable biases of experts. Among many different ML techniques, Deep Neural Networks (DNNs) have achieved state-of-the-art performance~\citep{lecun2015deep} in areas such as computer vision, speech recognition, and text analysis and, in some cases, have even exceeded human performance. DNNs are especially powerful and effective in these domains because of their ability to automatically extract features that may be previously unknown or highly unlikely for human experts in the field to grasp~\citep{Bengio-representation-2012}. This flexibility allows for the development of ML models that can accept raw data and thereby increase the productiveness of the transit search and vetting process.  

The availability of many \kepler\ and TESS pipeline products has provided researchers with a wide spectrum of input choices for machine-based transit signal classification, from the unprocessed raw pixels to the middle level pipeline products (e.g., processed flux data and centroid data) and final tabular datasets of diagnostic test values. Several transit signal classifiers have been previously developed and have each been designed to target inputs at different stages along this data processing pipeline.

\Robovetter~\citep{Coughlin_2016_robovetter} and \Autovetter~\citep{Jenkins-Autovetter-2014IAUS,McCauliff_2015} both utilize the final values generated via diagnostic statistics as inputs but differ from each other in their respective model designs: \Robovetter\ relies heavily on domain knowledge to create an expert-type classifier system while \Autovetter\ exploits a fully automated ML approach to build a random forest classifier. 
The main drawback of these classifiers is their high level of dependency on the pipeline and its final products, i.e., the features that are engineered and built from the diagnostic tests and transit fit. 

More recent works use the flux data directly, including ~\citep{shallue_2018, Dattilo_2019, Ansdell_2018, Osborn-deeplearning-2020} that train DNN models using flux and/or centroid time series, and~\citet{armstrong-2020-exoplanet} that combined diagnostics scalar values and flux data to train more accurate non-DNN models. Such models not only learn how to automatically classify transit signals but are also capable of extracting important features that allow the machine to classify these signals (unlike human-extracted features, which may not be suitable or sufficient for use by a machine). 
While the final Kepler data release DR25 applied a rule-based system, \Robovetter~\citep{Coughlin_2016_robovetter,Thompson_2018}, to generate the final planet candidate catalog, ML models are gaining trust to be employed in practice to help identify planet candidates. For example, \AstroNet\ is used to assist in the identification of TOIs for MIT's Quick-Look Pipeline, which operates on TESS Full Frame Image (FFI) data~\citep{guerrero2021TOI}, and it was also used to validate two new exoplanets~\citep{shallue_2018}. More recently, ~\citet{armstrong-2020-exoplanet} utilized ML methods to systematically validate 50 new exoplanets. Thus far, the primary reasons for the remaining hesitancy about utilizing fully data-driven DNN approaches for vetting and validating transit signals are the models' lack of explainability and acceptable accuracy.

The inefficacy of existing DNN classifiers used to vet transit signals has signaled researchers that the flux data alone may not be enough as inputs for these models, demonstrating the need to potentially use the minimally processed raw pixel data. Although we agree that the flux data alone is insufficient, we argue that using
raw pixels as inputs may prove to be impractical due to the magnitude of the \kepler\ dataset (about 68,000 
images $\times\;32\;\text{pixels/image on average}$  $\approx 2.2$ million pixel values for each transit signal for the \KeplerMission)\footnote{[$15\;\text{quarters}\times 90 \;\text{days} + 2 \;\text{quarters} \times 30 \;\text{days}] \times 2 \;\text{cadences/hour} \times 24 \;\text{hours/day} = 67,680 \;\text{cadences}$} and the lack of enough labeled data. Learning an effective classifier from data in such a high volume space requires exponentially more data than are available \citep[curse of dimensionality --][]{Bishop:2006:PRM:1162264}. For the final \kepler\ Data Release (Q1-Q17 DR25), for example, there are only approximately 34,000 labeled TCEs available (and approximately 200,000 transit signals if non-TCEs are included). 

To understand the difficulty of developing a model based on the raw pixel data, note that a complex standalone DNN intended to classify raw pixels must learn all of the steps in the raw pixel processing pipeline, which includes extracting photometry data from the pixel data, removing systematic errors, extracting other useful features from pixel and flux data, and generating final classifications from the ground up with a limited amount of supervision in the form of labeled data. Considering that there is a whole pipeline dedicated to processing these data, learning each of these complex steps in a data-driven manner is a difficult task. Domain-specific techniques such as specialized augmentation to generate more labeled data or self-supervised learning~\citep{Dosovitskiy-selfsupervised-2018} could be employed to partially mitigate the curse of dimensionality; however, we believe that more accurate DNN classifiers can be developed if the abundant amount of domain knowledge utilized in the design of the pipeline is used as a guide for designing the DNN architecture. Another advantage of this approach is that it helps to build DNN models that are more explainable, as we will show in this paper. By examining DV reports and how they are used by experts to vet transit signals, we introduce a new DNN model that can be used to build catalogs of planet candidates and mine new exoplanets. We show that our model is highly accurate in classifying transit signals of the  \KeplerMission\ and can be transferred to classify TESS signals.

This paper is organized as follows: We start Section~\ref{relatedwork} by discussing the manual (Section~\ref{sec:manual-vetting}) and machine (Section~\ref{sec:automatic-classification}) classification of transit signals and finish it with Section~\ref{sec:validation_related_work} that discusses exoplanet validation and Section~\ref{sec:our-contribution} that summarizes the contributions/novelties of this work. In order to justify the choices made in the design of our ML approach, we first provide a brief background of machine classification in Section~\ref{sec:machine-classification}, DNNs in Section~\ref{sec:DNNs}, and explainability in Section~\ref{sec:explainability}. This helps us to provide the details of the proposed DNN, \ExoMiner, in Section~\ref{sec:proposed-dnn}, relevant data preparation in Section~\ref{sec:datainput}, and optimizing the model in Section~\ref{sec:data-driven}. The second half of the paper is focused on experimental studies. After discussing the \kepler\ dataset used in this paper, baseline classifiers, and evaluation metrics in Section~\ref{sec:experimental_setup}, we study the performance of \ExoMiner\ in Section~\ref{sec:evaluation}.  Section~\ref{sec:stable_reliable} is focused on reliability and stability of \ExoMiner and Section~\ref{sec:exominer-explainability} introduces a framework for explainability of \ExoMiner. We prepare the ground for validating new exoplanets using \ExoMiner\ in Section~\ref{sec:exominer_validation} and validate over 301 new exoplanets in Section~\ref{sec:planetvalidation}. We report a preliminary study of applying \ExoMiner\ on TESS data in Section~\ref{sec:TESS}. Finally, Section~\ref{sec:discussions} concludes the paper by listing the caveats of this work and plotting the future directions. 

\section{Related Work}
\label{relatedwork}

In this section, we discuss the manual and automatic classification of transit signals. Because the design of the proposed model in this work is greatly influenced by the manual vetting process, we first review how experts vet transit signals. We then review the related works on automatic classification and validation of new exoplanets. We conclude this section by enumerating our contributions.

\subsection{Manual Classification of TCEs}
\label{sec:manual-vetting}
The \kepler/TESS pipeline might detect False Positive (FP) transit events due to sources such as eclipsing binaries (EBs), background eclipsing binaries (BEBs), planets transiting background stars, stellar variability, and instrument--induced artifacts~\citep{Borucki_2011_KOI2}. Manual classification of TCEs consists of the time-consuming evaluation of multiple diagnostic tests and fit values to determine if a TCE is a Planet Candidate (PC) or FP. The FP category may be further subdivided into Astrophysical FP (AFP) and Non-Transiting Phenomena (NTP). 

Several tests have been developed to diagnose different types of FPs. Combined with the flux data in multiple forms, these diagnostic tests are the main components of the 1-page DV summary and the longer DV report used by experts to vet transit signals~\citep{Twicken_2018_DV, twickenSDPDD2020}. The 1-page DV summary report includes: (1) stellar parameters regarding the target star to help determine the type of star associated with the TCE (Figure~\ref{DVsummaryreport}.0), (2) unfolded flux data to understand the overall reliability of the transit signal (Figure~\ref{DVsummaryreport}.1), (3) the phase-folded full-orbit and transit view of the flux data to determine the existence of a secondary eclipse and shape of the signal (Figure~\ref{DVsummaryreport}.2,4), (4) Phase-folded transit view of the secondary eclipse to check the reliability of the eclipsing secondary (Figure~\ref{DVsummaryreport}.3 ), (5) Phase-folded whitened transit view to check how well a whitened transit model is fitted to the whitened light curve (Figure~\ref{DVsummaryreport}.5), (6) phase-folded odd and even transit view to check false positive due to a circular EB target or a BEB when the detected period is half of the period of EB (Figure~\ref{DVsummaryreport}.6 ), (7) Difference image to check for the false positive due to a background object being the source of the transit signature (Figure~\ref{DVsummaryreport}.7), (8) and an analysis table that provides values related to the model fit and various DV diagnostic parameters to better understand the signal (Figure~\ref{DVsummaryreport}.8 ). 

The information in the odd \& even and weak secondary tests are both used in detecting EB false positives. The odd \& even depth test is useful when the pipeline detects only one TCE for a circular binary system; in this case, the depth of the odd and even views could be different. The weak secondary test is used when the pipeline detects a TCE for the primary eclipses and a TCE for the secondary in an eccentric EB system, or when there is a significant secondary that does not trigger a TCE. However, secondary events could also be caused by secondary eclipses of a giant, short-period transiting planet exhibiting reflected light or thermal emission. To distinguish between these two types of secondary events, experts use the secondary geometric albedo, planet effective temperature, and Multiple Event Statistic (MES) of the secondary event~\citep{Twicken_2018_DV, Jenkins2020KeplerHandbook}. \

\begin{figure}[htb!]
	\centering
	\subfigure[Kepler-1604 b Secondary Test]{\label{fig:wrong-period-secondary}\includegraphics[width=\columnwidth]{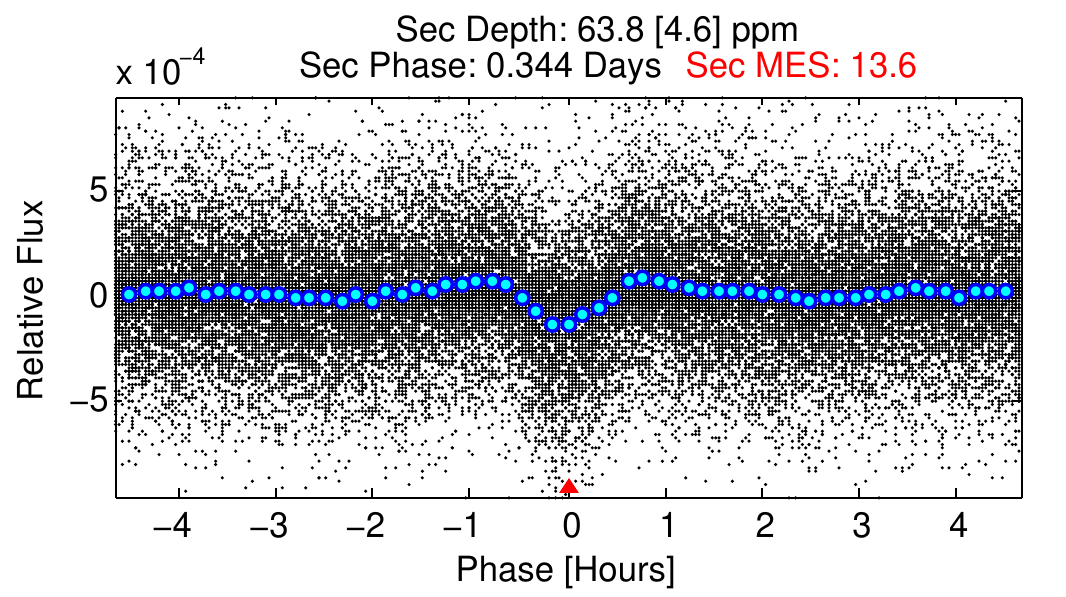}}
	\subfigure[Kepler-458 b Odd \& Even Test]{\label{fig:wrong-period-oddeven}\includegraphics[width=\columnwidth]{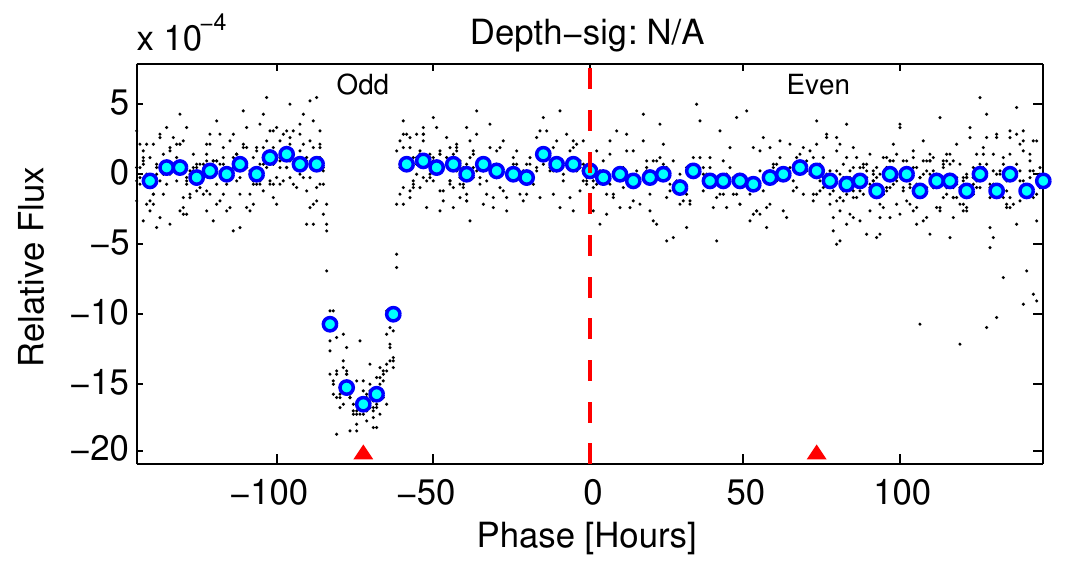}}
	\caption{An incorrect period can lead to the misclassification of a TCE. In the top panel, TCE KIC 8416523.1 was detected at twice the actual period of the transiting planet, resulting in half the transits being considered secondary events. In the bottom panel a TCE KIC 9663113.2 was detected at half the true period, resulting in no transit events for the even numbered transit phases.} 
\label{fig:wrong-period-confirmed}
\end{figure}

Moreover, note that a real exoplanet could be confused with an EB if the pipeline incorrectly measures the period for the corresponding TCE. Depending on whether the pipeline's returned period for an exoplanet is twice\footnote{Two TCEs can be generated for an exoplanet when its orbital period is less than the minimum orbital period in the \kepler\ pipeline transit search (0.5 days).} or half\footnote{The period can be one-half of the true value when there are less than three observed transit events for a given exoplanet. A minimum of three transits is required to define a TCE in the \kepler\ pipeline transit search.} the real period, the weak secondary or odd \& even tests might get flagged. Examples of exoplanets with incorrectly returned periods that resulted in statistically significant weak secondary or odd \& even tests are Kepler-1604~b and Kepler-458~b, respectively, as shown in Figure~\ref{fig:wrong-period-secondary} and Figure~\ref{fig:wrong-period-oddeven}. Such incorrect period can lead models to erroneously classify the corresponding exoplanet as an FP. Therefore, it is difficult for the machine to correctly classify a TCE as a PC without the correct orbital period \citep[available for these cases in the Cumulative \kepler\ Object of Interest (KOI) catalog --][]{Thompson_2018}\footnote{https://exoplanetarchive.ipac.caltech.edu/docs/data.html}. Furthermore, circumbinary transiting planets and planets with significant transit timing variations are difficult to classify correctly, since the pipeline was not designed to detect their aperiodic signatures.

After examining the DV summary, experts may also have to check the full DV report, which provides a more thorough analysis, in order to classify a TCE. An initial PC label obtained using the manual vetting process should then be subsequently submitted to a follow-up study by the rest of the community (using ground-based telescope observations, for example) in order to confirm the classification. 

The manual vetting process has several flaws: not only is it very slow and a bottleneck to efficiently identifying new planets, but it is also not always consistent or effective because of the biases that inevitably come with human labeling. Moreover, in order to make the manual process feasible and reduce the number of transit signals to examine, the pipeline rejects many signals that do not meet certain conditions, including signals below a defined signal-to-noise ratio (S/N). Importantly, some of the most interesting transits of small long-period planets may be just below the threshold but could be successfully vetted with the inclusion of information in the image data that is complementary to the information in the extracted light curve. We risk missing these planets because they are discarded by the pipeline before all the available information in the observed data has been considered. Unlike the manual vetting process, automatic classification is fast and can be used to rapidly vet transit signals potentially reducing the biases and inconsistencies that come with human vetting. Furthermore, automatic classification can be used to vet all transits, including low-S/N signals near the threshold, in order to identify more small planet candidates. 

\begin{table*}[htb]
\footnotesize
 \centering
\caption{Comparison of different models. \CheckmarkBold~means that the model uses the original non-scalar time series or image as its input. \PlusThinCenterOpen~ means that the model uses only a few scalar values summarizing the time series or image as its input.}
\label{table:classifiers-summary}
\begin{threeparttable}
\begin{tabularx}{\linewidth}{@{}Y@{}}
\begin{tabular}{cc|cccccccc}
\toprule
& & \vespa & \Robovetter & \Autovetter & \AstroNet & \ExoNet & \GPC & \RFC & \ExoMiner  \\
\midrule
\multicolumn{2}{c|}{Objective} & Val.\tnote{1} & 3C-Vet.\tnote{2} & 3C-Vet. & 2C-Vet.\tnote{3} & 2C-Vet. & 2C-Vet. & 2C-Vet. & 2C-Vet. \& Val. \\
\multicolumn{2}{c|}{Model Construction} & Gen.\tnote{4} & Exp.\tnote{5} & Dis.\tnote{6} & Dis. & Dis. & Dis. & Dis. & Dis. \\
\midrule
\multirow{7}{*}{Scalar inputs}
& Primary properties\tnote{7} & \CheckmarkBold &\CheckmarkBold & \CheckmarkBold & & & \CheckmarkBold &\CheckmarkBold &\CheckmarkBold\\
& Transit fit model & \CheckmarkBold &\CheckmarkBold  & \CheckmarkBold &  &  & \CheckmarkBold & \CheckmarkBold &\\
& Stellar parameters & &  & \CheckmarkBold & & \CheckmarkBold & \CheckmarkBold &\CheckmarkBold &\CheckmarkBold\\
& Optical ghost test & & \CheckmarkBold & \CheckmarkBold & & & \CheckmarkBold &\CheckmarkBold &\CheckmarkBold\\
& Bootstrap fap & & \CheckmarkBold & \CheckmarkBold & & & \CheckmarkBold &\CheckmarkBold &\CheckmarkBold\\
& Rolling band-fgt & & \CheckmarkBold & \CheckmarkBold & & & & &\CheckmarkBold\\
& Secondary properties\tnote{8} & & \CheckmarkBold & \CheckmarkBold & & & \CheckmarkBold & \CheckmarkBold&\CheckmarkBold\\
\midrule
\multirow{5}{*}{Non-scalar inputs}
& Unfolded flux & & \PlusThinCenterOpen &  &  &  &  & &\\
& Phase-folded flux & & \PlusThinCenterOpen & \PlusThinCenterOpen & \CheckmarkBold & \CheckmarkBold & \CheckmarkBold & \PlusThinCenterOpen  & \CheckmarkBold\\
& Odd \& even views & & \PlusThinCenterOpen & \PlusThinCenterOpen &  &  & \PlusThinCenterOpen &  \PlusThinCenterOpen &\CheckmarkBold\\
& Weak secondary flux & & \PlusThinCenterOpen & \PlusThinCenterOpen & &  & \PlusThinCenterOpen &\PlusThinCenterOpen&\CheckmarkBold\\
& Difference image & & \PlusThinCenterOpen & \PlusThinCenterOpen & & & \PlusThinCenterOpen  & &\PlusThinCenterOpen\\
& Centroid motion test & & \PlusThinCenterOpen & \PlusThinCenterOpen & & \CheckmarkBold & & &\CheckmarkBold\\
\bottomrule
\end{tabular}
\end{tabularx}
\begin{tablenotes}[para,flushleft]\scriptsize
  \item [1] Validation
  \item [2] Three classes vetting
  \item [3] Two classes vetting
  \item [4] Generative
  \item [5] Expert System
  \item [6] Discriminative
  \item [7] Includes MES, planet radius, depth, duration, etc.
  \item [8] Geometric albedo, planet effective temperature, and MES for secondary
\end{tablenotes}
\end{threeparttable}
\end{table*}

\subsection{Existing Methods for Automatic Classification of TCEs}
\label{sec:automatic-classification}

Some of the well known existing methods for automatic classification of transit signals are summarized in Table~\ref{table:classifiers-summary}. Even though they all can be considered machine classifiers, they differ in their design objective, inputs, and construction. Because these different models aim to address different problems, they should be compared with their objective in mind. Nonetheless, given that all these models are able to generate dispositions and disposition scores, they can be studied from the vetting and validation perspective. These approaches can be categorized into three groups: 
\begin{enumerate}[noitemsep,nolistsep,partopsep=1pt,topsep=1pt, leftmargin=4mm]
\item The expert system category includes only \Robovetter~\citep{Coughlin2017robovetter} and a ported version of \Robovetter\ for the TESS Mission called \TEC~\citep{guerrero2021TOI}. \Robovetter\ was designed to build a KOI catalog by leveraging manually incorporated domain knowledge in the form of if-then conditions. These if-then rules check different types of diagnostic and transit fit values. The developer of such an expert system must recognize special cases and completely understand the decision-making process of an expert in order to develop effective systems. As such, these systems are not fully automated data-driven models. They cannot also easily accept complex data inputs such as time series or images (the non-scalar inputs in Table~\ref{table:classifiers-summary}).   

\item The generative category includes models such as \vespa~\citep{Morton_2011_Vespa, Morton_2012_vespa, Morton-2016-vespa} that require the class priors $p(y=1)$ and $p(y=0)$ and likelihoods $P(X|y=1)$ and $P(X|y=0)$ to estimate the posterior $P(y=1|X)$ according to the Bayes' theorem:
\begin{equation}
p(y=1|X)=\frac{p(X|y=1)p(y=1)}{\sum_y p(X|y)p(y)} 
\end{equation}
where $y=1$ represents an exoplanet, $y=0$ represents a false positive, and $X$ is a representation of the transit signal. Such generative approaches require the detailed knowledge of the likelihood, $P(X|y)$, and prior, $P(y)$, for each class (exoplanet vs false positive), and class scenario (e.g., BEB). While in general both the likelihood and priors can be learned in a data-driven approach (using ML), \vespa\ estimates them by simulating a representative population with ``physically and observationally informed assumptions''~\citep{Morton_2012_vespa, Morton-2016-vespa}. \vespa\ fpp has been used to validate exoplanets~\citep{Morton-2016-vespa} when the posterior probability $P(y=1|X)$ is $>$0.99. 

\item The discriminative approach directly calculates the posterior probability by considering a functional form for $P(y|X)$. It includes all other models listed in Table~\ref{table:classifiers-summary}, i.e., \Autovetter~\citep{{Jenkins-Autovetter-2014IAUS,McCauliff_2015}}, \AstroNet~\citep{shallue_2018}, \ExoNet~\citep{Ansdell_2018}, \GPC/\RFC~\citep{armstrong-2020-exoplanet}, and \ExoMiner, the model proposed in this work. These models differ in the input data and function they use to estimate $P(y|X)$. The functional form of $P(y|X)$ in \Autovetter\ and \RFC\ is a random forest classifier. As such, they are limited to accepting only scalar values. The functional form for \AstroNet, \ExoNet, and \ExoMiner\ is a DNN which can handle different types of data input. \AstroNet\ only uses phase folded flux data. \ExoNet\ adds stellar parameters and centroid motion time series as input. \ExoMiner\ uses the unique components of data from the 1-page DV summary report, including scalar and non-scalar data types. \GPC\ is a Gaussian process classifier able to use both scalar and non-scalar data types as input once a kernel (similarity) function between data points is defined. \GPC\ combines different scalar data types with the phase folded flux time series. The scalar and non-scalar input types used for all these classifiers are summarized in Table~\ref{table:classifiers-summary}. Similar to the generative approach, one can validate exoplanets using discriminative models if the model is calibrated and $P(y=1|X)$ is sufficiently large. 
\end{enumerate}

The advantage of generative approaches is that they can also be used to generate new transit signals, hence the generative name. However, the accuracy of a generative model is directly affected by the validity of the assumptions made about the form of the likelihood function and the values of priors. These can be obtained in a fully data-driven manner (through ML methods) or using the existing body of domain knowledge, or a combination of both (e.g., \vespa). For example, to make the calculation of likelihood tractable, \citet{Morton_2012_vespa} represented a transit signal $X$ by a few parameters: transit depth, transit duration, and duration of the transit ingress and egress, 
and generated the likelihood $P(X|y)$ using (1) a continuous power law for exoplanet and (2) simulated data from the galactic population model, TRILEGAL~\citep{Girardi-TRILEGAL-2005}, for false positive scenarios. Such an approach relies on the existing body of knowledge regarding the form of the likelihood and the values of priors, which could be paradoxical, as mentioned in~\citet{Morton_2012_vespa}. For example, the occurrence rate calculation requires the posterior probability which itself requires the occurrence rate values.  Moreover, reducing the shape to only a few parameters defining the vertices of a trapezoid fitted to the transit or transit model is very limiting. The shape of a transit is continuous and cannot be defined by a small number of parameters when you consider the effect of star spots. More generally, a transit signal $X$ is summarized by only a few scalars, whereas to fully represent a transit signal for the task of classification, more data about the signal is needed, as we listed in Section~\ref{sec:manual-vetting}. Full representation of a transit signal using the approach taken by~\cite{Morton_2012_vespa} is not tractable. To be precise,~\cite{Morton_2012_vespa} answers the following question: what is the false positive probability (fpp) of a signal that has a specific depth, duration, and shape. Adding more information will change this probability. Refer to~\citet{Morton_2011_Vespa, Morton_2012_vespa} for the list of other assumptions.

Even though both the likelihood $p(X|y)$ and priors $p(y)$ can be learned using data with a more flexible representation for transit signal $X$ (e.g., flux) and modern generative approaches to ML (e.g., DNNs), discriminative classifiers are preferred if generating data is not needed. The compelling justification of \citet{Vapnik-statisticallearningtheory-1998} to propose the discriminative approach was that, ``one should solve the [classification] problem directly and never solve a more general problem as an intermediate step [such as modeling $p(X|y)$].'' 

The performance of existing discriminative approaches for transit signal classification depends on their flexibility in accepting and including multiple data types (as provided in DV reports) and on their ability to learn a suitable representation (or feature learning) for their input data. Among the existing machine classifiers, the ones introduced in~\citet{armstrong-2020-exoplanet}, e.g., \RFC\ and \GPC\ are the most accurate ones, as we will see in the experiments. Existing DNN models such as \AstroNet\ and \ExoNet\ are not as successful as the former since they do not use many diagnostic tests required to identify different types of false positives (check Table~\ref{table:classifiers-summary}). In contrast, \RFC\ and \GPC\ are relatively inclusive of different diagnostic tests (except the centroid test) required for the classification of transit signals. However, these classifiers (1) are not able to learn a representation for non-scalar data automatically; \GPC\ needs to be provided with a kernel function (similarity) prior to training, and \RFC\ is not able to directly receive non-scalar data such as flux, and (2) these classifiers are not inclusive of non-scalar (time series) data types and only use the results of statistical tests for most time series data such as odd \& even, weak secondary, and centroid. 

Similar or small modifications have been made to the aforementioned classifiers to perform classification/evaluation tasks for K2~\citep{Dattilo-2019-K2}, TESS~\citep{Armstrong-2017-SOM, Yu-2019-TESS, Osborn-deeplearning-2020}, and The Next-Generation Transit Survey \citep[NGTS --][]{Chaushev-2019-NGTS}.

\subsection{Validation of Exoplanets}
\label{sec:validation_related_work}
Given that traditional confirmation of planet candidates was not possible or practical due to the increase in the number of candidates and their specifics (e.g., small planets around faint stars), multiple works were developed aimed at statistically quantifying the probability that signals are false positives~\citep{Torres-2004-blender, Sahu_2006_FP, Morton_2012_vespa}. Technically, any classifier that can classify transit signals into PC/non-PC can be used for validation. However, such classifiers need to be employed with extra scrutiny to make sure their validation is reliable.

Existing approaches to planet validation use extra vetoing criteria in order to improve the precision of the generated catalog of validated planets. These criteria are often model specific and designed to address issues related to each model. For example, \vespa~\citep{Morton-2016-vespa} is not designed to capture FPs due to stellar variability or instrumental noise, and it might fail under other known scenarios such as offset blended binary FPs and EBs that contaminate the flux measurement through instrumental effects, such as charge transfer inefficiency ~\citep{Coughlin_contamination_2014}. Based on~\citet{Morton-2016-vespa}, \vespa\ is only reliable for KOIs that passed existing Kepler vetting tests. To avoid FPs due to systematic noise, \vespa\ might not work well for low-MES region and this is why KOIs in low-MES region are rejected for validation.

In order to improve the reliability of validating new exoplanets, ~\citet{armstrong-2020-exoplanet} required the new exoplanets to satisfy the following conditions: 1) the corresponding KOI has MES$\;>10.5\,\sigma$, as recommended by~\citet{Burke_2019}, 2) its \vespa\ fpp score is lower than $0.01$, 3) all of their four classifiers assign a probability score higher than 0.99 to that TCE, and 4) the corresponding KOI passes two outlier detection tests.

\subsection{Contributions/Novelty of this work}
\label{sec:our-contribution}
To classify transit signals and validate new exoplanets in this work, we propose a new deep learning classifier called \ExoMiner\ that receives as input multiple types of scalar and non-scalar data as represented in the 1-page DV summary report and summarized in Table~\ref{table:classifiers-summary}. \ExoMiner\ is a fully automated, data-driven approach that extracts useful features of a transit signal from various diagnostic scalar and non-scalar tests. The design of \ExoMiner\ is inspired by how an expert examines the diagnostic test data in order to identify different types of false positives and classify a transit signal. 

Additionally, \ExoMiner\ benefits from an improved ML data preprocessing pipeline for non-scalar data (e.g., flux data) and an explainability framework that helps experts better understand why the model classifies a signal as a planet or FP. 

We perform extensive studies to evaluate the performance of \ExoMiner\ on different datasets and under different imperfect data scenarios. Our analyses show that \ExoMiner\ is a highly accurate and robust classifier. We use \ExoMiner\ to validate 301 new exoplanets from a subset of KOIs that are not confirmed planets nor certified as FPs to date.

\section{Methodology}

\subsection{Machine Classification}
\label{sec:machine-classification}

Belonging to the general class of supervised learning in ML and statistics, machine classification refers to the problem of identifying a class label of a new observation by building a model based on a pre-labeled training set. Let us denote by $x$ and $y$ the instance (e.g., a transit signal) and its class label (e.g., exoplanet or false positive), respectively. $x$ and $y$ are related through function $f$ as $y=f(x)+\epsilon$ with noise $\epsilon$ having mean zero and variance $\sigma^2$. Provided with a labeled training dataset $\MD={(x_1, y_1),\;(x_2, y_2),\;...(x_N, y_N)}$, the task of classification in ML is focused on learning $\hat{f}(x,D)$ that approximates $f(x)$ to generate label $y$ given instance $x$.

A good learner, discriminative or generative, should be able to learn from the training set and then generalize well to unseen data. In order to generalize well, a supervised learning algorithm needs to trade-off between two sources of error: bias and variance~\citep{Bishop:2006:PRM:1162264}. To explain this, note that the expected error of $\hat{f}(x,\mathcal{D})$ when used to predict the output $y$ for $x$ can be decomposed as~\citep{Bishop:2006:PRM:1162264}:
\begin{align}
\ME_\MD\big[y-\hat{f}(x,\MD)\big]&=\big(\ME_\MD[\hat{f}(x,\MD)]-f(x)\big)^2\\
\nonumber
&+\ME_\MD[\ME_\MD[\hat{f}(x,\MD)]-\hat{f}(x,\MD)^2]+\sigma^2
\end{align}
where the expectation $\ME$ is taken over different choices of training set $\MD$.  The first term on the right hand side, i.e., $\big(\ME_\MD[\hat{f}(x,\MD)]-f(x)\big)^2$, is called bias and refers to the difference between the average prediction of the model learned from different training sets and the true prediction. A learner with high bias error learns little from the training set and performs poorly on both training and test sets; this is called underfitting. The second term on the right hand side of this equation, i.e., $\ME_\MD[\ME_\MD[\hat{f}(x,\MD)]-\hat{f}(x,\MD)^2]$, is called variance and is the prediction variability of the learned model when the training set changes. A learner with high variance overtrains the model to the provided training set. Therefore, it performs very well on the training set but poorly on the test set; this is called overfitting. The last term, i.e., $\sigma^2$, is the irreducible error.  

A successful learner needs to find a good trade-off between bias and variance by understanding the relationship between input and output variables from the training set. Such learners are able to train models that perform equally well for both training and unseen test sets. There are three factors that determine the amount of bias and variance of a model: 1) the size of the training set, 2) the size of $x$ (its complexity or number of features), and 3) the size of the model in terms of the number of parameters. Complex models with many learnable variables, relative to the size of training set, can always be trained to perform very well on the training set but poorly on unseen cases. Different regularization mechanisms are designed in ML to reduce the complexity of the model automatically when there are not enough training data. This includes 1) early stopping of the learning process by monitoring the performance of the model on a held-out part of the training set, called validation set, and 2) penalizing the effective number of model parameters in order to reduce the model complexity. We will utilize both in this work.  
 
\begin{figure*}[ht!]
	\centering
	\subfigure[A single neuron]{\label{fig:neuron}\includegraphics[width=55mm]{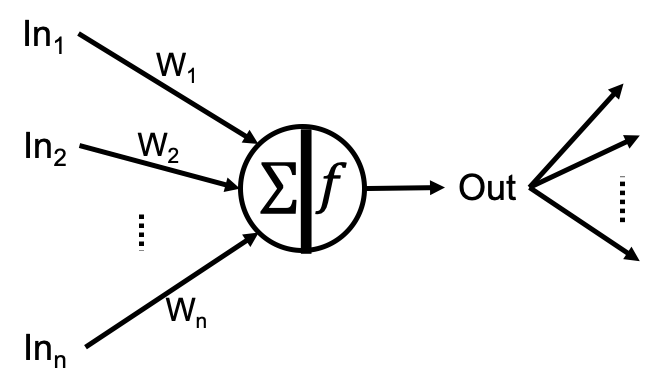}}
	\hskip 0.2in
	\subfigure[FC and dropout layers]{\label{fig:FClayer}\includegraphics[width=55mm]{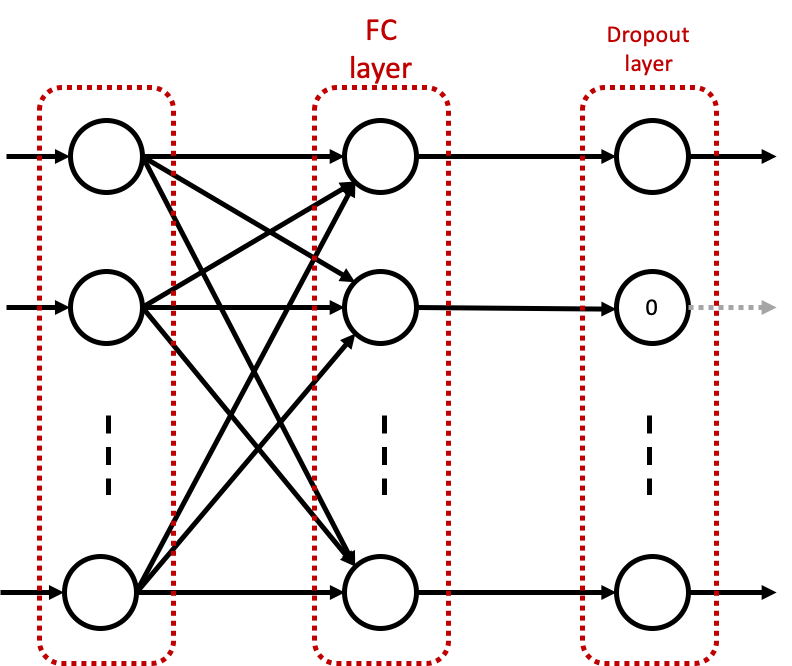}}
	\subfigure[Convolutional and pooling layers]{\label{fig:convolutionallayer}\includegraphics[width=55mm]{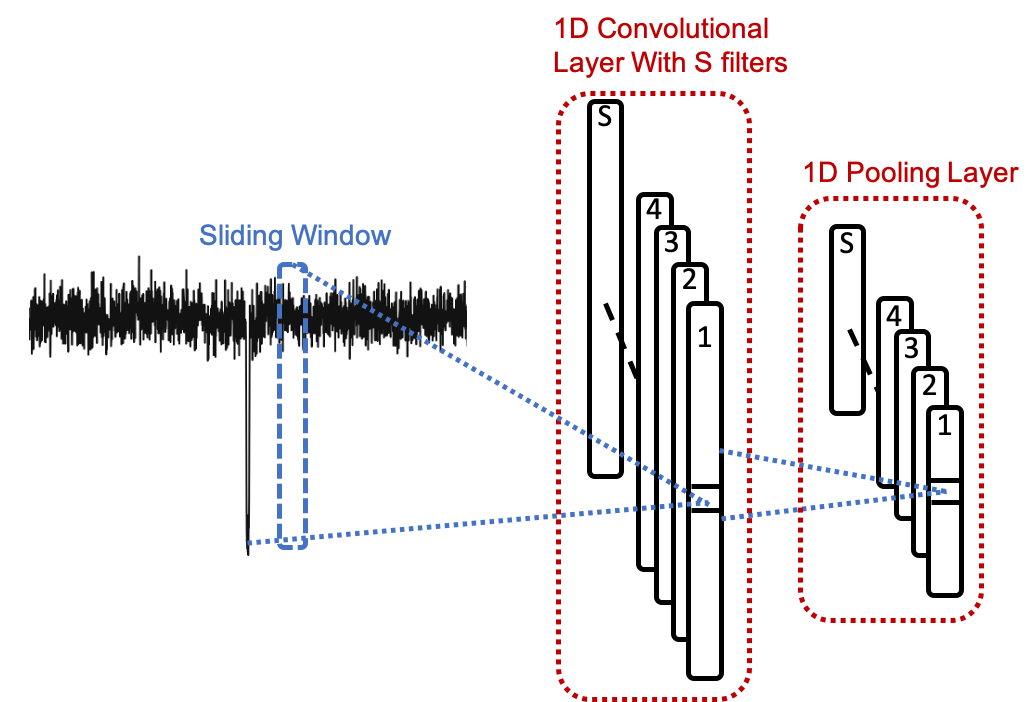}}
	\caption{Main elements of the NN used in this work.} 
\label{Convolutional-concepts}
\end{figure*}

\newpage
\subsection{Deep Neural Networks}
\label{sec:DNNs}

Generally considered as discriminative models~\citep{NG-2002}, Neural Networks (NNs) are nonlinear computing systems that can approximate any given function~\citep{Bishop:2006:PRM:1162264}.  An NN consists of many non-linear processing units, called neurons. A typical neuron computes a weighted sum of its inputs and then passes it through a non-linear function, called an activation function (Figure~\ref{fig:neuron}), to generate its output:
\begin{equation}
O=f\left(\sum_i w_iI_i\right).
\label{eq:neuron}
\end{equation}
 
\begin{figure}[htb!]
	\centering
	\subfigure[]{\label{fig:convolutionalbranch}\includegraphics[width=27mm]{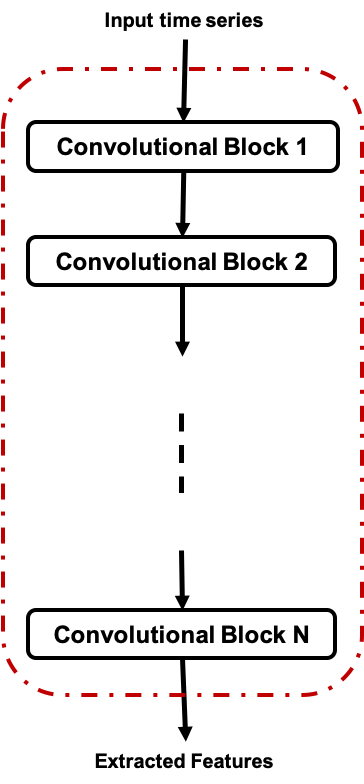}}
	\subfigure[]{\label{fig:convolutionalblock}\includegraphics[width=27mm]{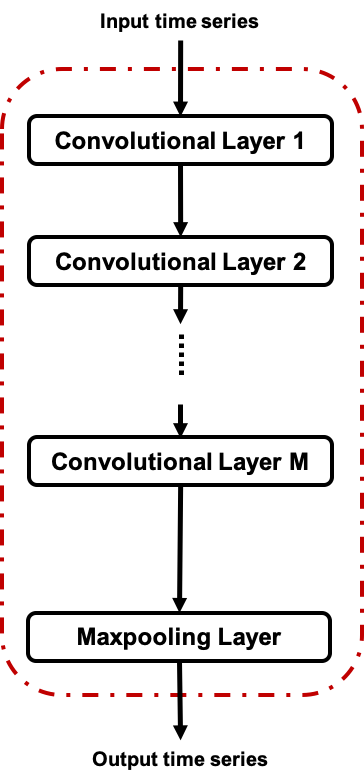}}
	\subfigure[]{\label{fig:FClblock}\includegraphics[width=27mm]{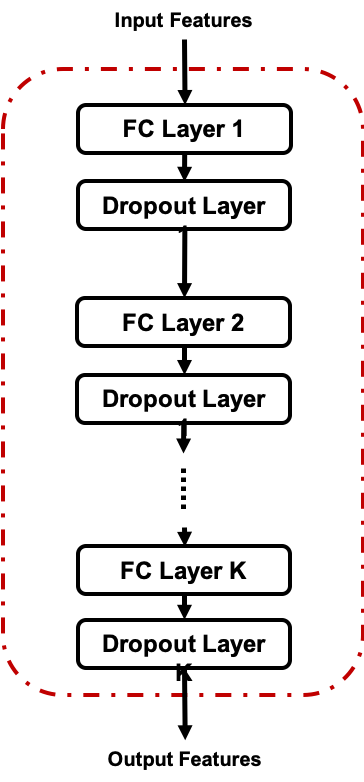}}
	\caption{a) Each convolutional branch consists of N convolutional blocks, b) each convolutional block consists of M convolutional layers and one maxpooling layer, and c) the FC block consists of multiple FC layers, each followed by a dropout layer.} 
\vskip -0.2in
\label{fig:Convolutional-elements}
\end{figure}

Neurons in an NN are organized in layers. In the most common NNs, neurons in each layer are of a similar type, and the output of neurons in one layer is the input to the neurons in the next layer. Neuron types are characterized by the restrictions on their weights (how many non-zero weights) and the type of activation function, $f$ (Equation~\ref{eq:neuron}), that they use. In this work, we use the following layers: 
\begin{enumerate}[noitemsep,nolistsep,partopsep=1pt,topsep=1pt, leftmargin=4mm]
\item{Fully Connected (FC) layers:} Shown in Figure~\ref{fig:FClayer}, a neuron in an FC layer receives the outputs of all neurons from the previous layer (no restriction on the weights). We use the Parametric Rectified Linear Unit (PReLU) activation function for the neurons in the FC layer because it is well behaved in the optimization process. PReLU, defined as follows, is a variation of the ReLU function ($f(x)=x^+=max(0,x)$) that avoids the dying neuron\footnote{When using ReLU function, some neurons get stuck on the negative side because the slope of ReLU is zero on that side. A neuron stuck in the negative side is called a dying neuron.} problem:
\begin{equation}
  f(x) =
    \begin{cases}
      ax & \text{if $x<0$}\\
      x  & \text{if $x\geq0$},\\
    \end{cases}       
\end{equation}
where $a$ is a parameter learned from the data.

\item{Dropout layers:} Shown in Figure~\ref{fig:FClayer}, these layers are a form of regularization designed to prevent overfitting. The number of neurons in this layer is equivalent to the number of neurons from the previous layer. This layer is designed to drop a percentage of processing neurons (based on a rate called the dropout rate) in the previous layer to make the model training noisier (by not optimizing all weights in each iteration\footnote{This is the essence of any regularization technique.}) and so more robust. 
\item{One dimensional (1-d) convolutional layers:} As shown in Figure~\ref{fig:convolutionallayer}, the neurons in convolutional layers, which are called filters, operate only on a small portion of the output of the previous layer (i.e., only a subset of weights are non-zero in Equation~\ref{eq:neuron}). The neurons in a convolutional layer extract features locally using small connected regions of the input. They are designed specifically for temporal or spatial data where there is locality information between inputs. Each neuron extracts features locally on many patches of the input (e.g., sliding windows on a time series).
Filters in a convolutional layer produce a feature map from the previous layer. Similar to the FC layer, we use Parametric ReLU as the activation function for the neurons in this layer.
\item{Pooling layers:} Shown in Figure~\ref{fig:convolutionallayer}, pooling layers are a form of regularization and designed to downsample feature maps generated by the convolutional layers. These layers perform the same exact operation over patches of the feature map. Normally, the operation performed by the pooling layer is fixed and not trainable. The maximum and mean are the most common operations used in pooling layers. In this paper, we use one dimensional maxpooling layers that calculate the maximum values of sliding windows over the input. 
\item{Subtraction layers:} These layers receive two sets of input variables and perform element-wise subtraction of one set from the other. The weights in this layer are fixed and each neuron has two inputs: the weight for one is $+1$ and for the other one is $-1$. 
\end{enumerate}
A DNN is an NN with a larger number of layers. Each layer of neurons extracts features on top of the features extracted from the previous layer.  A set of layers combined together to follow a specific purpose is called a block. In this paper, we use two types of blocks: a convolutional block that consists of one or multiple convolutional layers with a final maxpooling layer (Figure~\ref{fig:convolutionalblock}), and an FC block that consists of one or more FC layers, each followed by a dropout layer (Figure~\ref{fig:FClblock}). A DNN with convolutional layers is called a CNN \citep[][]{LeCun-1989-CNN}.

\begin{figure*}[htb!]
\begin{center}
\centerline{\includegraphics[width=\textwidth]{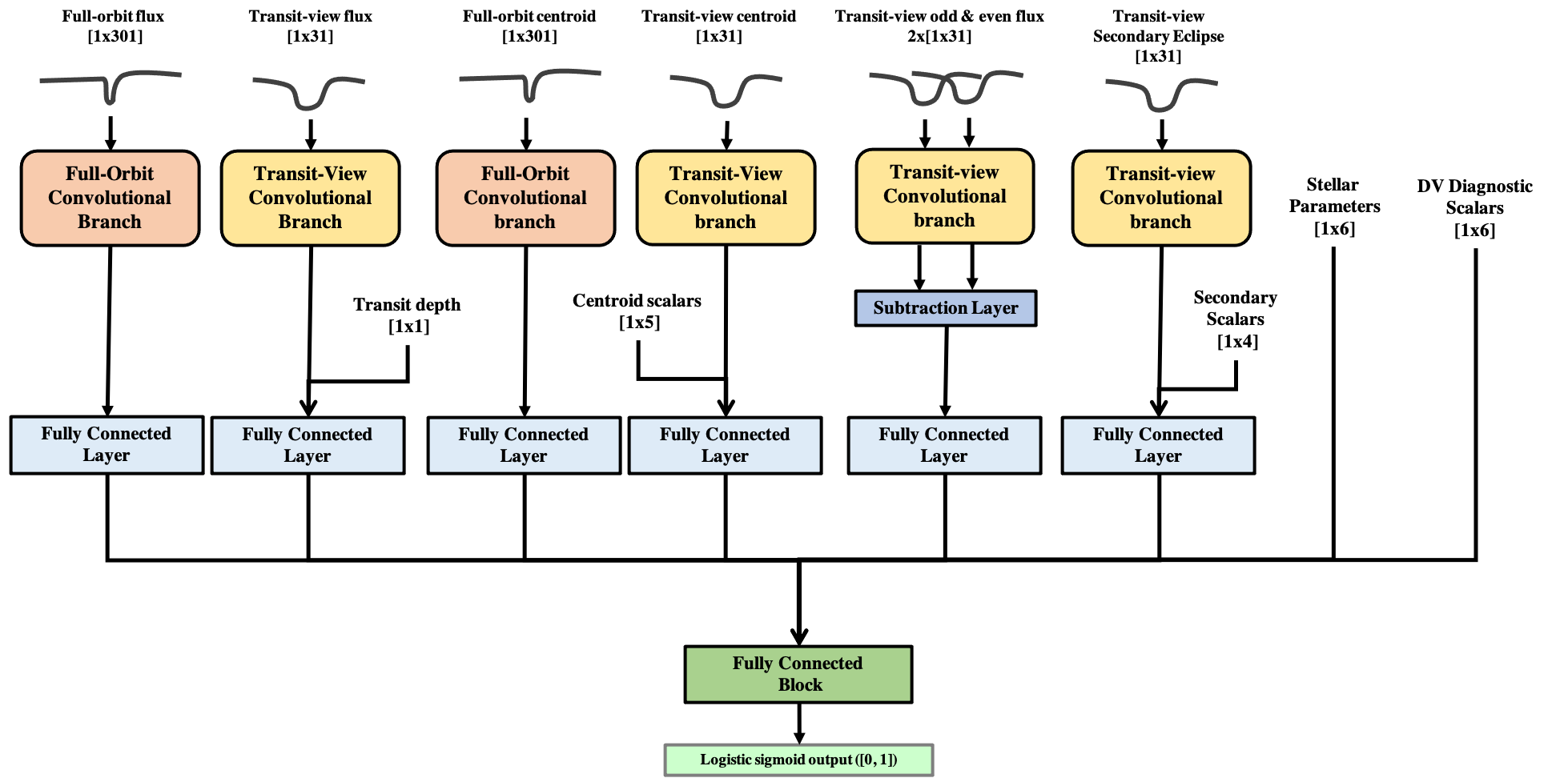}}
\caption{ \ExoMiner\ architecture.}
\label{fig:dnn-architecture}
\end{center}
\end{figure*}

\subsection{Explainability of DNNs}
\label{sec:explainability}
Explainability remains a central challenge for adopting and scaling DNN models across all domains and industries. Various perturbation- and gradient-based methods have been developed for providing insight into the predictions made by DNN classifiers. 

Occlusion sensitivity mapping is among the simplest of these methods, and it provides a model agnostic approach for localizing regions of the input data that have the greatest impact on a model\textsc{\char13}s classifications~\citep{Zeiler-2014-explainability}. This region can then be considered in the context of the classification problem to illuminate how the model makes its predictions. For time series data, occlusion sensitivity mapping involves systematically masking the points of the input data that fall within a sliding window and analyzing the changes in the model’s predictions as different sliding windows are occluded. 

Noise sensitivity mapping is a similar perturbation-based explainability method where noise is added to the points of the input data that fall within each sliding window, and a heat map is generated to identify the input regions where the addition of noise has the greatest effect on the model’s predictions~\citep{greydanus-2018-visualizing}. The advantage of this technique is that no sharp artifacts or edges that may bias the classification are introduced into the input data when the sliding window is applied. 

Finally, Gradient-weighted Class Activation Mapping (Grad-CAM) uses the gradients of the final convolutional layer, rather than the input data, to develop class-specific heat maps of the regions most important to the model’s classifications~\citep{Selvaraju_2019_explainability}. Compared to perturbation-based explainability methods, Grad-CAM is much less computationally expensive because it does not require repeatedly making predictions as the sliding window is moved, and it eliminates the subjectivity of choosing the size of the sliding window.

\subsection{Proposed DNN Architecture}
\label{sec:proposed-dnn}

To specify a DNN model, one needs two sets of parameters: 1) parameters related to the general architecture of DNNs (e.g., how to feed inputs to the model, the number of blocks and layers, the type of layers/filters, etc.). These parameters are usually either decided using domain knowledge, hyper-parameter optimization tools, or a combination of both; and 2) the particular weights between neurons. The weights are variables optimized using a learning algorithm that optimizes a data-driven objective function \citep[using e.g., gradient descent --][]{Bishop:2006:PRM:1162264}. The parameters in the first set and those related to learning (such as batch size, type of optimizer, and loss function) are called hyper-parameters and must be fixed prior to learning the second set.

We utilize domain knowledge to preprocess and build representative data to feed to our DNN model and also to decide about its general architecture. Given that an expert vets a transit signal by examining its DV report or similar diagnostics, we will use unique components of the DV report as inputs to our DNN model. This includes full-orbit and transit-view flux data, full-orbit and transit-view centroid motion data, transit-view secondary eclipse flux data, transit-view odd \& even flux data, stellar parameters, optical ghost diagnostic (core and halo aperture correlation statistics), bootstrap false alarm probability (bootstrap-pfa), rolling band contamination histogram for level zero (rolling band-fgt), orbital period, and planet radius~\citep{Twicken_2018_DV}. We also include scalar values related to each test to make sure the test is effective. These scalar values include: 
\begin{enumerate}[noitemsep,nolistsep,partopsep=0pt,topsep=0pt, leftmargin=4mm]
\item Secondary event scalar features: geometric albedo comparison (Section~\ref{sec:manual-vetting}) and planet effective temperature comparison statistics (Section~\ref{sec:manual-vetting}), weak secondary maximum MES, and weak secondary transit depth. 
\item Centroid motion scalar features: flux-weighted centroid motion detection statistic, centroid offset to the target star according to the \kepler\ Input Catalog~\citep[KIC;][]{Brown-2011-KIC}, centroid offset to the out-of-transit centroid position (OOT)\footnote{The out-of-transit centroid position for the DV OOT centroid offsets is determined by a Pixel Response Function (PRF) fit to the time-averaged, background-corrected pixel values in the vicinity of the observed transit(s) for the given quarter (or sector).}, and the respective uncertainties for these two centroid offsets. These scalar values provide complementary information to the transit and full-orbit views of the centroid motion time series, which are helpful for correctly classifying false positives (mainly BEBs). 
\item Transit depth, which is important to understand the size of the potential planet. The depth information is lost during our normalization of the flux data\footnote{Similar to~\citep{shallue_2018, Ansdell_2018, armstrong-2020-exoplanet}}. We explain the details of how we create these data inputs in Section~\ref{sec:datainput}. 
\end{enumerate}

\begin{figure}[htb!]
\begin{center}
\centerline{\includegraphics[width=0.8\columnwidth]{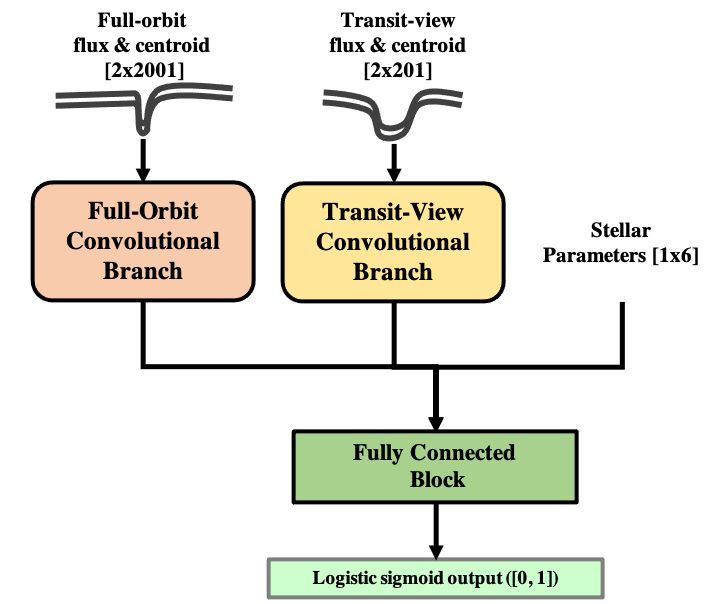}}
\caption{\ExoNet: DNN architecture used in~\citep{Ansdell_2018}.}
\label{fig:fdl-dnn-architecture}
\end{center}
\end{figure}

The general architecture of the proposed DNN model is depicted in Figure~\ref{fig:dnn-architecture} in which we use a concept called convolutional branch to simplify the discussion. A convolutional branch (Figure~\ref{fig:convolutionalbranch}) consists of multiple convolutional blocks (Figure~\ref{fig:convolutionalblock}). After each convolutional branch (yellow and orange boxes for full-orbit and transit views, respectively), an FC layer (light blue boxes) is utilized to combine extracted features and scalar values related to a specific diagnostic test to produce a final unified set of features. These final sets of features are then concatenated with other scalar values (stellar parameters and other DV diagnostic tests) to feed into an FC block (dark green box) that extracts useful, high-level features from all these diagnostic test-specific features. The results are fed to a logistic sigmoid output to produce a score in $[0,1]$ that represents the DNN's likelihood that the input is a PC. A default threshold value of 0.5 is used to determine the class labels but can be changed based on the desired level of precision/recall. Below, we enumerate some more details about the \ExoMiner\ architecture:
\begin{itemize}
\item The sizes of the inputs to the full-orbit and transit-view convolutional branches are 301 and 31, respectively. This is different from the previous works~\citep{shallue_2018, Ansdell_2018, armstrong-2020-exoplanet} that use input sizes of 2001 and 201. Our design is inspired by the binning scheme utilized in the DV reports, which yields smoother plots. Note that the binning for the full-orbit and transit views on the DV 1-page summary reports employs five bins per transit duration (width = duration $\times~0.2$). Although this generates 31 bins for transit-views of most TCEs, this may produce many more bins than 301 for the full-orbit view. However, we fixed the number to 301 because the size of the input to the classifier must be the same for all TCEs.
\item Inspired by the design of the DV reports, which only include transit-views for the secondary and odd \& even tests, we only used the transit-views for these tests; note that the full-orbit view does not provide useful information for these diagnostic tests. 
\item  Unlike \ExoNet, which stacked the flux and centroid motion time series into different channels of the same input and fed them to a single branch (Figure~\ref{fig:fdl-dnn-architecture}), we feed time series related to each test to separate branches because the types of features extracted from each diagnostic test are unique to that test. 
\item For the odd \& even transit depth diagnostic test, we use the odd and even views as two parallel inputs to the same processing convolutional branch. This design ensures that the same exact features are extracted from both views. The extracted features from the two views are fed to a non-trainable subtraction layer that subtracts one set of features from the other. The use of a subtraction layer is inspired by the fact that domain experts check the odd and even views for differences in the transit depth and shape. 
\item The FC layer after each convolutional branch is designed with two objectives in mind: 1) it allows us to merge the extracted features of a diagnostic time series with its relevant scalar values, and 2) by setting the same number of neurons in this layer, we ensure that the full-orbit and transit-view branches have the same number of inputs to the FC block. This is different from~\citep{shallue_2018, Ansdell_2018} in which the output sizes of the full-orbit and transit-view convolutional branches are different by design. 
\end{itemize}

So far, we utilized domain knowledge to design the general structure of our DNN. However, the total number of convolutional blocks per branch, the number of convolutional layers per block, and the number of FC layers in the FC block are hyper-parameters of the model that need to be decided prior to training the model. Other hyper-parameters include the kernel size for each convolutional layer, the number of kernels for the convolutional layers, the number of neurons for the FC layers, the type of optimizer, and the learning rate (step size in a gradient descent algorithm). We use a hyper-parameter optimizer called BOHB~\citep[Bayesian Optimization and HyperBand,][]{hpo-Falkner-2018} to set these parameters prior to training.

The above process of combining domain knowledge and ML tools, such as hyper-parameter optimization, provides us with a DNN model architecture that can be trained in a data-driven approach to classify transit signals. In a sense, this is similar to \Robovetter, which leverages domain knowledge in order to design the general classification process in the form of if-then rules with if-then condition threshold variables that are partially optimized in a data-driven approach. The difference is the amount of automation in the process. 
\Robovetter\ relies fully and comprehensively on the domain knowledge not only for feature extraction (the results of diagnostic tests) and classification rules (if-then conditions) but also to set the values for most of the variables. By contrast, our ML process limits the use of domain knowledge to the choice of inputs, the preprocessing of those inputs, and the general architecture of the DNN. Then, by optimizing many different aspects of the architecture through hyper-parameter optimization and the optimization of the model's connection weights in a data-driven training process, our DNN learns useful features from raw diagnostic tests to classify transit signals. 

The complete ML process ensures that the model is able to correctly classify the training data and also generalize to future unseen transit signals. This results in a machine classifier that can be quickly and easily used to classify other datasets (e.g., a model learned on \kepler\ data can be used to classify transit-like signals in TESS data). Overall, \Robovetter\ is much closer to the manual vetting process than the machine classifier presented in this paper.

\begin{algorithm}[htb!]
\SetAlgoLined
	\KwInput{Time Series with Colored Noise}
	\KwOutput{Detrended Time Series }
	\caption{Detrending: Removing Colored Noise}
  \begin{algorithmic}[1]
	\STATE Remove missing values from the time series.
	\STATE Split the array into new arrays if the time interval between adjacent cadences is larger than 0.75 days. 
	\STATE Remove the transits using adjusted padding $\min\Big(3\times transit\;duration,\; \lvert weak\;secondary\;phase\rvert,\; period\Big)$ around the center of the transits and then linearly interpolate across the transits\footnote{This ensures that we remove all the transits for the primary TCE while minimizing changes to the rest of the time series (e.g., deformation of a potential secondary event).}. 
    \STATE Fit a spline for each gapped and linearly interpolated array.
	\STATE Generate detrended arrays by dividing each array in the original time series of step 2 by the fitted spline.
	\STATE Return the detrended arrays.
  \end{algorithmic}
   \label{alg:whitening}
\end{algorithm}

\subsection{Data Preparation}
\label{sec:datainput}
In this section, we describe the steps performed to build the diagnostic tests required as inputs to the proposed DNN in Figure~\ref{fig:dnn-architecture}. The process includes the following two major steps:  
\begin{itemize}
\item In the first step, we create each diagnostic test time series from the Presearch Data Conditioning flux or the  Moment of Mass centroid time series in the Q1-Q17 DR25 light curve FITS files~\citep{Thompson-2016-handbook} available at the Mikulski Achive for Space Telescopes (MAST)\footnote{https://archive.stsci.edu}. This process consists of: 1) a preprocessing and detrending step (Algorithm~\ref{alg:whitening}) that removes low-frequency variability usually associated with stellar activity while preserving the transits of the given TCE and 2) a phase-folding and binning step for each test in order to generate full-orbit and transit views (Algorithm~\ref{alg:phasefolding}). Phase-folding uses the period signature of the signal to convert the time series to the phase domain, while binning creates a more compact and smoother representation of the transit signal by averaging out other signals present in the time series that are not close to the harmonic frequencies. Both of these steps are similar to~\citep{shallue_2018}; however, there are some minor but important differences in our algorithms. In detrending, we use a variable padding as a function of the period and duration of the transit to make sure we remove the whole transit prior to the spline fit (Algorithm~\ref{alg:whitening}). In phase-folding and binning, we use fewer bins for the transit and full-orbit views (31 and 301 instead of 201 and 2001, respectively) to obtain smoother time series. Similar to~\citep{shallue_2018}, we use  ($bin width=duration\times 0.16$) for the transit-views\footnote{The bin width of 0.16 is obtained in~\citep{shallue_2018} through hyper-parameter optimization.}. However, for the full-orbit view we adjust the bin width to make sure the bins cover the whole time series and converge to the bin width of the transit-views when possible (Algorithm~\ref{alg:phasefolding}).

\item In the final step, we rescale different diagnostic tests to the same range to improve the numerical stability of the model and reduce the training time. The rescaling also helps to ensure that different features have an equal and unbiased chance in contributing to the final output; however, if the range of values provides critical information, rescaling leads to the loss of information. In what follows, we introduce rescaling treatments specific for different time series data and provide a solution for cases where the rescaling leads to the loss of some critical information. Algorithm~\ref{alg:scalarnormalization} describes the general normalization approach for most scalar features that follow a normal distribution. As we explain later in this section, we introduce a different normalization when the normal distribution assumption does not hold for a given scalar feature. 
\end{itemize}
 
Here, we detail the specific preprocessing steps for each diagnostic test fed to the DNN:

\begin{itemize}
\item{\textit{Full-orbit and transit-view flux data:}} We first perform standard preprocessing in Algorithm~\ref{alg:whitening} on the flux data and then run Algorithm~\ref{alg:phasefolding} to generate both full-orbit and transit views. To normalize the resulting views of each TCE, we subtract the median $med_{f}=median(flux)$ from each view and then divide it by the absolute value of the minimum: $min_f=\lvert min(flux)\rvert$. This generates normalized views with median 0 and minimum value -1 (fixed depth value for all TCEs). As a result of this re-scaling, the depth information of the transit is lost. Thus, we add the transit depth as a scalar input to the FC layer after the transit-view convolutional branch (Figure~\ref{fig:dnn-architecture}). We normalize the transit depth values using Algorithm~\ref{alg:scalarnormalization}.

\item{\textit{Transit-view of secondary eclipse flux:}} First, we use the standard preprocessing outlined in Algorithm~\ref{alg:whitening} on the flux data. We then remove the primary transit by using a padding size that ensures the whole primary is removed without removing any part of the secondary using the following padding strategy: 
if the $transit\; duration<\lvert secondary\;phase\rvert$, we remove $3\times transit\; duration$ around the center of the primary transit. Otherwise, we take a more conservative padding size of $\lvert secondary\;phase\rvert + transit\;duration$, which removes the primary up to the edge of the secondary event. This padding strategy is large enough to make sure that the whole transit is removed, however it is not so large that it removes part of the secondary. Also note that this padding is different from the one we used in Algorithm~\ref{alg:whitening}; here, we aim to gap the primary without damaging the secondary event. For spline fitting in Algorithm~\ref{alg:whitening}, we aim to make sure we completely gap the transits associated with the TCE. 

\begin{algorithm}
	\KwInput{Original Time Series}
	\KwOutput{Phase-Folded Time Series }
	\caption{Generating Full-orbit and Transit views\footnote{Note that the bins may or may not overlap depending on the transit duration and period.}}
  \begin{algorithmic}[1]
	\STATE Fold the data over the period for each TCE with the transit centered.
	\STATE Use 301 bins of $width=\max \left (period/301, transit\;duration \times 0.16\right )$ to bin data for full-orbit view. 
	\STATE Use 31 bins of $width=transit\;duration*.16$ to bin data for transit-view. The transit-view is defined by $5\times transit\;duration$ centered around the transit.
	\STATE Return full-orbit and transit views. 
  \end{algorithmic}
   \label{alg:phasefolding}
\end{algorithm} 

\begin{algorithm}
	\KwInput{$X$: A single scalar feature of all TCEs in training set}
	\KwOutput{$X'$: Normalized feature of all TCEs}
	\caption{Normalizing scalar feature}
  \begin{algorithmic}[1]
	\STATE Compute the median $med_X$ and standard deviation $\delta_X$ from the training set.
	\STATE Compute $X_{temp}$ by subtracting $med_X$ from values in $X$ and divide them by $\delta_X$ (i.e., $X_{temp}=\frac{X-med_X}{\delta_X}$).
	\STATE Remove the outliers from $X_{temp}$ by clipping the values outside $20\delta$ to produce $X'$.
	\STATE Return (1) $X'$ and (2) $med_X$ and  $\delta_X$ for normalizing future test data.
  \end{algorithmic}
   \label{alg:scalarnormalization}
\end{algorithm} 

We obtain the phase for the secondary from the \kepler\ Q1-Q17 DR25 TCE catalog available at the NASA Exoplanet Archive\footnote{https://exoplanetarchive.ipac.caltech.edu}. Using the phase in conjuntion with the ephmerides for the primary transit we then generate the transit-view using Algorithm~\ref{alg:phasefolding}. We normalize this view by subtracting the $med_f$ and dividing by $max(min_f, \lvert min(secondary)\rvert)$. This normalization method preserves the relative depth of the secondary with respect to the primary event. As with the flux view, we add the scalar transit depth value of the secondary as an input to the FC layer. Besides the transit depth of the secondary event, three other scalar values that help distinguish hot Jupiters from eclipsing binaries are added. These include the geometric albedo and planet effective temperature comparison statistics, and the weak secondary maximum MES. We normalize these four scalar values using Algorithm~\ref{alg:scalarnormalization}. 

\item{\textit{Transit-view of odd \& even:}} We perform standard preprocessing using Algorithm~\ref{alg:whitening} for the flux data. We then use Algorithm~\ref{alg:phasefolding} separately for the odd and even transit time series to obtain the odd and even transit-views. We normalize each odd and even time series by subtracting $med_f$ and dividing by $min_f$. This rescaling using the full flux time series rescaling parameters ensures that the relative size of the odd and even time series are preserved. 

\item{\textit{Full-orbit and transit views of centroid motion data:}}  Unlike~\citep{Ansdell_2018}, which used heuristics to compute the centroid motion time series, we follow~\citep{Jenkins_2010_kepler8b,Bryson_2013} to generate centroid motion data that preserve the physical distance (e.g., in arcseconds) of the transiting object from the target. Assuming centroids are provided in celestial coordinates $\alpha$ (R.A.) and $\delta$ (Dec.), we denote the R.A. and Dec. components of centroid at cadence $n$ by $C_\alpha(n)$ and $C_\delta(n)$, and their out-of-transit averages by $\bar{C}_\alpha^{out}$ and $\bar{C}_\delta^{out}$, respectively. If the observed depth of the transit is $d_{obs}$, the location of the transit source can be approximated as:

\begin{align}
&\alpha_{transit}(n)=\bar{C}_\alpha^{out}-\left( \frac{1}{d_{obs}}-1\right)\frac{C_\alpha(n)-\bar{C}_{\alpha}^{out} }{\cos{\delta_{target}}}\nonumber,\\
&\delta_{transit}(n)=\bar{C}_\delta^{out}-\left( \frac{1}{d_{obs}}-1\right)\left(C_\delta(n)-\bar{C}_\delta^{out}\right). 
\label{eq:transit-location}
\end{align}

By subtracting the location of the target stars, i.e., $\alpha_{target}(n)$ and $\delta_{target}(n)$, from the location of the transit source, we obtain the amount of shift for the R.A. and Dec. components of the centroid motion for each cadence as follows:
\begin{align*}
\Delta_\alpha(n)=&\left(\alpha_{transit}(n)-\alpha_{target}(n)\right)\cos{\delta_{target}}\;, \\
\Delta_\delta(n)=&\delta_{transit}(n)-\delta_{target}(n).
\end{align*}
The overall centroid motion shift can be computed as:
\begin{align}
D(n)=&\sqrt{\Delta_\alpha^2(n)+\Delta_\delta^2(n)}.
\label{eq:overallmotion}
\end{align}

Algorithm~\ref{alg:centroid} summarizes the steps performed to obtain the final centroid motion time series used to feed into the DNN. To normalize this resulting motion time series for each transit, we subtract the median of the view (centering time series) and divide by the standard deviation across the training set.  

This specific normalization ensures that the relative magnitudes of different centroid times series, which provide critical information for diagnosing BEBs, are preserved. Centering the time series makes it easier for the DNN to find relevant patterns; however, it leads to the loss of the centroid shift related to the target star position. Thus, we feed the centroid offset information generated in DV to the FC layer after the transit-view convolutional branch (Fourth branch in Figure~\ref{fig:dnn-architecture}). The centroid shift scalar features utilized in this branch are the flux-weighted centroid motion detection statistic, difference image centroid offset from KIC position and OOT location, and associated uncertainties. 

\begin{algorithm}
	\KwInput{Raw Centroid Time Series}
	\KwOutput{Processed Centroid Time Series}
	\caption{Constructing Centroid Motion Full-Orbit and Transit Views}
  \begin{algorithmic}[1]
	\STATE 	To obtain $C_\alpha$ and $C_\delta$, convert row and column centroid time series from the CCD frame to R.A. and Dec.
	\STATE 	Compute $\bar{C}^{out}$ as the average out-of-transit centroid position. 
	\STATE  Remove colored noise in $C_\alpha$ and $C_\delta$ using Algorithm~\ref{alg:whitening}. 
	\STATE 	Multiply the whitened arrays by $\bar{C}^{out}$ to obtain $c(t)$ to recover the original scale. We perform this step because dividing by the spline fit in Algorithm~\ref{alg:whitening} removes the scale. 
	\STATE	Compute the transit source location using Equation~\ref{eq:transit-location}.
	\STATE Compute the overall centroid motion time series using Equation~\ref{eq:overallmotion}.
	\STATE Convert the overall centroid motion time series from units of degree to arcsec.
	\STATE Run Algorithm~\ref{alg:phasefolding} to obtain full-orbit and transit views.
	\STATE Normalize full-orbit and transit views by subtracting the median and dividing by the training set standard deviation.  
  \end{algorithmic}
   \label{alg:centroid}
\end{algorithm}

\item{\textit{Other DV metrics:}} Scalar features either related to stellar parameters or representing results of other diagnostic tests are another type of data provided in DV reports that experts use to classify TCEs. Similar to ~\citet{Ansdell_2018}, we use stellar effective temperature ($T_{\rm eff}$), radius ($R_*$), mass ($M_*$), density ($\rho_*$), surface Gravity ($\log g$), and metallicity ($[F_e/H]$) obtained from the Kepler Stellar Properties Catalog Update for Q1-Q17 DR25 Transit Search~\citep{mathur2016kepler} and Gaia DR2~\citep{brown2018gaia}, which revised stellar properties for \kepler\ targets in the KIC. Among several DV diagnostic metrics for Q1-Q17 DR25~\citep{Twicken_2018_DV}, we use six of them: optical ghost core aperture correlation statistic, optical ghost halo aperture correlation statistic, bootstrap-pfa, rolling band-fgt, fitted orbital period, and derived planet radius. These features are not provided in the other data inputs described above, or are lost through data preprocessing. Thus, we add them as inputs to our DNN model (referred to as DV Diagnostic Scalars in Figure~\ref{fig:dnn-architecture}) to complement the information that exists in other branches. Except for bootstrap-pfa, we normalize all scalar features using Algorithm~\ref{alg:scalarnormalization}. For bootstrap-pfa, we use Algorithm~\ref{alg:scalarnormalization} to normalize log(bootstrap-pfa) because bootstrap-pfa values do not follow a normal distribution and standardizing the raw data does not yield values in a comparable range to the other inputs. 

\item\textit{Data Correction:}
\label{sec:datacorrection}
ML algorithms are usually tolerant to some level of noise in the input/label data by resolving conflicting information (especially when a sizable amount of data is available). However, if the percentage of such noisy cases is high, the model is not able to learn the correct relationship between the input data and labels. Given that the size of our annotated data is limited and most diagnostic tests are only useful for a small subset of data, it is important that a data-driven approach receives the critical information in a correct format so that it can learn accurate concepts/patterns. Because of this, we make two corrections to the data. 

\begin{figure}[htb!]
	\centering
	\subfigure[Full-orbit view for TCE KIC~1995732.1]{\label{fig:full-orbit}\includegraphics[width=\columnwidth]{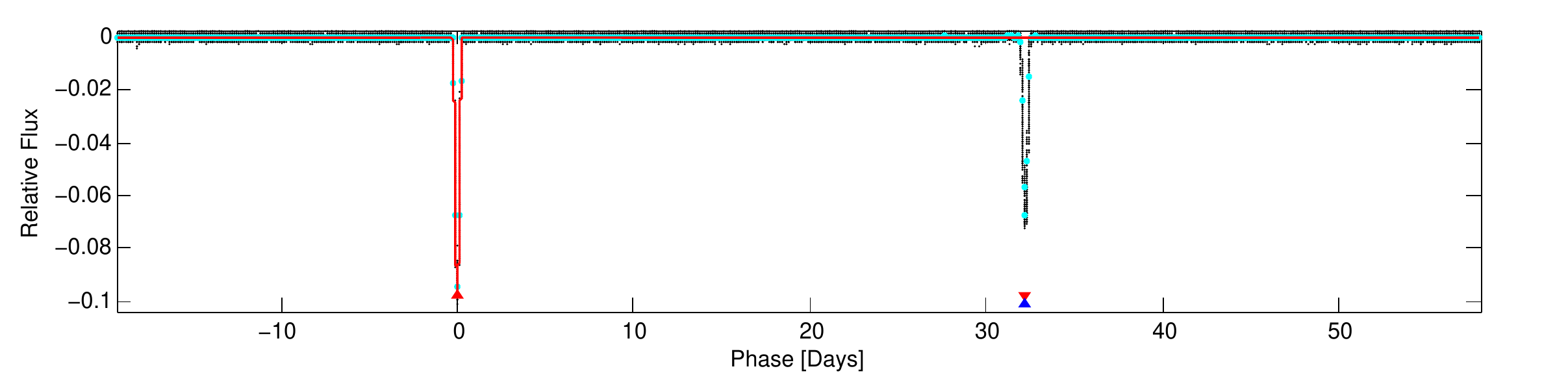}}
	\subfigure[Secondary event for TCE KIC~1995732.1]{\label{fig:firstsecondary}\includegraphics[width=\columnwidth]{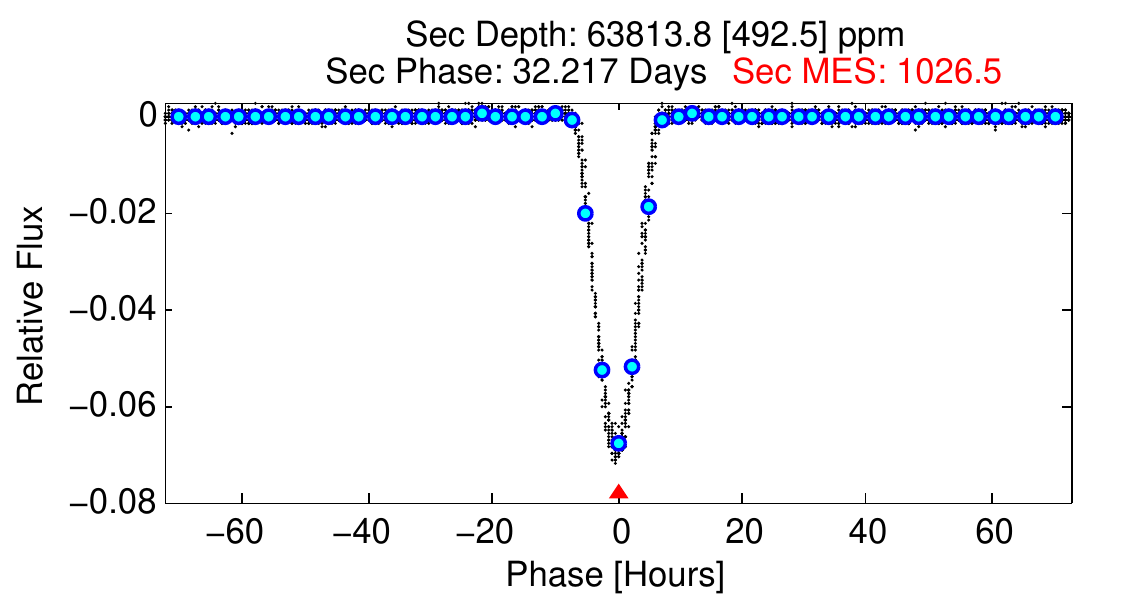}}
\hskip -0.05in
	\subfigure[Secondary event for TCE KIC~1995732.2]{\label{fig:secondsecondary}\includegraphics[width=\columnwidth]{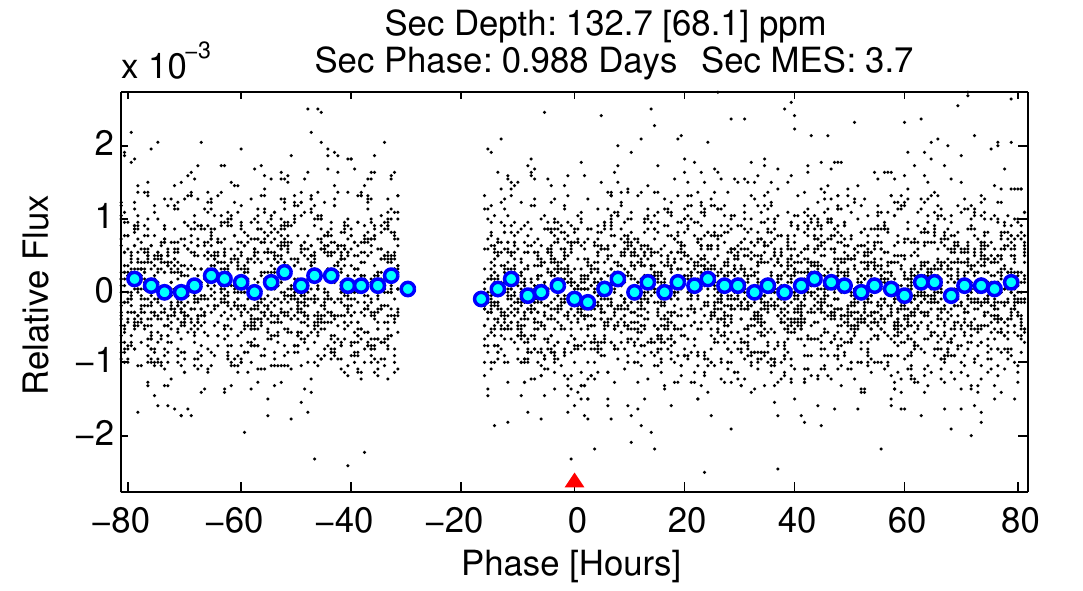}}
	\caption{KIC~1995732 -- an example of an EB with two TCEs having inconsistent secondary assignments.} 
	\label{fig:secondary-example}
\end{figure}

First, we adjust the secondary diagnostic information of a subset of TCEs whose secondary event metrics do not meet our needs. When two TCEs are detected for an EB system, we would like the secondary event to refer to the other TCE of that system. However, the pipeline removes the transits or eclipses of previously detected TCEs associated with a target star. For example, consider KIC $1995732$, which has two TCEs with similar period, 77.36 days, and are both part of the same EB system. As you can see from the full-orbit view of the first TCE of this system in Figure~\ref{fig:full-orbit}, there are two transits for this system. The secondary of the first TCE, depicted in Figure~\ref{fig:firstsecondary}, points to the second TCE. However, the pipeline removes the transits associated with the first TCE to detect the second TCE of this system, which results in a failed secondary test for the second TCE (Figure~\ref{fig:secondsecondary}). To correct the secondary information of such systems, we search the earlier TCEs of each given TCE to make sure the secondary is mutually assigned; i.e., if a TCE represents the secondary of another TCE, the secondary of that TCE should represent the first TCE as well. There were a total of 2089 such TCEs in the Q1-Q17 DR25 TCE catalog. After correcting the secondary event for such TCEs, we also corrected the geometric albedo and planet effective temperature comparison statistics, weak secondary maximum MES, and transit depth. These corrections are important because without the correct secondary event assignment, the model does not have enough information to classify these TCEs correctly. Note that the correcting procedure we explain above might fail if (1) the phase of the weak secondary or the epoch information of the two TCEs of the EB are not accurate, e.g., TCE KIC $10186945.2$ and (2) if the period of the first detected TCE is double that of the second detected TCE, e.g., TCE KIC $5103942.2$. The secondary of such eclipsing binaries are difficult for a data-driven model to classify correctly if other diagnostic tests are good. 

\begin{figure}[htb!]
	\centering
	\subfigure[Orbital period for all TCEs]{\label{fig:periodmistakes}\includegraphics[width=\columnwidth]{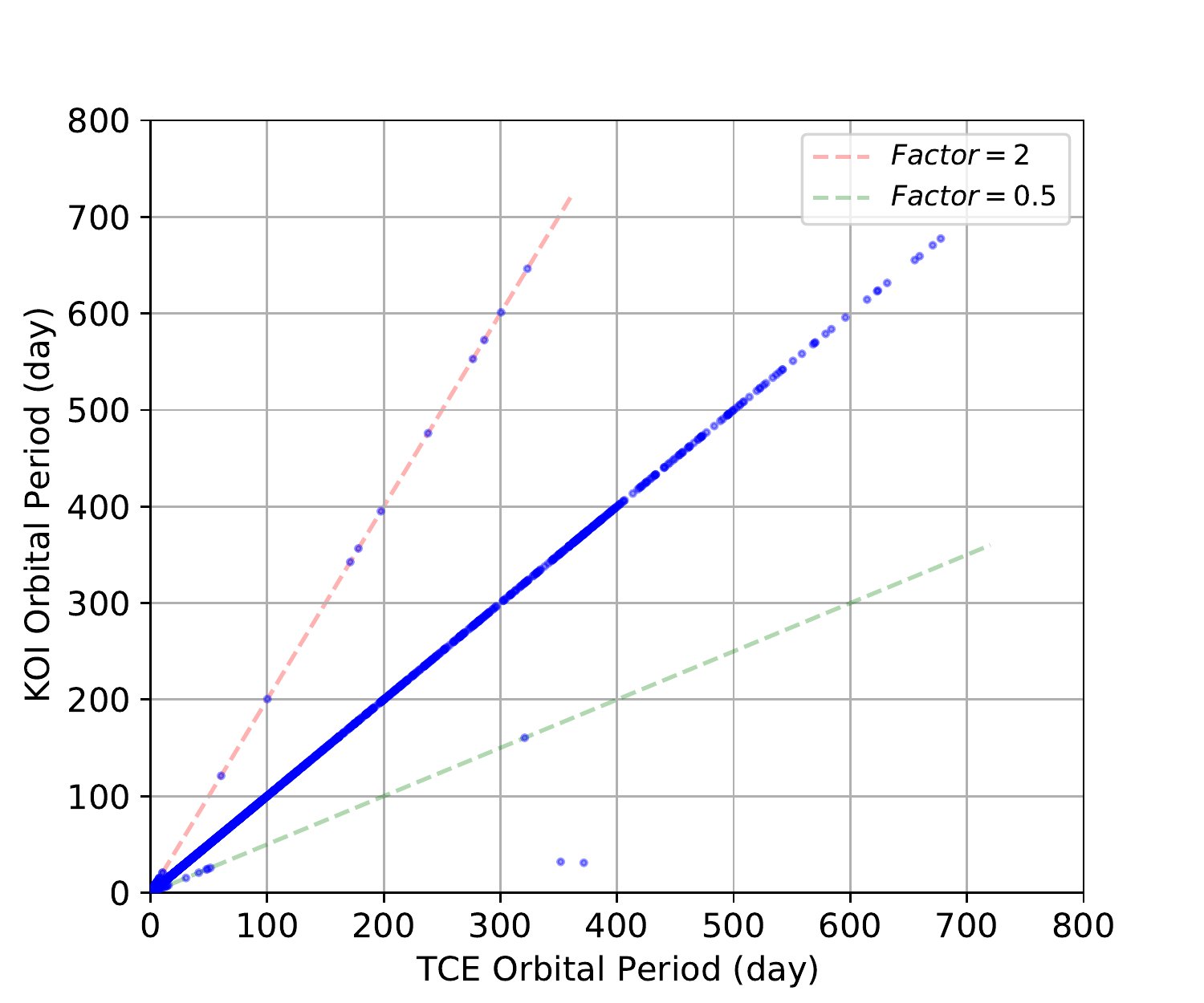}}
	\subfigure[Closeup for short-period TCEs]{\label{fig:periodmistakeszoom}\includegraphics[width=\columnwidth]{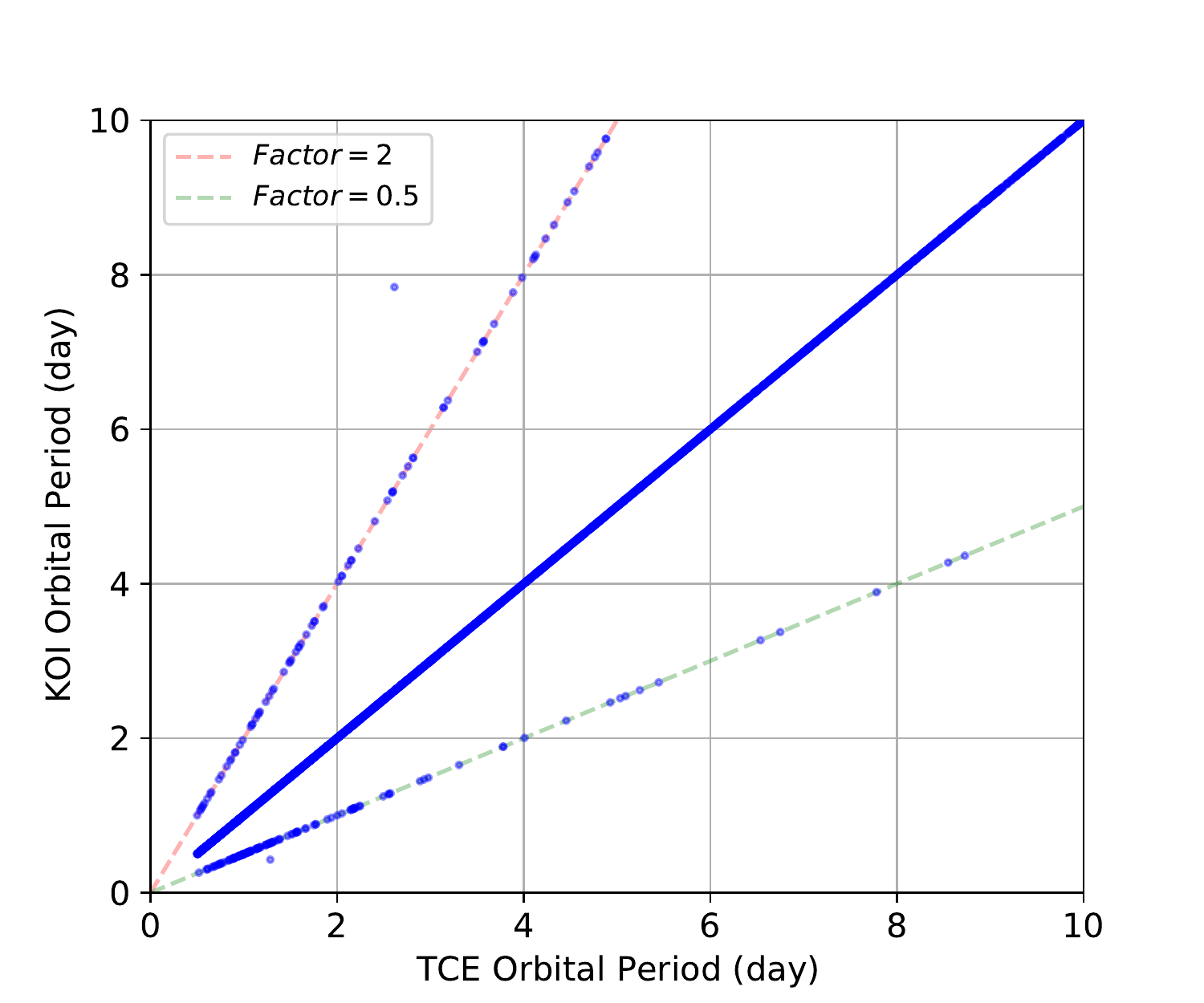}}
	\caption{Period in the Q1-Q17 DR25 TCE catalog versus Cumulative KOI catalog.} 
\label{fig:period-example}
\end{figure}

\begin{figure*}[htb!]
\begin{center}
\centerline{\includegraphics[width=\textwidth]{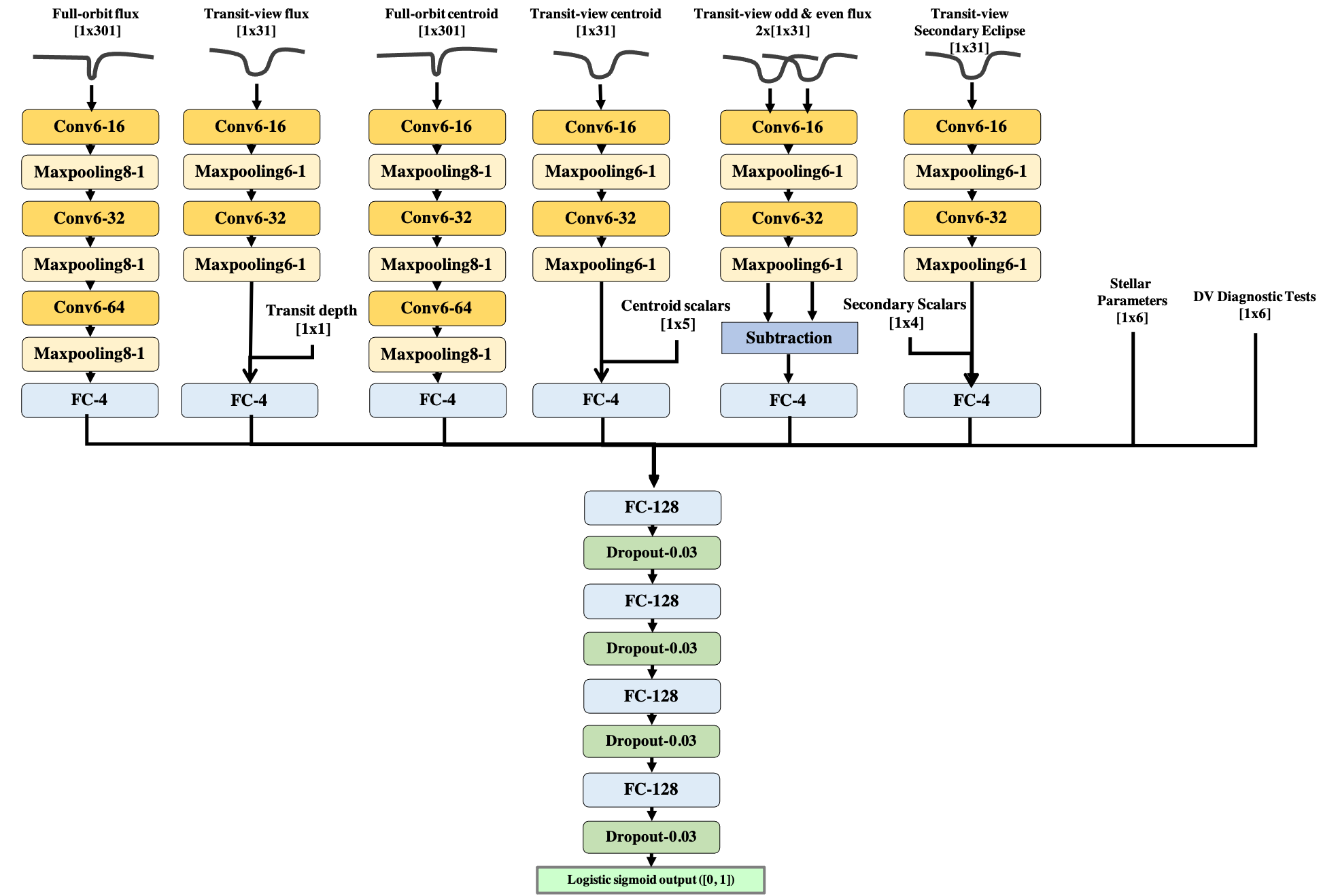}}
\caption{Optimized DNN Architecture using hyperparameter optimizer; Convolutional layers: Conv$<$kernel size$>$-$<$number of filters$>$, maxpooling layers: Maxpooling$<$pool size$>$-$<$stride size$>$, FC layer: FC$<$number of neurons$>$, Dropout layer: Dropout$<$dropout rate$>$.}
\label{fig:hpo-dnn-architecture}
\end{center}
\end{figure*}

Second, the period values listed in the Q1-Q17 DR25 TCE catalog differ from the period reported in the Cumulative KOI catalog for some TCEs, as seen in Figure~\ref{fig:period-example}. The periods reported for most of these TCEs in the Q1-Q17 DR25 TCE catalog are double or half that of the correct periods reported in the Cumulative KOI catalog. For eclipsing binaries, this does not create a problem as there are two tests created for these different scenarios: either the secondary test or odd \& even test, depending on whether the period is half or double that of the true period. For exoplanets, however, an incorrectly reported period can create an artificial secondary event or a failed odd \& even test when the period is double or half that of the true period. Examples of such scenarios are Kepler-1604~b and Kepler-458~b that are described in Section~\ref{sec:manual-vetting} (Figure~\ref{fig:wrong-period-secondary} and Figure~\ref{fig:wrong-period-oddeven}). Therefore, we corrected the period of CPs when the values differ between the Q1-Q17 DR25 TCE and Cumulative KOI catalogs. There were a total of 5 such TCEs: $3239945.1$, $6933567.1$, $8561063.3$, $8416523.1$, and $9663113.2$.
\end{itemize}

\subsection{Data-Driven Approach}
\label{sec:data-driven}

We use a data-driven approach to optimize two sets of parameters: the hyper-parameters and model parameters. 

\begin{itemize}
\item{\textbf{Optimizing hyper-parameters:}} We use the BOHB algorithm~\citep{hpo-Falkner-2018} to optimize hyper-parameters related to 1) the architecture, i.e., the number of convolutional blocks for full-orbit and transit-view branches, the number of convolutional layers for each convolutional block, the filter size in convolutional layers, the number of FC layers, the number of neurons in FC layers (one for the FC layer after the convolutional branch and one for the FC layers in the FC block), the dropout rate, the pool size, the kernel and pool strides, and 2) the training of the architecture: choice between Adam~\citep{Kingma-2014-adam} or Stochastic Gradient Descent (SGD) optimizers\footnote{Adam and SGD are two widely used stochastic gradient descent algorithms for optimizing DNNs.}, and the learning rate (step size in gradient descent). The optimized architecture found using the BOHB algorithm is shown in Figure~\ref{fig:hpo-dnn-architecture}. 
Note that we use a shared set of hyper-parameters for the full-orbit branches (flux and centroid) and another set of shared hyper-parameters for the transit-view branches (flux, centroid, secondary, and odd-even) to reduce the hyper-parameter search space. The full-orbit view branches have three convolutional blocks and the transit-view branches have two convolutional blocks each with only one convolutional layer. The FC block has four FC layers each with 128 neurons followed by a dropout layer with dropout rate=0.03. The FC layer after the convolutional branch has only 4 neurons which means that only a few features for each diagnostic test are used for classification.  
\item{\textbf{Optimizing model parameters:}} After optimizing the hyper-parameters and fixing the DNN architecture, the connection weights of the DNN are learned in a data-driven approach using the Adam optimizer with a learning rate=6.73e-05, using the Keras~\citep{chollet2015} deep learning API on top of TensorFlow~\citep{tensorflow2015-whitepaper}. 
\end{itemize}

\section{Experimental setup}
\label{sec:experimental_setup}
\subsection{Dataset}
\label{sec:dataset}
\begin{figure}[ht!]
\begin{center}
\centerline{\includegraphics[width=\columnwidth]{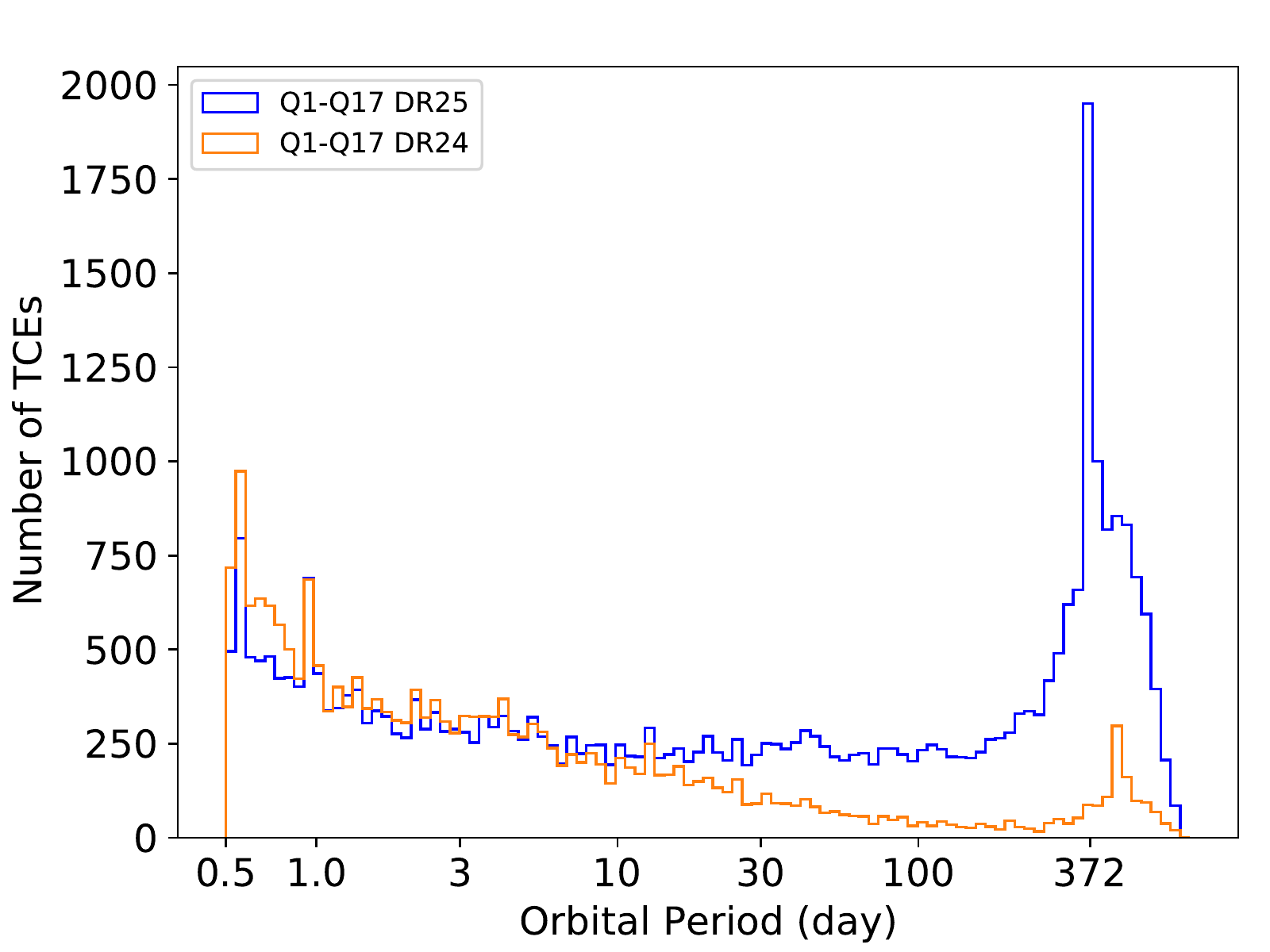}}
\caption{Histograms of orbital period for the Q1-Q17 DR25 and Q1-Q17 DR24 TCE catalogs.}
\label{fig:DR25vsDR24period}
\end{center}
\end{figure}

Previous deep learning studies for classifying Kepler\textsc{\char13}s TCEs~\citep{shallue_2018, Ansdell_2018} have used \Autovetter\ training labels for Q1-Q17 DR24 TCE catalog~\citep{Seader_2015}. As seen in Figure~\ref{fig:DR25vsDR24period}, the data distribution has changed significantly from Q1-Q17 DR24 TCE catalog to Q1-Q17 DR25 TCE catalog, and there are more TCEs with longer periods in Q1-Q17 DR25 than in Q1-Q17 DR24 because the statistical bootstrap was not used as a veto in the Transiting Planet Search (TPS) component of the \kepler\ pipeline for the former~\citep{Twicken_2016-autovetterlabels}. 
To obtain a more reliable and updated training set, similar to~\citet{armstrong-2020-exoplanet}, we utilized the Kepler Q1-Q17 DR25 TCE catalog and generated a working dataset as follows: first, we removed all rogue TCEs from the list. Rogue TCEs are three-transit TCEs that were generated by a bug in the \kepler\ pipeline~\citep{Coughlin_2016_robovetter}. For the PC category, we used the TCEs that are listed as confirmed planets (CPs) in the Cumulative KOI catalog\footnote{The Cumulative KOI catalog was downloaded in Februry 2020, so it does not include the newest CPs including those validated by~\citet{armstrong-2020-exoplanet}.}. 
For the AFP class, we used the TCEs in the Q1-Q17 DR25 list that are certified as false positives (CFP) in the \kepler\ Certified False Positive table ~\citep{Bryson-2015-certifiedlist}. According to~\citet{Bryson-2015-certifiedlist}, the CFP disposition in the CFP table is very accurate, and a change of disposition for this set is expected to be rare. For NTPs, while \Autovetter\ uses a simpler test to obtain non-transit like signals, \Robovetter\ performs a more comprehensive set of tests. Thus, for the NTP class, we combine TCEs vetted as non-transits  by \Robovetter\ (any TCE from Q1-Q17 DR25 TCE catalog that is not in the Cumulative KOI catalog) and TCEs reported as certified false alarms (CFAs) in the CFP table. 

\begin{table}[htb]
 \centering
\caption{Datasets used in this paper compared to~\citep{shallue_2018, Ansdell_2018} and~\citep{armstrong-2020-exoplanet}}
\label{table:dataset}
\begin{threeparttable}
\begin{tabularx}{\linewidth}{@{}Y@{}}
\begin{tabular}{c|p{0.22\linewidth}p{0.22\linewidth}p{0.22\linewidth}}
\toprule
Class & Our Dataset & \citet{armstrong-2020-exoplanet} &\citet{shallue_2018} and ~\citet{,Ansdell_2018} \\
\midrule
 PC  & 2291 & 2274 & 3600 \\

 AFP  & 3538 & 3100 & 9596 \\

 NTP  & 24779 & 2959 & 2541 \\
 Total  & 30609 & 8333 & 15737 \\
 \bottomrule
\end{tabular}
\end{tabularx}
\end{threeparttable}
\end{table}

Table~\ref{table:dataset} compares the datasets used in~\citet{shallue_2018}, ~\citet{Ansdell_2018}, ~\citet{armstrong-2020-exoplanet}, and this paper. Note that there are a lot more PCs and AFPs in~\citet{shallue_2018} and~\citet{Ansdell_2018} compared to our dataset. This is because we only used CPs from the Cumulative KOI catalog and excluded PCs that are not yet confirmed in order to have a more reliable (less noisy) labeled set, similar to~\citet{armstrong-2020-exoplanet}. However, unlike~\citet{armstrong-2020-exoplanet}, which only used a subset of NTPs to create a more balanced dataset, we included all NTPs. Similar to the previous ML works, we use a binary classifier to classify TCEs into PC and non-PC classes. Therefore, the classifier is indifferent to a training set with mixed labels between AFPs and NTPs\footnote{Our specific way of annotation, described in the previous paragraph, may confuse some NTPs with AFPs. For example, the second TCE belonging to an EB usually does not become a KOI, but technically it is an AFP.}. Given that our training set has a lower rate of PCs (8.5\% vs. 22.8\% in~\citet{shallue_2018} and~\citet{ Ansdell_2018}, and 33.3\% in~\citet{armstrong-2020-exoplanet}, which makes it more imbalanced), building an accurate classifier that can correctly retrieve PCs is more difficult. 

Similar to ~\citet{shallue_2018, Ansdell_2018, armstrong-2020-exoplanet}, we take an 80\%, 10\%, 10\% split for training, validation and test sets, respectively. However, in order to study the behavior of different models when the training/test split changes, we perform a 10-fold cross validation (CV), i.e., we split the data into 10 folds, each time we take one fold for test and the other 9 folds for training/validation (8 folds for training and one fold for validation). Also, unlike the existing studies that split TCEs into training, validation, and test sets, we split TCEs by their respective target stars. TCEs detected for the same target star have the same light curve and stellar parameters. By splitting the dataset at the TCE level, one shares data between the training, validation, and test sets, and produces a biased and unrealistic test set that the model is not supposed to be exposed to during training and validation.  

Splitting a dataset based on target stars is more realistic given that the model needs to classify the TCEs of new target stars in the future. We do not want to train the model with test data so that we can more accurately evaluate the model. This new setup creates a more difficult ML task, but ensures that the model can be used across missions (e.g., a model trained using the \kepler\ data can be employed to classify TESS data).

\subsection{Baseline Classifiers}
\label{sec:baseline}
We compare the following classifiers:

\begin{enumerate}[itemsep=-1mm]
\item{\textbf{Robovetter:}} We used the \Robovetter\ scores for \kepler\ Q1-Q17 DR25 TCE catalog~\citep{Coughlin_2016_robovetter}. That version of \Robovetter\ utilized all known CPs and CFPs to fine-tune its rules. To correctly understand its performance, however, it must be fine-tuned using only the training set and then evaluated on the test set. Given that it has used the data in the test set, the performance metrics we report on the test set here are an overestimate of the true performance. The correct numbers would be lower if the test set were hidden from \Robovetter\ during parameter setting.

\item{\textbf{AstroNet:}} We used the \AstroNet\ code available on github~\citep{shallue_2018}, preprocessed and trained the model using the same setup and DNN architecture as provided in~\citet{shallue_2018} on the dataset used in this paper. We also optimized the architecture for their model using our own data and hyper-parameter optimizer BOHB. This led to slightly worse result, which was expected because the authors used more resources to optimize their architecture. Thus, we report only the results using their optimized architecture. Following~\citet{shallue_2018}, we trained ten model for each of the ten folds and report the average score for each fold in a 10-fold CV.  

\item{\textbf{ExoNet:}} The original code of \ExoNet\ is not available.
Thus, we implemented it on top of the code available for \AstroNet\ based on the details (including the DNN general architecture) provided in~\citet{Ansdell_2018}. We optimized the details of the architecture for \ExoNet\ using our own data and BOHB. This resulted in a better performance which was expected, since~\citet{Ansdell_2018} did not optimize their architecture for the new set of inputs (flux and centroid motion time series data, and stellar parameters) and just used the one optimized by~\citet{shallue_2018}. Therefore, we report the results of the optimized architecture using BOHB. Following~\citet{Ansdell_2018}, we train ten different models and report the average score for each fold in a 10-fold CV.

\item{\textbf{Random Forest Classifier (\RFC) in~\citet{armstrong-2020-exoplanet}:}} We used the scores provided by the authors. Note that~\citet{armstrong-2020-exoplanet} used the Q1-Q17 DR25 TCE dataset, similar labeling to us (CPs, CFPs/CFAs, and NTPs), and provided the scores for all TCEs using cross-validation. However, \citet{armstrong-2020-exoplanet} split the data based on the TCE instead of the target star used use in this work.

\item{\textbf{Gaussian Process Classifier (\GPC) in~\citet{armstrong-2020-exoplanet}:}} We used the scores provided by the authors, similar to \RFC\ above. We report \GPC\ because it is the only classifier in~\citet{armstrong-2020-exoplanet} that receives light curves directly as input. 

\item{\textbf{ExoMiner:}} This is the \ExoMiner\ classifier shown in Figure~\ref{fig:hpo-dnn-architecture}. All hyper-parameters, including the DNN architecture were optimized using our hyper-parameter optimizer. Similar to \AstroNet\ and \ExoNet, we trained \ExoMiner\ ten times and report the average score for each fold in a 10-fold CV. 

\item{\textbf{ExoMiner-TCE:}} This is the \ExoMiner\ classifier shown in Figure~\ref{fig:hpo-dnn-architecture} when trained on a dataset split into training, validation, and test sets based on TCEs instead of target stars. All hyper-parameters are optimized using our hyper-parameter optimizer. We trained \ExoMiner\ ten times and report the average score for each fold in a 10-fold CV. The motivation for this baseline is to have a fair comparison with \GPC\ and \RFC, whose results are based on splitting the data at the TCE level.
\end{enumerate}

\subsection{Evaluation Metrics}
\label{sec:metrics}

To study and compare the performance of the above models, we use different classification and ranking metrics listed below:
\begin{itemize}
\item{\textbf{Accuracy:}} This is the fraction of correctly classified TCEs in the test set. Given that our dataset (and generally any realistic transit dataset) is very imbalanced, with about 7.5\% PCs, this metric is not very informative; To see this, note that a classifier that classifies all transits into the FP class has an accuracy of 92.5\%. However, accuracy provides some insights when studied in conjunction with other metrics.

\item{\textbf{Precision:}} Also called positive predictive value in ML, this is the fraction of TCEs classified as PC that are indeed PC; i.e.
\begin{equation}
\nonumber
Precision=\frac{true\;positives}{true\;positives+false\;positives}
\end{equation}
Note that as we make the dataset on which we measure precision more imbalanced, the precision decreases for a fixed model. As an example, suppose we reduce the size of the positive class to $\frac{1}{k}$ of the original size ($k>1$), the new precision would become
\begin{equation}
\label{eq:imabalnced_precision}
\frac{true\;positives}{true\;positives+k\times false\;positives}.    
\end{equation} 

\item{\textbf{Recall:}} Also called the true positive rate, this is the fraction of PCs correctly classified as PC; i.e.
\begin{equation}
\nonumber
Recall=\frac{true\;positives}{true\;positives+false\;negatives}
\end{equation}

\item{\textbf{Precision-Recall (PR) curve:}} The PR curve summarizes the trade-off between precision and recall by varying the threshold used to convert the classifier's score to a label. PR AUC is the total area under the PR curve. An ideal classifier would have an AUC of 1.

\item{\textbf{Receiver Operating Characteristic (ROC) curve:}} The ROC curve summarizes the trade-off between the true positive rate (recall) and false positive rate (fall-out) when varying the threshold used to convert the classifier score to label. ROC AUC is the total area under the ROC. 

\item{\textbf{Recall@p:}} This is the value of recall for precision=$p$. We are specially interested at $p=0.99$ to study the behavior of the model for planet validation. 

\item{\textbf{Precision@$k$ (P@$k$):}} P@$k$ is the fraction of the top $k$ returned TCEs that are true PCs. P@$k$ is a commonly used metric in recommendation and information retrieval systems in which the top $k$ retrieved items matter the most. For the classification of TCEs, this metric is valuable because it provides a sensible view of interpreting the performance of the classification model in ranking the TCEs. Overall, P@$k$ for a model shows how reliable that model is for recommending a PC for follow-up study or for validating new exoplanets. For example, if P@100 is 0.95, it means that there are only five non-PCs in the top 100 TCEs ranked by the model. 
More formally, P@$k$ is defined as:
\begin{equation}
\label{eq:precision_k}
P@k=\frac{true\;positives\;in\;top\;k\;TCE}{k}
\end{equation}
Note that if a model gives the same scores to the top $m>k$, for all such $m$ and $k$, we have:
\begin{equation}
\label{eq:precision_k_same}
P@k=\frac{true\;positives\;in\;top\;m\;TCE}{m}
\end{equation}
This is because the model does not provide any particular ordering for the top such $m$ TCEs. To understand this, consider a model that gives a score of 1.0 for 100 TCEs out of which 50 are PCs. If we are only interested in knowing how many of the top 10 TCEs are PC, i.e., P@10, we have $50/100$ chance of success in average. 
\end{itemize}

A few notes about the above metrics: 

First, note that the ML classifiers we compare in this paper generate a score between 0 and 1 for each TCE. The higher the score generated for a TCE, the more confident the model is that the TCE is indeed a PC. In order to classify a TCE, one must use a threshold to convert the generated score into a label. The values of accuracy, precision, and recall are a function of this threshold, which can be set to obtain a desired precision/recall trade-off. Unless noted otherwise, we use the standard value of 0.5 for this threshold. When we report the results later in Table~\ref{table:generalperformance} to investigate the performance of different models, we put precision and recall in the same column (denoted Precision \& Recall) because we can always change the threshold to trade-off one versus the other. The idea is to have a model whose performance is better overall. 

Second, note that the ROC AUC and PR AUC are not threshold dependent and are thus good metrics for evaluating the performance of different classifiers. The difference between ROC AUC and PR AUC is that the former measures the performance of the classifier in general, and the latter measures how good the classifier is with an emphasis on the class of interest, in our case PC. In other words, the PR curve is more useful for needle-in-a-haystack type problems where the positive class is rare and interesting. 

Third, note that precision and recall are called reliability and completeness in exoplanet research terminology~\citep{Coughlin2017robovetter}.

Finally, note that P@$k$ is indifferent to the specific scores of ranked items and only cares about the rank and relevancy. For example, the score for the item ranked at position $l$ ($l\leq k$) could be very low for P@$k$. P@$k$ just measures how relevant the top items are independent of how large their scores are.

\begingroup
\setlength{\tabcolsep}{2.5pt} 
\begin{table}[htb]
 \centering
\caption{Scores of different classifiers trained in this work. This table describes the available columns. The full table is available online.}
\label{table:classification_results}
\begin{threeparttable}
\begin{tabularx}{\linewidth}{@{}Y@{}}
\begin{tabular}{ll}
\toprule
Column & Description \\
\midrule
target\_id & KIC ID\\
tce\_plnt\_num & TCE planet number\\
tce\_period & TCE period\\
mes & TCE MES\\
original\_label & NTP, AFP, or PC\\
binary\_label & 0 for NTP or AFP, 1 for PC\\
astronet & Score assigned by \AstroNet \\
exonet & Score assigned by \ExoNet \\
exominer & Score assigned by \ExoMiner \\
exominer\_tce & Score assigned by \ExoMiner-TCE \\ 
\bottomrule
\end{tabular}
\end{tabularx}
\end{threeparttable}
\end{table}
\endgroup

\section{Performance Evaluation}
\label{sec:evaluation}
The results of the classifiers trained in this work is reported in  Table~\ref{table:classification_results}. Using these scores reported in this table, we study the performance of different classifiers in this Section. 
\begingroup
\setlength{\tabcolsep}{2.5pt} 
\begin{table*}[!htbp]
\footnotesize
 \centering
\caption{Performance of different classifiers using 10-fold CV. The numbers are bold when the corresponding model outperforms other classifiers for an evaluation metric. For each metric, we report 1) on the first row ,the value computed over the full dataset for each classifier, and 2) on the second row and in parentheses, the 99\% confidence intervals using Student's t-test around the mean of the 10 folds in the CV for each classifier. Note that the values of P@k are computed for different values of $k$ for these two options because the number of PCs for each fold is much less than that in the full dataset.}
\label{table:generalperformance}
\begin{threeparttable}
\begin{tabularx}{\linewidth}{@{}Y@{}}
\begin{tabular}{cccccccccccc}
\toprule
Model/Metric & Precision \& Recall & Accuracy &  PR AUC & ROC AUC & P@100 & P@1000 & P@2200\tnote{1} \\
 &   &   &  &  & (P@50) & (P@100) & (P@200) \\
\midrule
  \multirow{2}{*}{\Robovetter\tnote{2}} & 0.951 \& 0.975 & 0.994 &  0.958 & 0.994 & 0.966\tnote{3} & 0.966\tnote{3} & 0.966\tnote{3}\\
   & (0.950$\pm$0.015) \& (0.975$\pm$0.000) & (0.994$\pm$0.001) & (0.958$\pm$0.019) & (0.994$\pm$0.003) & (0.967$\pm$0.019) & (0.967$\pm$0.019) & (0.969$\pm$0.017)\\
\addlinespace[.2cm]

 \multirow{2}{*}{\AstroNet} & 0.861 \& 0.885 & 0.981  & 0.925 & 0.993 & 0.950 & 0.967 & 0.882\\
   & (0.862$\pm$0.022) \& (0.885$\pm$0.000) & (0.981$\pm$0.003) & (0.927$\pm$0.014) & (0.993$\pm$0.003) & (0.976$\pm$0.013) & (0.967$\pm$0.016) & (0.906$\pm$0.024)\\
\addlinespace[.2cm]
 
 \multirow{2}{*}{\ExoNet} & 0.925 \& 0.864 & 0.985 & 0.956 & 0.995 & 0.990 & 0.989 & 0.916 \\
  & (0.925$\pm$0.008) \& (0.865$\pm$0.000) & (0.985$\pm$0.002) & (0.956$\pm$0.007) & (0.995$\pm$0.001) & (0.996$\pm$0.009) & (0.990$\pm$0.008) & (0.936$\pm$0.016)\\
\addlinespace[.2cm]
 
 \multirow{2}{*}{\GPC\tnote{2}} & 0.921 \& 0.964 & 0.991 & 0.982 & 0.998 & \textbf{1.000} & \textbf{0.997} & 0.954\\
 & (0.921$\pm$0.011) \& (0.964$\pm$0.000) & (0.991$\pm$0.002) & (0.982$\pm$0.007) & (0.998$\pm$0.001) & (0.996$\pm$0.009) & (0.997$\pm$0.005) & (0.972$\pm$0.013)\\
\addlinespace[.2cm]

 \multirow{2}{*}{\RFC\tnote{2}} & 0.929 \& 0.955 & 0.991 & 0.979 & 0.998 & 0.991 & \textbf{0.991} & 0.954\\
  & (0.930$\pm$0.015) \& (0.956$\pm$0.000) & (0.991$\pm$0.002) & (0.979$\pm$0.008) & (0.998$\pm$0.001) & (0.992$\pm$0.011) & (0.992$\pm$0.011) & (0.970$\pm$0.011)\\
\addlinespace[.2cm]

 \multirow{2}{*}{\ExoMiner} & \textbf{0.968} \& \textbf{0.974} &  \textbf{0.996} & \textbf{0.995} & \textbf{1.000} & \textbf{1.000} & \textbf{0.999} & \textbf{0.985} \\
  & (0.968$\pm$0.011) \& (0.974$\pm$0.000) & (0.996$\pm$0.001) & (0.995$\pm$0.002) & (1.000$\pm$0.000) & (1.000$\pm$0.000) & (1.000$\pm$0.000) & (0.990$\pm$0.009)\\
\addlinespace[.2cm]

 \multirow{2}{*}{\ExoMiner-TCE} & \textbf{0.965 \& 0.977} &  \textbf{0.996} & \textbf{0.995} & \textbf{1.000} & \textbf{1.000} & \textbf{0.998} & \textbf{0.987} \\
   & (0.964$\pm$0.011) \& (0.978$\pm$0.000) & (0.996$\pm$0.001) & (0.996$\pm$0.002) & (1.000$\pm$0.000) & (1.000$\pm$0.000) & (1.000$\pm$0.000) & (0.991$\pm$0.006)\\
\bottomrule
\end{tabular}
\end{tabularx}
\begin{tablenotes}[para,flushleft]\scriptsize
  \item [1] There are a total of $2291$ confirmed exoplanets in the dataset.
  \item [2] The confidence intervals for \Robovetter, \GPC, and \RFC\ are only approximations. This is because \Robovetter\ does not use cross-validation, and \GPC\ and \RFC\ use different folds in CV. 
  \item [3] \Robovetter\ assigns a score of 1.0 to 1312 TCEs of which there are a total of 1268 PCs. Thus, P@100=P@1000=0.966. The value for P@2200 is also 0.966 by chance.  
\end{tablenotes}
\end{threeparttable}
\end{table*}
\endgroup

\subsection{Comparison on Kepler Q1-Q17 DR25 data}
Table~\ref{table:generalperformance} shows the performance of different classifiers using 10-fold CV. For each evaluation metric, we report its value over the Kepler Q1-Q17 DR25 dataset after combining the scores for all test sets (ten non-overlapping folds). We also report the mean and 99\% confidence interval using Student's t-test  over ten folds in parentheses. Given that for \Robovetter, \RFC, and \GPC, the data splits for training, test, and validation sets are different from the ones we used here, the means and confidence intervals reported for these methods are approximations. However, because these two estimates (for the complete dataset and for the average of the 10 folds) do not differ significantly for other classifiers, we can assume that the approximation is reasonable. 

Overall, the performance of \ExoMiner\ is better than all baselines, in terms of classification and ranking metrics, particularly when looking at existing deep learning models, i.e., \AstroNet\ and \ExoNet. 

There are a few points to note here. 

First, one should keep in mind that there is some level of label noise in the dataset. Thus, any performance comparison needs to be done with caution. We would like to note that the label noise in the NTP and AFP classes is minimal. The major portion of the label noise is in the exoplanet (PC) category. Because the label noise in the positive class does not affect the performance of a model negatively in terms of metrics such as precision and $P@k$, it is hard to attribute the superior performance of \ExoMiner to the label noise.  

Second, note that there is little difference between the performance of \ExoMiner\ and \ExoMiner-TCE. As we see in the reliability analysis, Section~\ref{sec:stable_reliable}, the size of the training set is large enough to cover different data scenarios even when we split by target stars. Thus, splitting by TCE or target star does not change the results significantly. Nonetheless, splitting by target star is the correct approach because of potential data leakage between training and test sets, as mentioned before. 

Third, note how different classifiers compare in terms of P@$k$. In order to understand how many PCs exist in the top TCEs ranked by each model in the full dataset, simply multiply P@$k$ by $k$. For example, \ExoMiner\ returns $999$ PCs in the top 1000 ranked TCEs. These numbers are $991$ and $966$ for \RFC\ and \Robovetter, respectively. A model that can rank transit signals by putting PCs in the top of the list is highly desirable for occurrence rate calculations and follow-up studies to confirm new exoplanets.

Finally, note that \ExoMiner\ has lower variability (smaller confidence intervals) compared to the baselines. This implies that \ExoMiner\ is more stable with regards to changes in the training and test sets.

\begin{figure}[htb!]
	\centering
	\subfigure[PR curve]{\label{fig:PR-curve}\includegraphics[width=.77\columnwidth]{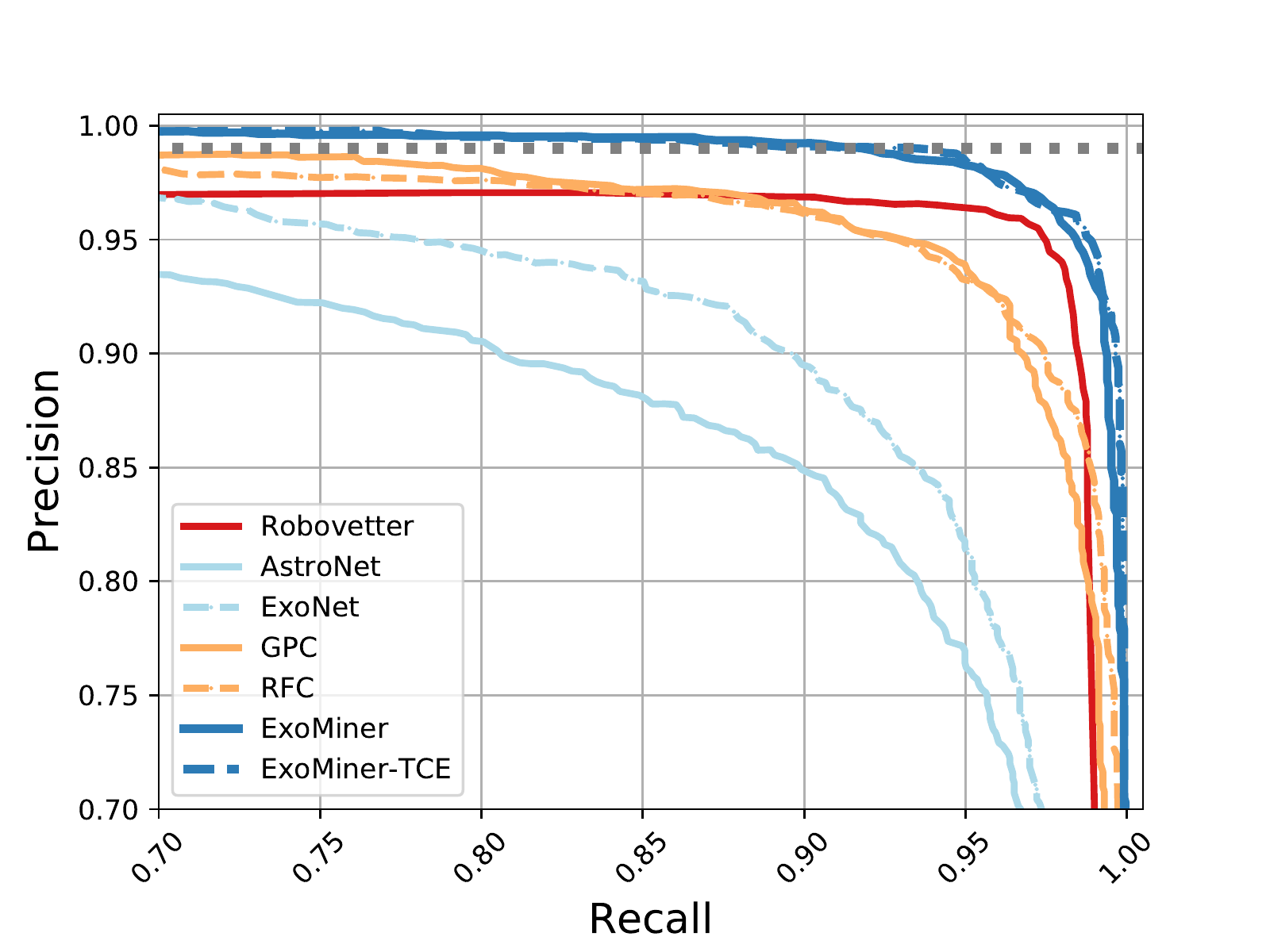}}
	\vskip -0.07in
	\subfigure[Precision@k curve]{\label{fig:Precision_k}\includegraphics[width=.77\columnwidth]{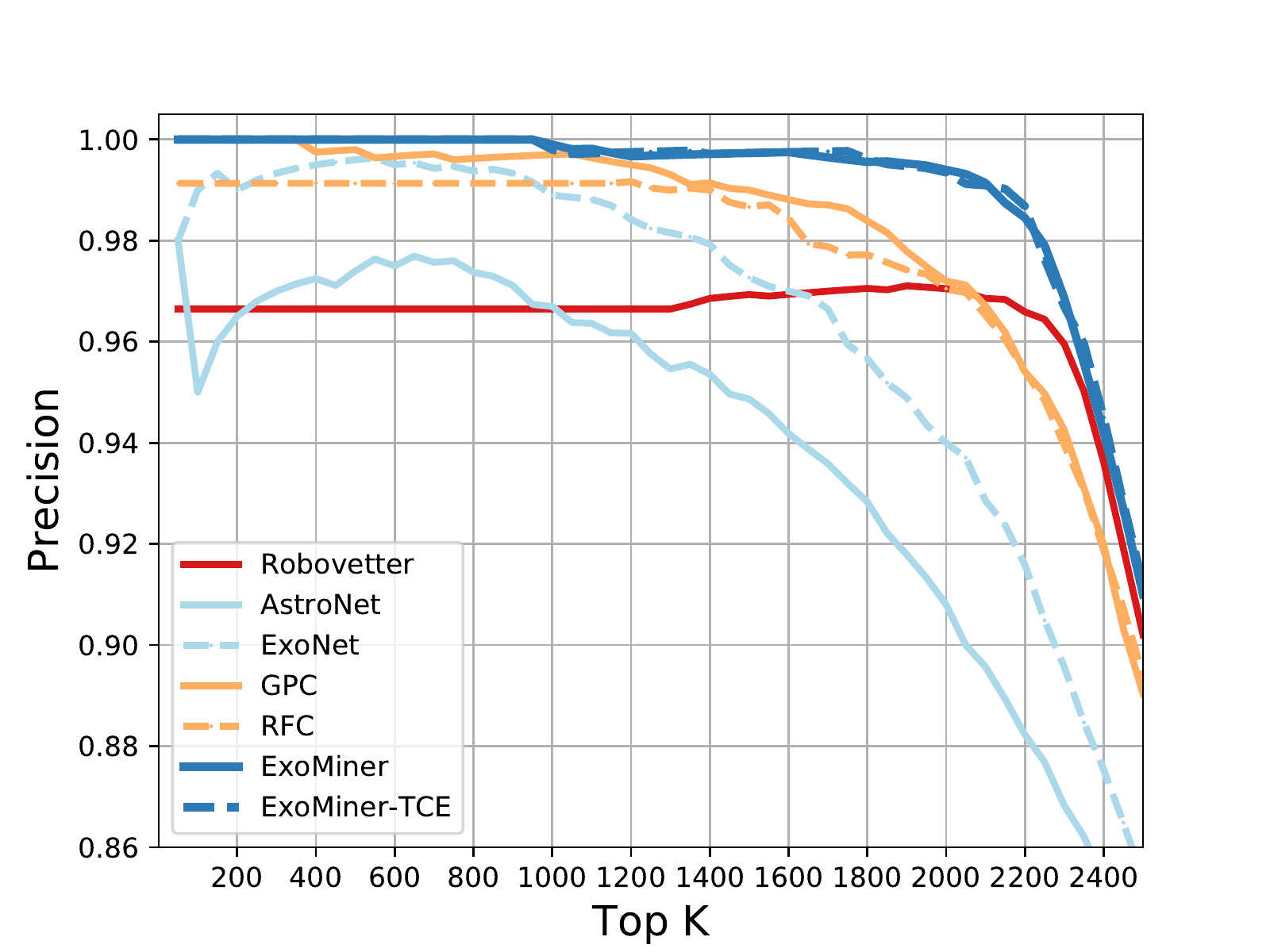}}
	\vskip -0.07in
	\subfigure[Score of TCE at position k]{\label{fig:Score_k}\includegraphics[width=.77\columnwidth]{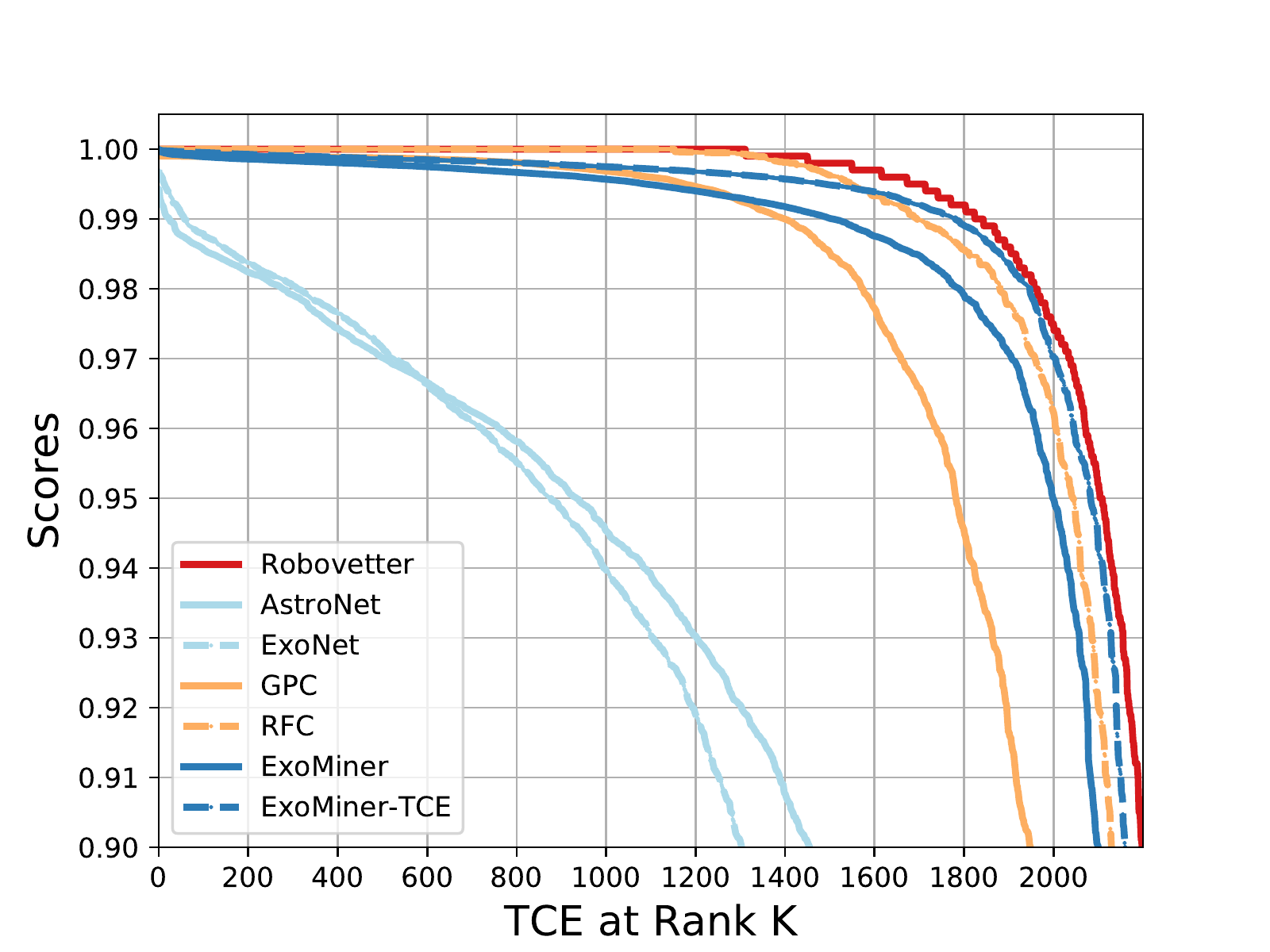}}
	\vskip -0.07in
	\subfigure[Distribution of scores]{\label{fig:Scores_Hist}\includegraphics[width=.77\columnwidth]{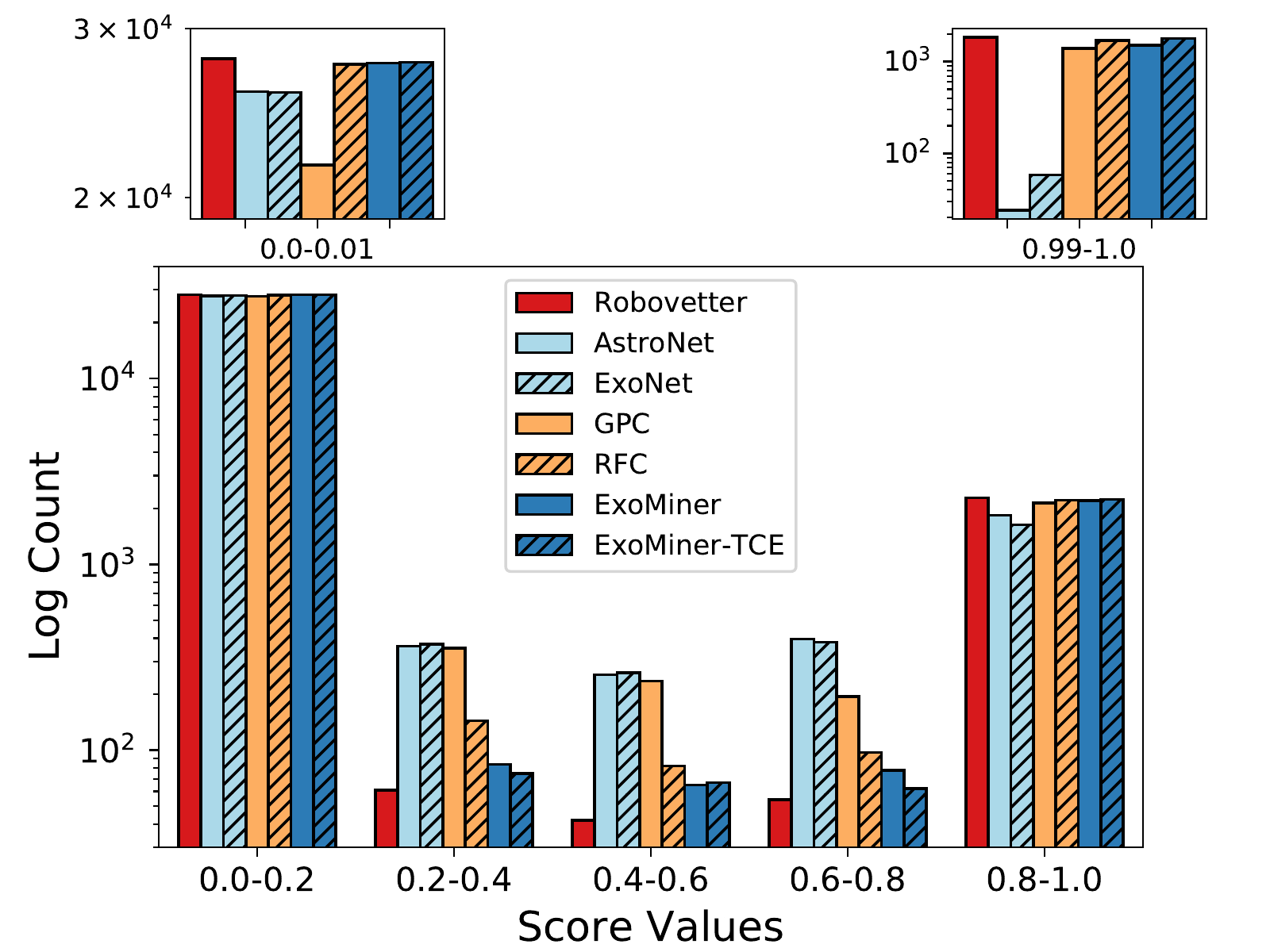}}
\caption{Precision-Recall (PR), Precision@k, Score@$k$, and Score Distribution of different models.}
\label{fig:Performance-curve}
\end{figure}

\begin{table}[!htbp]
\footnotesize
 \centering
\caption{Recall value at a desirable precision value (Recall@$p$). The numbers are bold when the corresponding classifier outperforms other classifiers. The numbers in parentheses show the 99\% confidence interval around the mean of 10-fold CV using Student's t-test. The confidence intervals for \Robovetter, \GPC, and \RFC\ are only approximations.}
\label{table:precision-recall-values}
\begin{threeparttable}
\begin{tabularx}{\linewidth}{@{}Y@{}}
\begin{tabular}{cccc}
\toprule
 Method & Recall@$0.99$ & Recall@$0.975$ & Recall@$0.95$ \\
\midrule
 \multirow{2}{*}{\Robovetter}  &  0.0\tnote{1} & 0.0\tnote{1} & 0.976\\
   &  0.364$\pm$0.484 & 0.433$\pm$0.485 & 0.974$\pm$0.000\\
\addlinespace[.2cm]

 \multirow{2}{*}{\AstroNet}  & 0.001\tnote{2} & 0.333 & 0.643\\
  & 0.198$\pm$0.174 & 0.343$\pm$0.143 & 0.642$\pm$0.012\\
\addlinespace[.2cm]

 \multirow{2}{*}{\ExoNet}  & 0.506 & 0.619 & 0.803\\
  & 0.479$\pm$0.190 & 0.643$\pm$0.09 & 0.806$\pm$0.001\\
\addlinespace[.2cm]

 \multirow{2}{*}{\GPC}  & 0.763 & 0.825 & 0.943\\
  &  0.716$\pm$0.156 & 0.848$\pm$0.101 & 0.94$\pm$0.001\\
\addlinespace[.2cm]

 \multirow{2}{*}{\RFC}  &   0.677 & 0.807 & 0.937\\
  & 0.563$\pm$0.314 & 0.776$\pm$0.115 & 0.935$\pm$0.001\\
\addlinespace[.2cm]

 \multirow{2}{*}{\ExoMiner}  & \textbf{0.936} & \textbf{0.966} & \textbf{0.986}\\
  & 0.939$\pm$0.030 & 0.964$\pm$0.009 & 0.988$\pm$0.000\\
\addlinespace[.2cm]
 
 \multirow{2}{*}{\ExoMiner-TCE}  & \textbf{0.949} & \textbf{0.960} & \textbf{0.990}\\
  & 0.95$\pm$0.020 & 0.961$\pm$0.015 & 0.991$\pm$0.000\\
 \bottomrule
\end{tabular}
\end{tabularx}
\begin{tablenotes}[para,flushleft]\scriptsize
  \item [1] \Robovetter\ assigned a score of 1.0 to 1312 TCEs from which there are 44 non-PCs. So no threshold can attain a precision of 0.99 or 0.975. 
  \item [2] There are FPs with very high AstroNet scores. 
\end{tablenotes}
\end{threeparttable}
\end{table}

To get a better sense of the performance of different methods, we plotted their precision-recall curves in Figure~\ref{fig:PR-curve}. It is clear that \ExoMiner\ outperforms the baselines significantly in all points. When validating new exoplanets, high values of precision are desirable, and \ExoMiner\ has high recall values at high precision values. As a matter of fact, at a fixed precision value of $0.99$, \ExoMiner\ is able to retrieve $93.6$\% of all exoplanets in the test set. This number is $76.3$\% for the second best model, \GPC. We provide the recall values of different methods for different precision values in Table~\ref{table:precision-recall-values}.  

We compare how different models perform in terms of ranking the top $k$ TCEs in Figure~\ref{fig:Precision_k}. Note that \ExoMiner\ is a better ranker compared to existing classifiers. To provide a clear comparison, there is no FP in the top 975 and 984 TCEs when ranked by \ExoMiner\ and \ExoMiner-TCE, respectively. This value is 0, 3, 26, 379, and 0 for \Robovetter, \AstroNet, \ExoNet, \GPC, and \RFC, respectively. The reason that this is zero for \Robovetter\ and \RFC\ is because these models assign a score of 1.0 to many TCEs including FPs (check Equation~\ref{eq:precision_k_same}), which prevents sorting the top TCEs.   

To further understand how the scores change at different rank positions and how the scores of different classifiers are distributed, we plot the scores of different models for the TCE at position $k$ in Figure~\ref{fig:Score_k} and the score histogram in Figure~\ref{fig:Scores_Hist}. Note that there is not much difference between the behavior of different classifiers in terms of the score distribution. The better performance of \ExoMiner\ is due to its better score assignment. However, \ExoMiner\ is generally more confident about the label of TCEs. Generally speaking, ML models that use a more comprehensive list of diagnostic tests are more confident about their label too. There are 1844, 24, 58, 1399, 1694, 1507, and 1781 TCEs with score $>0.99$ for \Robovetter, \AstroNet, \ExoNet, \GPC, \RFC, \ExoMiner, and \ExoMiner-TCE, respectively (upper right panel in Figure~\ref{fig:Scores_Hist}). ML classifiers such as \AstroNet\ and \ExoNet\ that do not use different diagnostic tests required to correctly distinguish exoplanets from FPs are less confident than ML classifiers such as \GPC, \RFC, and \ExoMiner\ that utilize a more comprehensive list of diagnostic tests. 
As we will show in Section~\ref{sec:training_set_sensitivity}, the confidence of the model is also dependent on the size and the level of label noise in the training set.

\subsection{Comparison on \kepler\ planets confirmed by Radial Velocity}
\label{sec:radial-velocity}
Given that some of the \kepler\ confirmed exoplanets could have the wrong disposition, in this section we report the performance of difference classifiers on a subset of TCEs that only includes those \kepler\ exoplanets confirmed using radial velocity. We obtained the list of \kepler\ exoplanets that are confirmed using radial velocity from the Planetary System table on the NASA Exoplanet Archive\footnote{https://exoplanetarchive.ipac.caltech.edu}. There are a total of $310$ such confirmed exoplanets in the Q1-Q17 DR25 dataset. Table~\ref{table:generalperformance_RV} reports the performance of different classifiers on this dataset that includes all NTPs and AFPs, but only PCs that are exoplanets confirmed using radial velocity. We report a single value for the full subset here because the results of averaging over ten folds are very similar. Note that the recall value is equivalent to the accuracy of a classifier on the positive class; in our case, it counts the percentage of exoplanets confirmed by radial velocity that the model classifies as PC. As can be seen, \ExoMiner\ consistently performs better than existing classifiers on this dataset. Also, compared to the performance on the full dataset, the precision for all classifiers decreases significantly. This is because there are significantly fewer confirmed planets in this dataset (almost 1/7 of the original dataset). As we discussed in Section~\ref{sec:metrics}, precision decreases for more imbalanced datasets (Check Equation~\ref{eq:imabalnced_precision}).

\begingroup
\setlength{\tabcolsep}{1.3pt} 
\begin{table}[!htbp]
\footnotesize
 \centering
\caption{Performance of different classifiers using 10-fold CV on the dataset that only includes those \kepler\ exoplanets confirmed by radial velocity.}
\label{table:generalperformance_RV}
\begin{threeparttable}
\begin{tabularx}{\linewidth}{@{}Y@{}}
\begin{tabular}{ccccccccc }
\toprule
Model/Metric & Precision \& Recall & Accuracy &  PR AUC & ROC AUC \\
\midrule
\Robovetter & 0.721 \& 0.968 & 0.996 & 0.800 & 0.992\\
\AstroNet &  0.438 \& 0.819 & 0.987 & 0.622 & 0.989\\
\ExoNet & 0.618 \& 0.839 & 0.993 & 0.816 & 0.994\\
\GPC & 0.601 \& 0.919 & 0.993 & 0.876 & 0.997\\
\RFC &  0.630 \& 0.913 & 0.993 & 0.825 & 0.998\\
\ExoMiner & 0.800 \& 0.955 & 0.997 & 0.952 & 0.999\\
\ExoMiner-TCE & 0.785 \& 0.968 & 0.997 & 0.961 & 1.000\\
\bottomrule
\end{tabular}
\end{tabularx}
\end{threeparttable}
\end{table}
\endgroup

\subsection{Comparison to~\citet{Santerne_independent_2016}}
\label{sec:santerne}
Similar to~\citet{Morton-2016-vespa} and~\citet{armstrong-2020-exoplanet}, we report the distribution of scores for \ExoMiner\ on those \kepler\ candidates for which~\citet{Santerne_independent_2016} provided dispositions. Table~\ref{table:comparing_saterne} reports the mean and median scores. The behavior of \ExoMiner\ on these disposition categories is very similar to existing classifiers. \ExoMiner\ generates high scores for both planet and BD categories and low scores for EB and CEB categories. However, note that the scores for BD are well below the validation threshold of 0.99.  

\begingroup
\setlength{\tabcolsep}{2.5pt} 
\begin{table}[htb]
 \centering
\caption{Scores of \ExoMiner\ for~\citet{Santerne_independent_2016} dispositioned KOIs }
\label{table:comparing_saterne}
\begin{threeparttable}
\begin{tabularx}{\linewidth}{@{}Y@{}}
\begin{tabular}{cccccc}
\toprule
Label && Number && Mean & Median \\
\midrule
planet && 44 && 0.773 & 0.861 \\
EB && 48 && 0.138 & 0.086 \\
Contaminating EB (CEB) && 15 &&  0.175 & 0.04 \\
Brown Dwarf (BD) && 3 && 0.86 & 0.84\\
Undetermined && 18 && 0.754 & 0.896 \\
\bottomrule
\end{tabular}
\end{tabularx}
\end{threeparttable}
\end{table}
\endgroup

\begin{figure}[htb!]
\begin{center}
\subfigure[Accuracy for different threshold values.]{\label{fig:accuracy-at-t_santerne}\includegraphics[width=85mm]{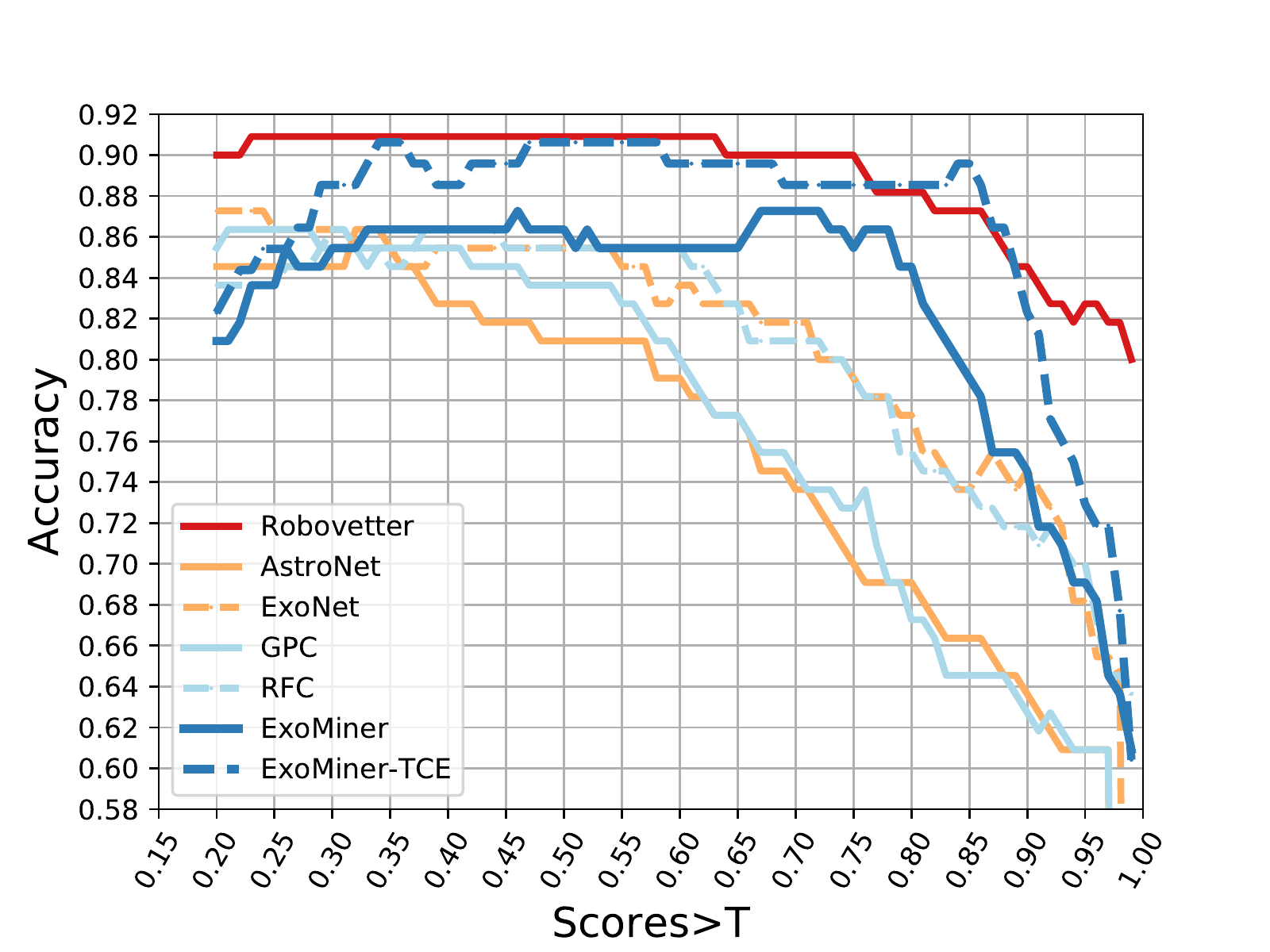}}
\subfigure[Number of planets for different threshold values.]{\label{fig:PCs-at-t_santerne}\includegraphics[width=85mm]{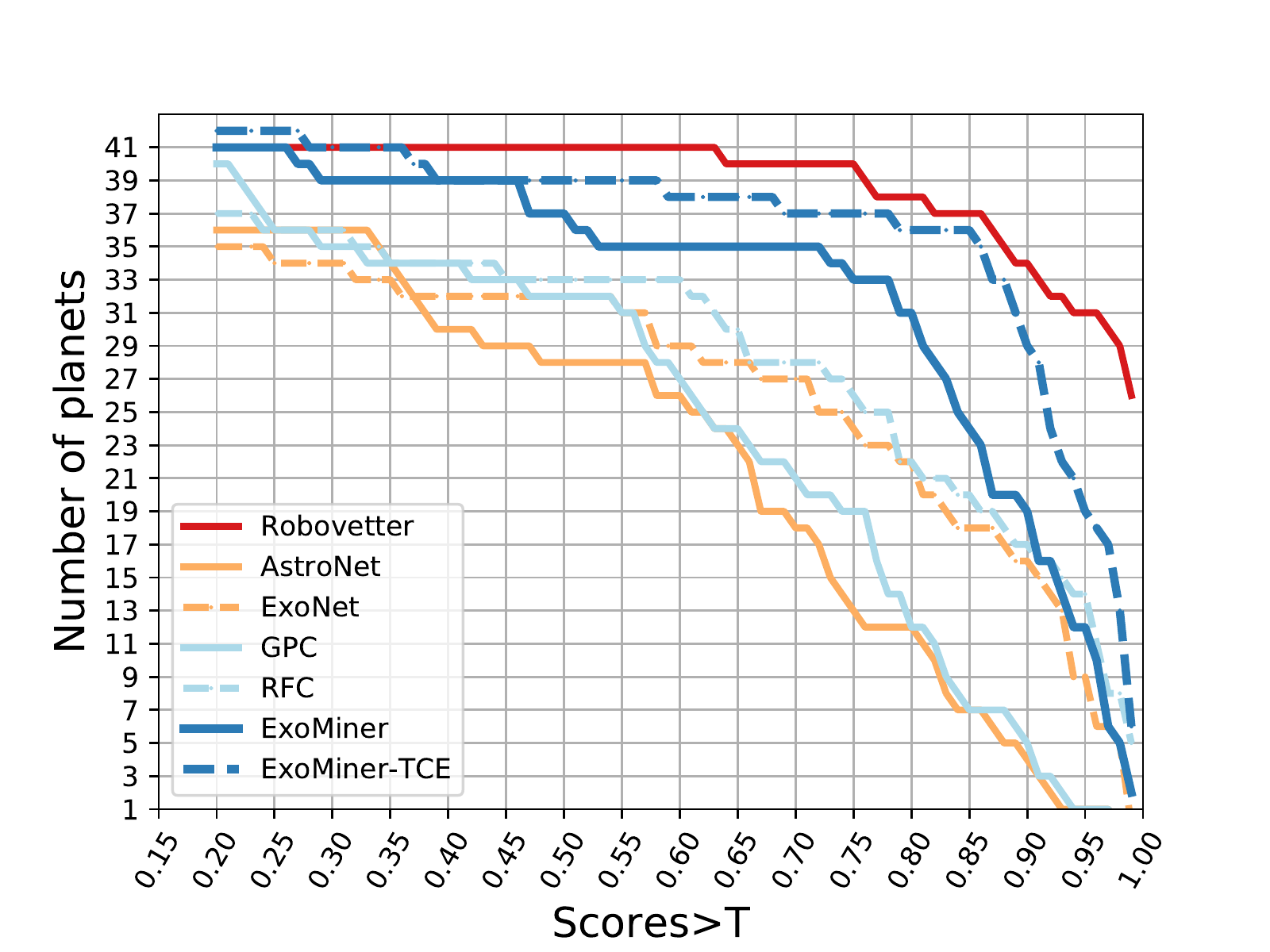}}
\caption{Accuracy and number of detected planets for different classifiers on Santerne dispositioned KOIs~\citep{Santerne_independent_2016} when a threshold $T$ is set for classification, i.e., we classify a signal as PC when the classifier score $>T$.}
\label{rejectionplots_santerne}
\end{center}
\end{figure}

The mean and median score values of each category do not provide much detailed information about the performance of the model or score distribution of the model. Given that there are only a total of 110 dispositioned KOIs in~\citet{Santerne_independent_2016} and it is balanced (44 planets and 56 FPs), we report the performance of different classifiers in terms of accuracy when different threshold values are used to label a KOI as planet. Note that accuracy is more informative for this small balanced dataset\footnote{Because the dataset is small, other metrics such as precision are very unstable for many thresholds. Precision can be 0.5 at one threshold when there are only two KOIs above threshold and become 1.0 for the next threshold when there is only one KOI.}. Figure~\ref{rejectionplots_santerne} reports the results of this experiment. As you can see, \ExoMiner\ and \ExoMiner-TCE outperform all classifiers, except \Robovetter, on this set of independently dispositioned KOIs. The fact that \Robovetter\ performs well on this data set could be because \Robovetter\ was fine-tuned on all known exoplanets and false positives at the time (including this data set), whereas the performance of all other comparing baselines are measured on a held-out test data set. Note that \ExoMiner-TCE performs better than \ExoMiner. Even though this is not a sizeable dataset, splitting by TCE seems to make the classification easier. 

Also, note that for higher threshold values, we observed that all methods perform poorly on the planet category, as seen in Figure~\ref{fig:PCs-at-t_santerne}. This is because the planets in this set of KOIs are in the large planet region for which classification is more difficult. Thus, the classifiers are less confident on this dataset.  

\begin{figure}[htb!]
	\centering
	\subfigure[\ExoMiner\ and \vespa\ 1-fpp scores for all KOIs in both evaluation and unused datasets.]{\label{fig:exominervsvespa_scatter1}\includegraphics[width=1.1\columnwidth]{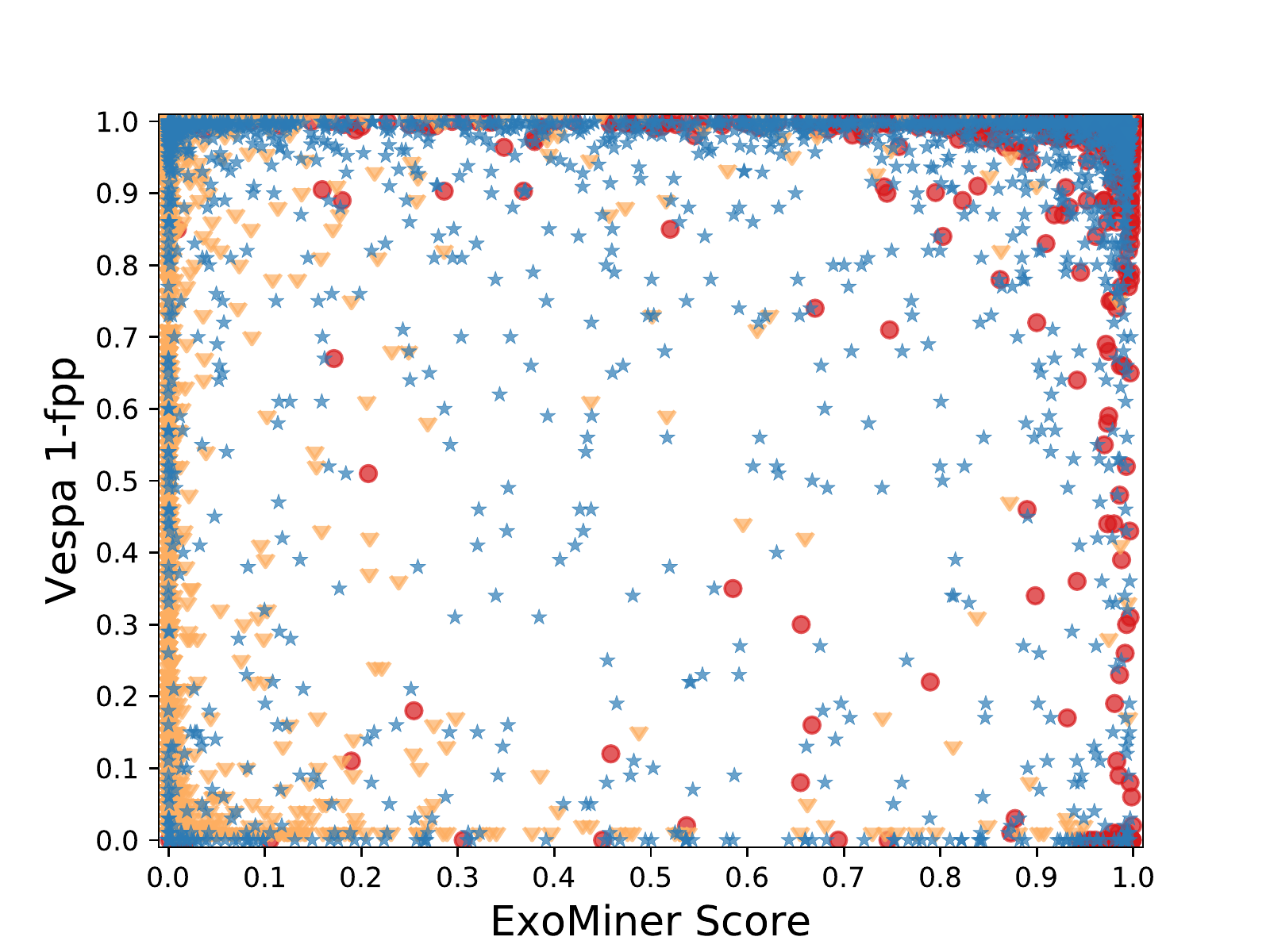}}
	\subfigure[Zoom for each corner. Note that the precision of \vespa\ fpp is two decimal places resulting in the banding pattern in lower left corner.]{\label{fig:exominervsvespa_scatter2}\includegraphics[width=1.05\columnwidth]{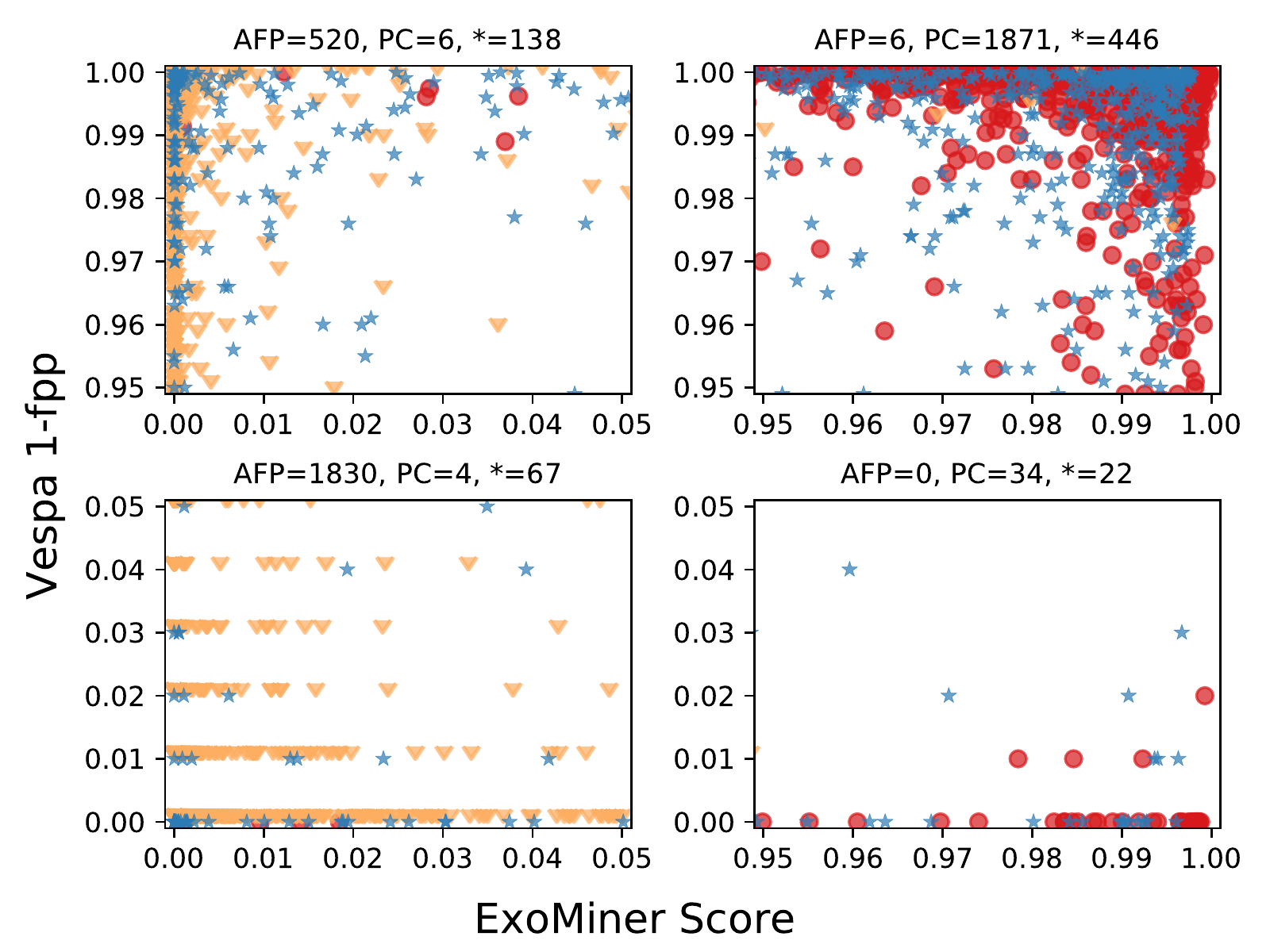}}
	\caption{Comparison with \vespa~\citep{Morton-2016-vespa}. Confirmed exoplanets are shown as red circles, AFPs as yellow triangles, and unused KOIs as blue stars. Note that \ExoMiner\ is often correct when \ExoMiner\ and \vespa\ disagree (upper left and lower right corners).}
\label{fig:exominervsvespa_scatter}
\end{figure}

\subsection{Comparison to \vespa}
\label{sec:comparison-vespa}
Figure~\ref{fig:exominervsvespa_scatter} compares the scores provided by \ExoMiner\ with those provided by \vespa\ for the KOIs. Note that we reported the scores for unused KOIs, i.e., those that are not part of our evaluation dataset (combined training, validation and test sets) using blue star markers. We will discuss this later in Section~\ref{sec:planetvalidation}. Overall, \ExoMiner\ and \vespa\ agree on 76\% of cases when a threshold of 0.5 is used. This number is 73\% for \GPC, as reported in~\cite{armstrong-2020-exoplanet}. \vespa\ fpp scores are generally not accurate when they are in disagreement with \ExoMiner\ scores. Similar behavior was observed in~\cite{armstrong-2020-exoplanet}. Figure~\ref{fig:exominervsvespa_scatter2} shows the score comparison for the corner cases. The numbers above each sub-figure show the number of cases for each category. As can be seen, there are a lot of FPs (520 FPs) with low \vespa\ fpp values (upper left corner of Figure~\ref{fig:exominervsvespa_scatter2}), which is the range where validation of exoplanets occurs. To correct this, \cite{Morton-2016-vespa} proposed multiple vetoing criteria in addition to low \vespa\ fpp for validation.  

For a validation threshold of 0.99, \ExoMiner\ gives scores $>0.99$ to nineteen dispositioned KOIs (strongly classifying them as exoplanets) while \vespa\ gives fpp scores $>0.99$ (strongly classifying them as FPs). These are all confirmed exoplanets. On the other corner, \ExoMiner\ gives scores $<0.01$ to 363 dispositioned KOIs (strongly believe they are FPs) for cases that \vespa\ fpp $<0.01$ (strongly believe they are planets). Except one KOI in this list, i.e., Kepler-452 b, they are also FP. Based on this analysis, \ExoMiner\ is generally more accurate than \vespa\ in the validation threshold region. 

\section{Stability and Reliability of ExoMiner}
\label{sec:stable_reliable}
In this section, we perform reliability analysis by 1) studying the sensitivity of \ExoMiner\ to training set size and label noise and 2) examining some examples of misclassified exoplanets and FPs.  

\subsection{Training Set Sensitivity}
\label{sec:training_set_sensitivity}
We perform two tests in order to understand the sensitivity of \ExoMiner\ with regard to imperfect training set scenarios, namely training set size and label noise.  

\begin{figure*}[htb!]
	\centering
	\subfigure[Performance for Threshold=0.5]{\label{fig:training-size1}\includegraphics[width=.35\textwidth]{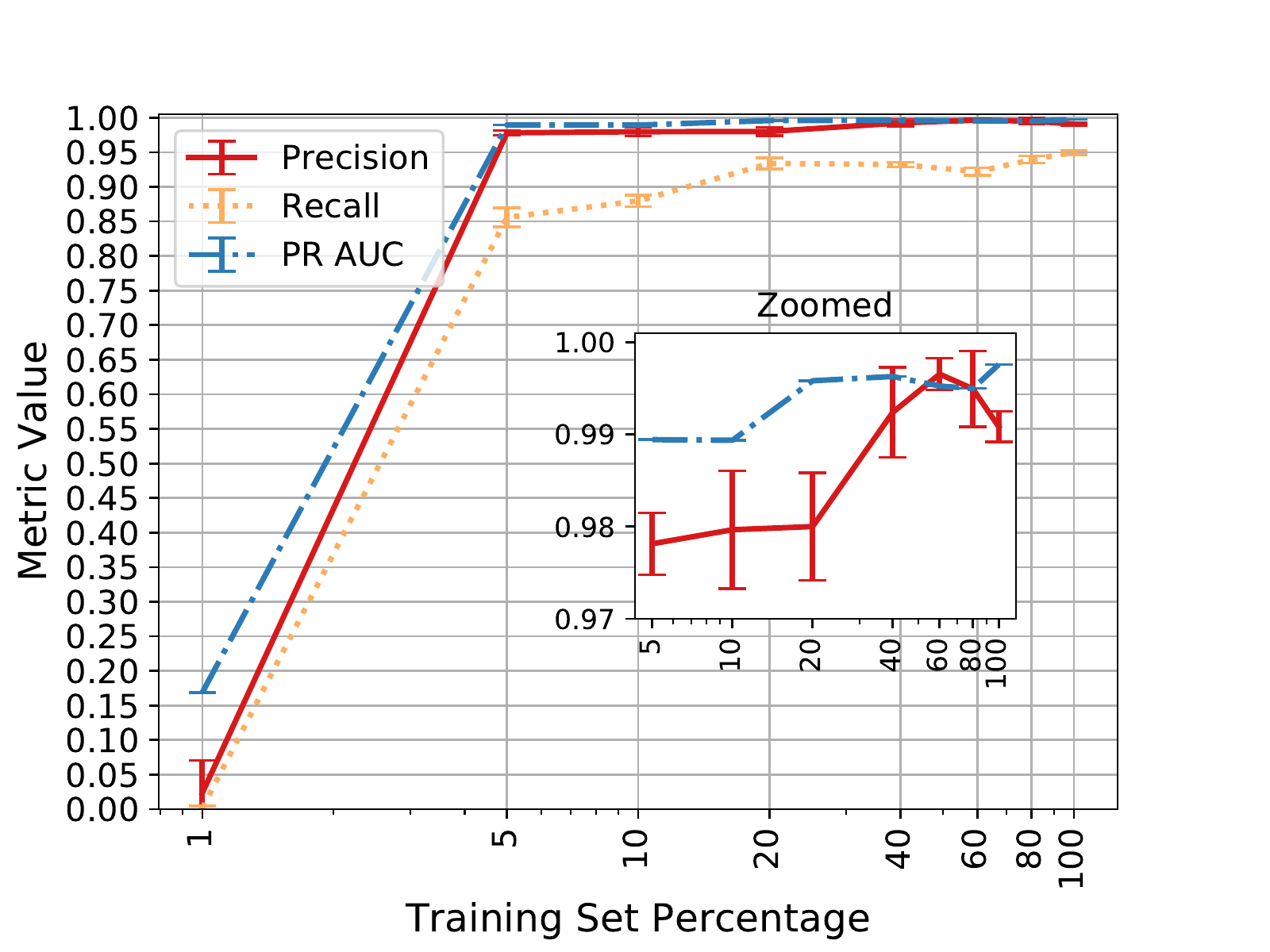}}
	\hskip -.25in
	\subfigure[Average Scores for Threshold=0.5]{\label{fig:training-size2}\includegraphics[width=.35\textwidth]{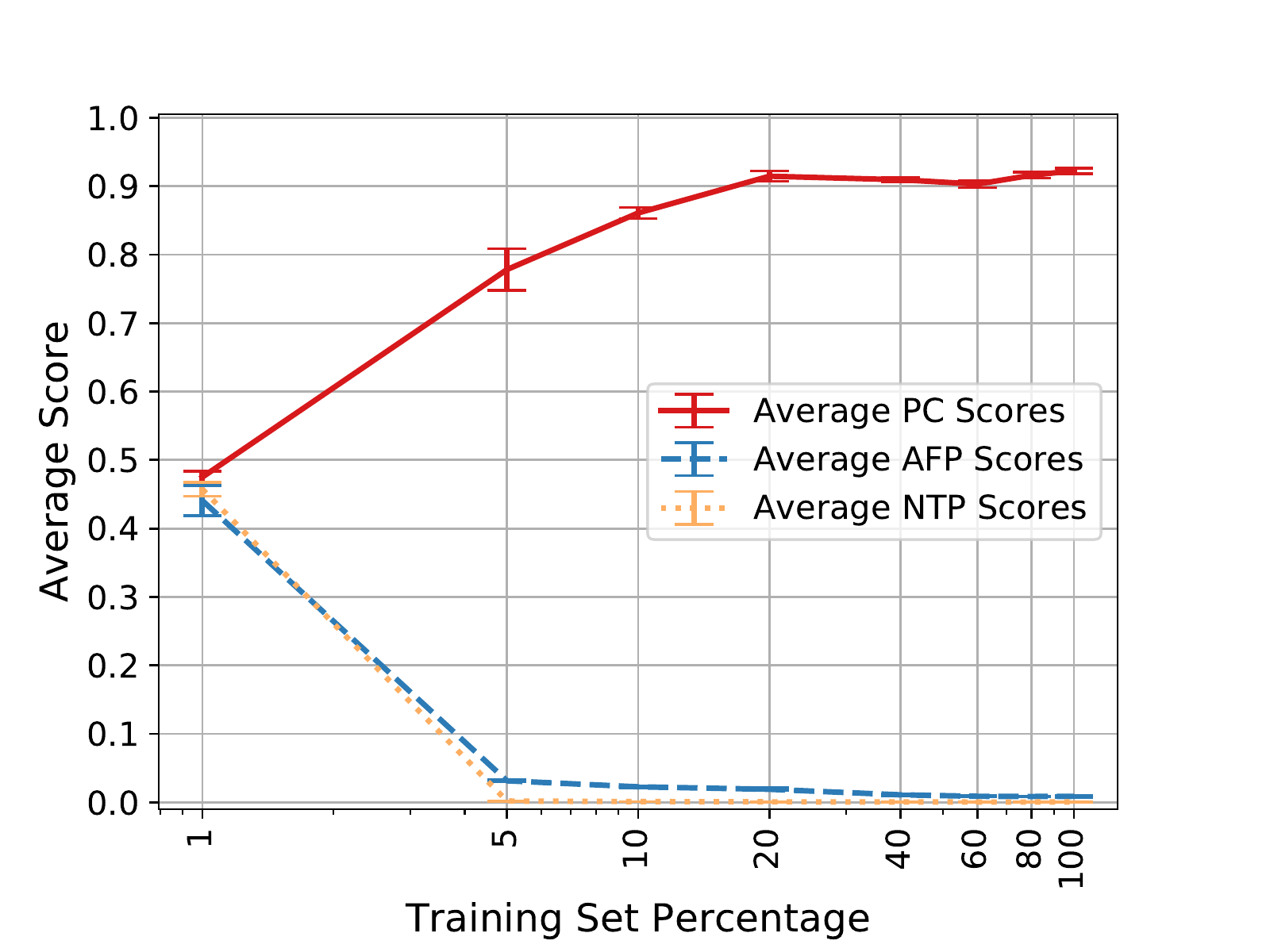}}
	\hskip -.25in
	\subfigure[Performance and Scores for Threshold=0.99]{\label{fig:training-size-0.99}\includegraphics[width=.35\textwidth]{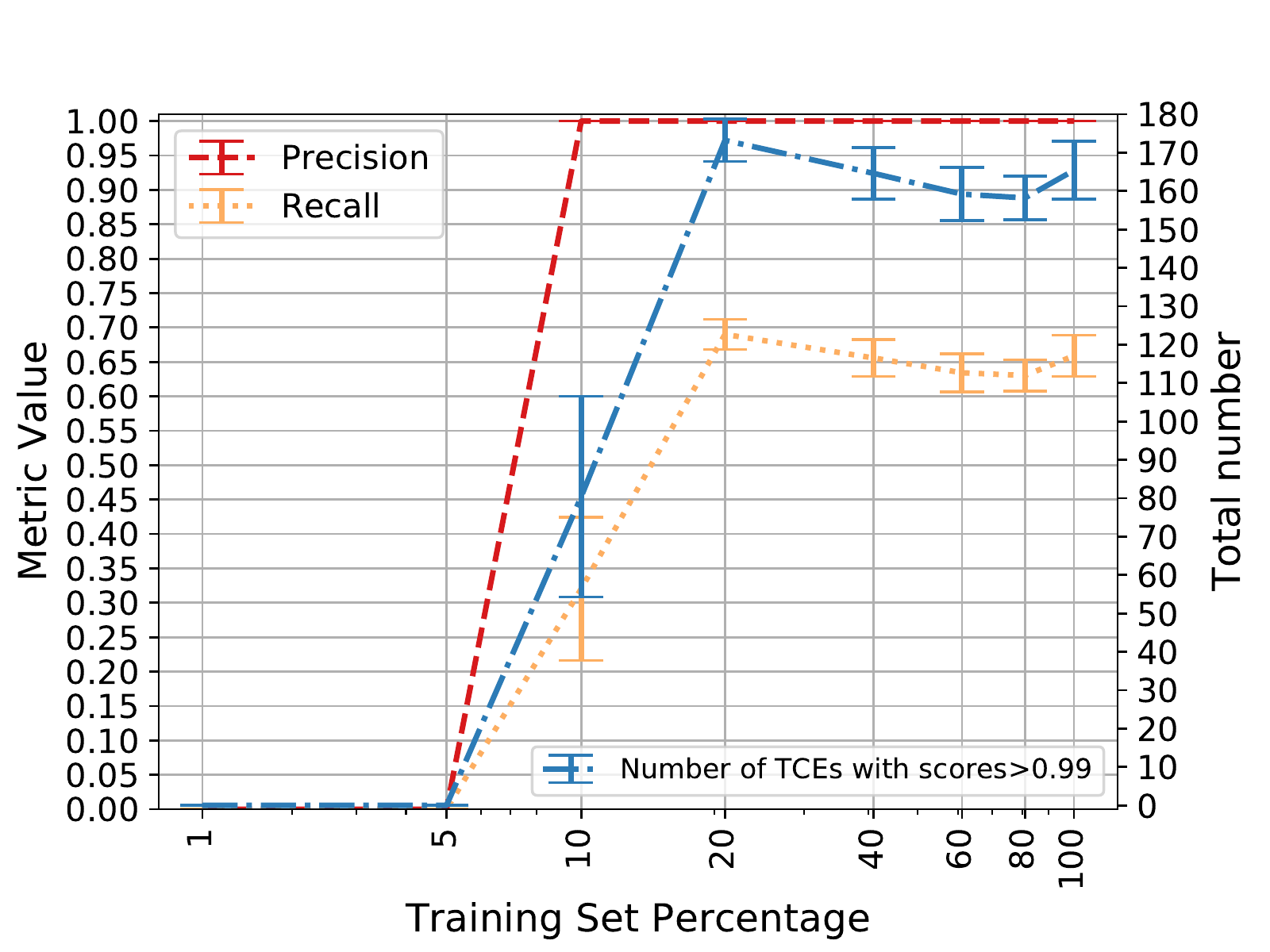}}
	\caption{Randomly selecting $q\%$ TCEs from the original training set to train the model. The plots show the mean value $\pm$ standard deviation of different test set metrics over the 5 runs for each value of $q$. The right y-axis in Figure c shows the total number of test set TCEs with scores $>0.99$.} 
\label{fig:training-size-analysis}
\end{figure*}

\begin{figure*}[htb!]
	\centering
	\subfigure[Performance for Threshold=0.5]{\label{fig:training-size1-AFP}\includegraphics[width=.35\textwidth]{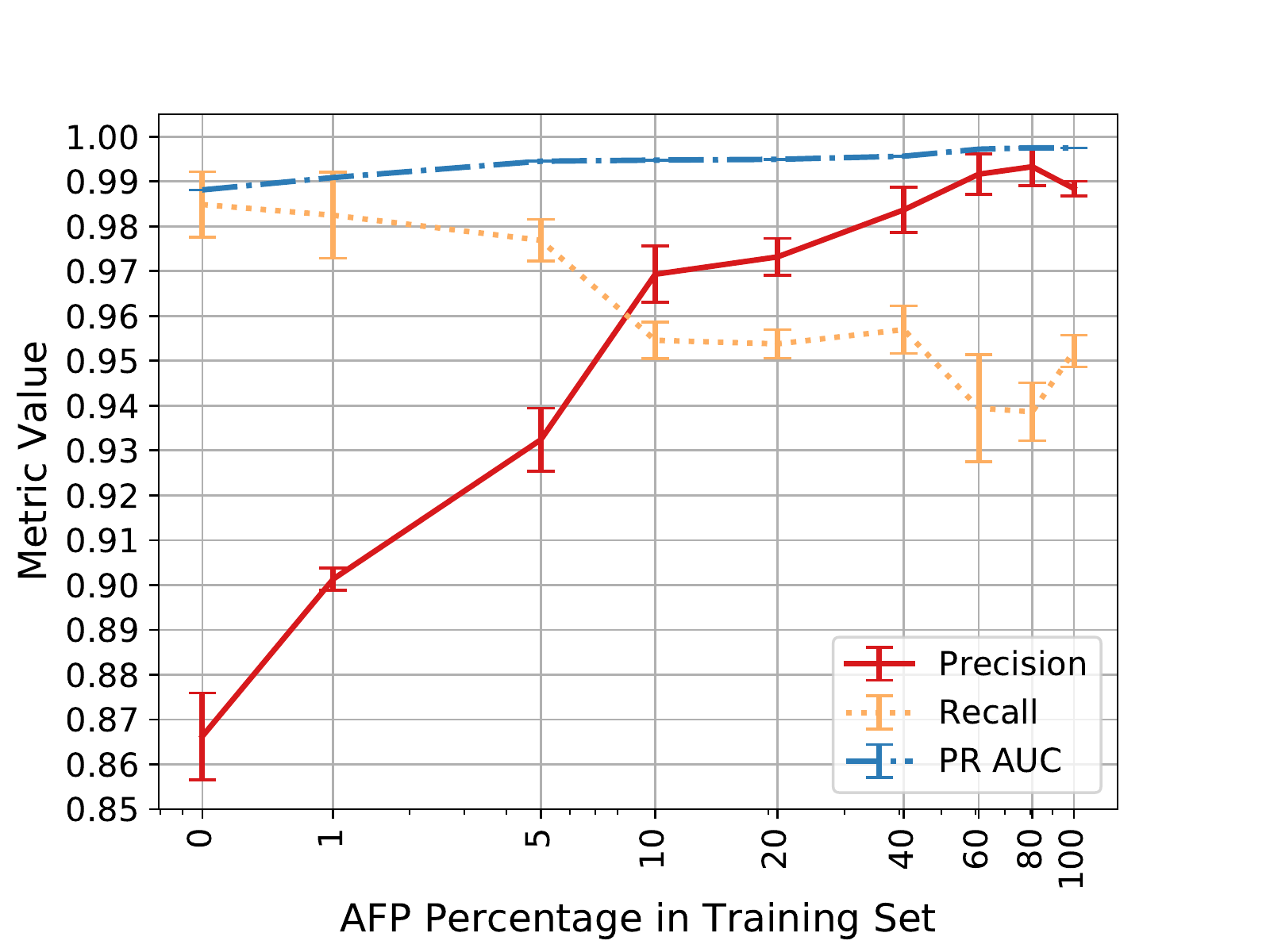}}
	\hskip -.25in
	\subfigure[Average Scores for Threshold=0.5]{\label{fig:training-size2-AFP}\includegraphics[width=.35\textwidth]{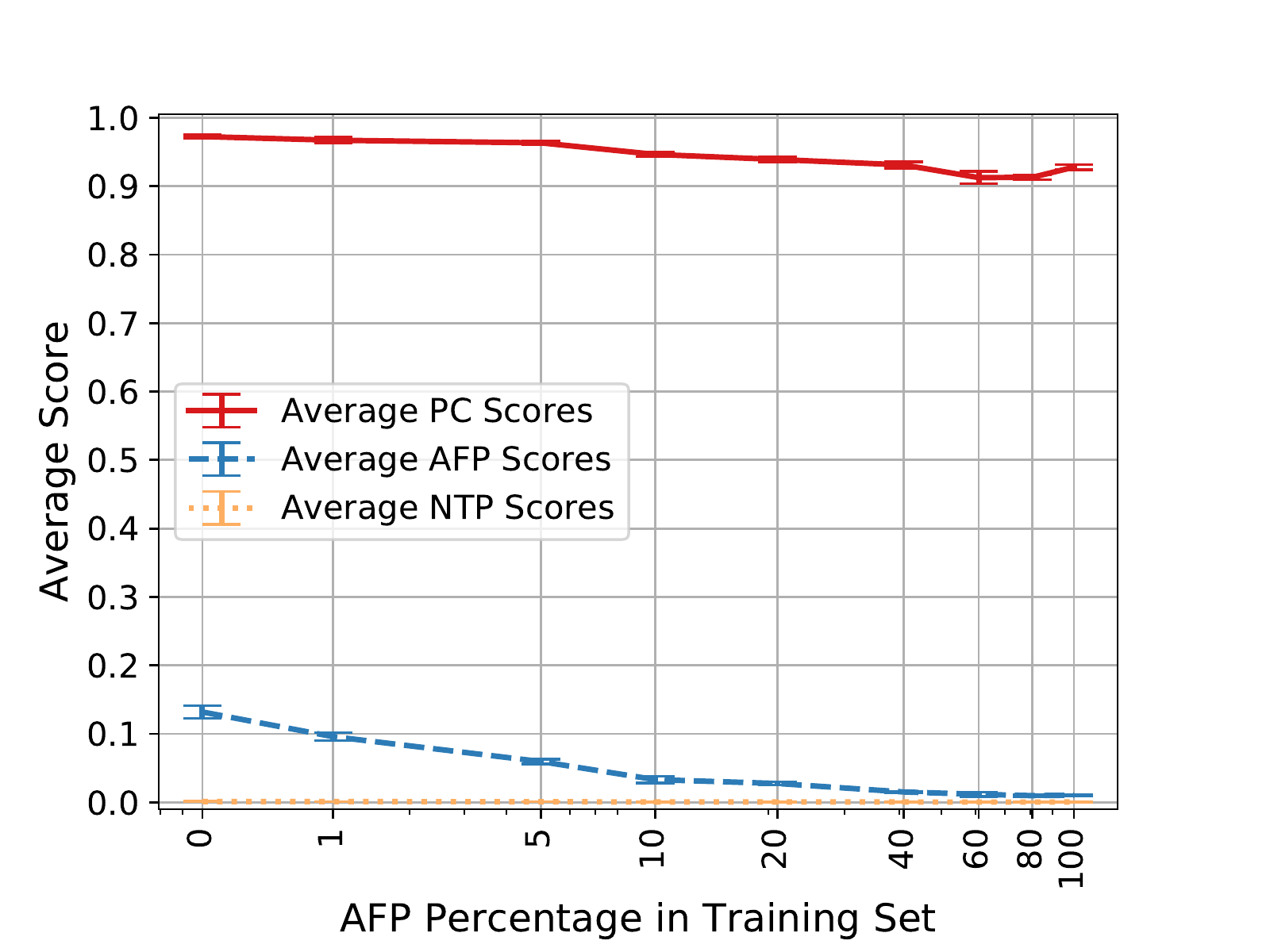}}
	\hskip -.25in
	\subfigure[Performance and Scores for Threshold=0.99]{\label{fig:training-size-0.99-AFP}\includegraphics[width=.35\textwidth]{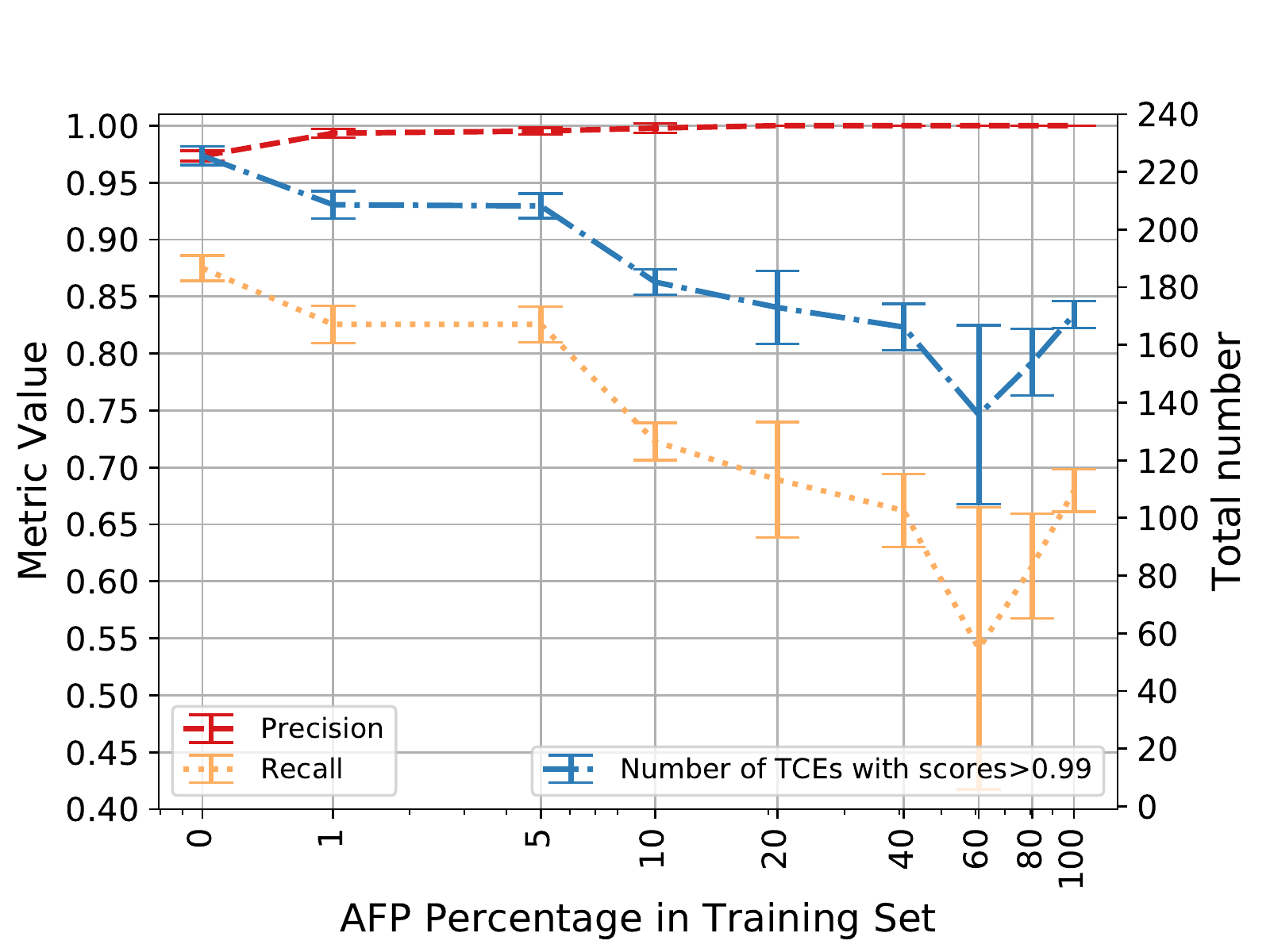}}
	\caption{Randomly keeping only $q\%$ AFPs from the original training set to train the model. The plots show the mean value $\pm$ standard deviation of different test set metrics over the 5 runs for each value of $q$. The right y-axis in Figure c shows the total number of TCEs in the test set with scores $>0.99$.} 
\label{fig:training-size-analysis-AFP}
\end{figure*}

First, we would like to test the hypothesis that the size of the training set for this problem might not be large enough in order to cover different data scenarios, i.e., the existence of underrepresented scenarios in the training set. As we keep reducing the size of the training set, we start removing some of the more uncommon scenarios in the dataset at some point, preventing the model from learning from these specific observations, and leading to a decrease in performance for the fixed test set. We study this using the following two settings:
\begin{itemize}
\item{\textbf{Full training set:}} In this first setting, we reduce the size of the full training set as follows: 1) we split 80\% of our data into training, 10\% into validation, and 10\% into test sets and 2) we randomly select $q\%$ of training and validation sets to train a new model and check its performance on the held-out fixed test set generated in step (1). Figure~\ref{fig:training-size1} shows the results of this experiment where we report precision, recall, PR AUC, average scores for each category (PC, AFP, and NTP) for $q\%$ varying between $1\%$ and $100\%$ over five random runs for each $q\%$ to curb stochastic effects due to the random selection of a subset of TCEs. As expected, the performance of \ExoMiner\ decreases in terms of all evaluation metrics as the size of the training set decreases. However, it is very robust even up to 5\%. The main effect is the decrease in the recall values, and the narrowing of the differences between scores of PCs and FPs, as shown in Figure~\ref{fig:training-size2}. \\

To study the effect on the validation of new exoplanets when a threshold $=0.99$ is used, we plotted in Figure~\ref{fig:training-size-0.99} the precision, recall and the total number of cases for this threshold on the test set. For $q=5\%$, no TCEs in the test set receive scores $>0.99$, meaning that for a small enough training set there are no validated TCEs in this set. By increasing the size of the training set, the total number of TCEs with score $>0.99$ grows up to $q=20\%$ and then does not change. \ExoMiner\ becomes more confident of the labels that assigns for the test set TCEs for a larger training set size. Interestingly, at $q=10\%$, there is a large variance over the five runs of experiments in terms of both the number of TCEs with scores $>0.99$ and recall value. The higher variance for $q=10\%$ means that 10\% of the training set is not large enough for the model to classify PCs with high confidence. Depending on the amount of information provided by 10\% of the training set to classify different scenarios in the test set, the total number of TCEs with scores $>0.99$ in the test set can vary largely.
\item{\textbf{Only AFPs:}} In this setting, we only reduce the number of AFPs in the training and validation sets in step 2) of the above procedure. This experiment tests the behavior of the \ExoMiner\ for the hypothesis that some AFP scenarios might not be well represented in the training set compared to the variation that exists in the PC category. Figure~\ref{fig:training-size-analysis-AFP} shows the performance when the percentage of AFPs is decreased down to zero percent. As mentioned before, because the relative rate of planets versus AFPs increases when lowering the percentage of AFPs in the training set, the precision decreases and recall increases in the test set. However, even at a one percent AFPs, precision is not significantly affected. At the extreme case of zero percent AFPs, the precision decreases to around 86.5\%, which is still reasonably good. The good performance observed at zero percent AFP might be due to the presence of AFPs in the NTP category. A subset of AFPs in the NTP (non-KOI) category corresponds to the secondary eclipses detected for EBs, as we discussed in Section~\ref{sec:datacorrection}. Removing those 2089 AFPs in the NTP category leads to another 2\% reduction of precision (precision=$0.842$). This shows there is an inherent label noise between these two categories, and that there are still more AFPs in the NTP category that allow the learner to understand the difference between PCs and AFPs. Furthermore, these results seem to show that the label noise between PC and non-PC categories is minimal. If the PC category was contaminated significantly with FPs, we would expect to see a significant decrease in performance in the test set as the number of AFPs in the training set is reduced.

More interestingly, precision at a validation threshold of $0.99$ is not much affected even at as low as one percent AFPs, as shown in Figure~\ref{fig:training-size-0.99-AFP}. Also, note that the average scores of AFPs in the test set increase with lower number of AFPs in the training set, but overall they are far from the classification threshold of $0.5$. 
\end{itemize}

Lower average scores (lower confidence of the model) for smaller training set size can be explained using bias/variance trade-off that we discussed in Section~\ref{sec:machine-classification}. When the model does not see enough examples, it can easily overfit to the training set due to small sample size. To prevent overfitting, the learner automatically stops optimizing the model parameters on the training set when the performance on the validation set starts degrading. This early stopping results in a less confident model that has lower PC scores and higher FP scores. At the extreme 1\% (or zero percent for the second experiment), the average PC scores become very similar to average FP scores on the test set and the model performs poorly. To understand this behavior in terms of bias/variance tradeoff, note that the variance of model increases with smaller training set, i.e., there is more variation of the model performance when the size of the training set is reduced. Thus more regularization, including early stopping, is required in order to reduce variance. This will increase the bias. One can also look at this from a Bayesian perspective; the model aims to label an unseen (test set) transit signal $x$ using the evidence provided by training set $\MD$. As we increase the size of $\MD$, the model becomes more confident about the label of $x$ because the model has seen more similar cases.

Based on these results, \ExoMiner\ shows robustness to the degree of representation of different sub-populations of PCs and FPs in the training set. Even in cases where those groups are misrepresented, \ExoMiner\ is able to learn useful representations to distinguish between PCs and non-PCs and attain a reasonable performance.

\begin{figure*}[htb!]
	\centering
	\subfigure[Performance at threshold=0.5.]{\label{fig:koi-noise1}\includegraphics[width=.35\textwidth]{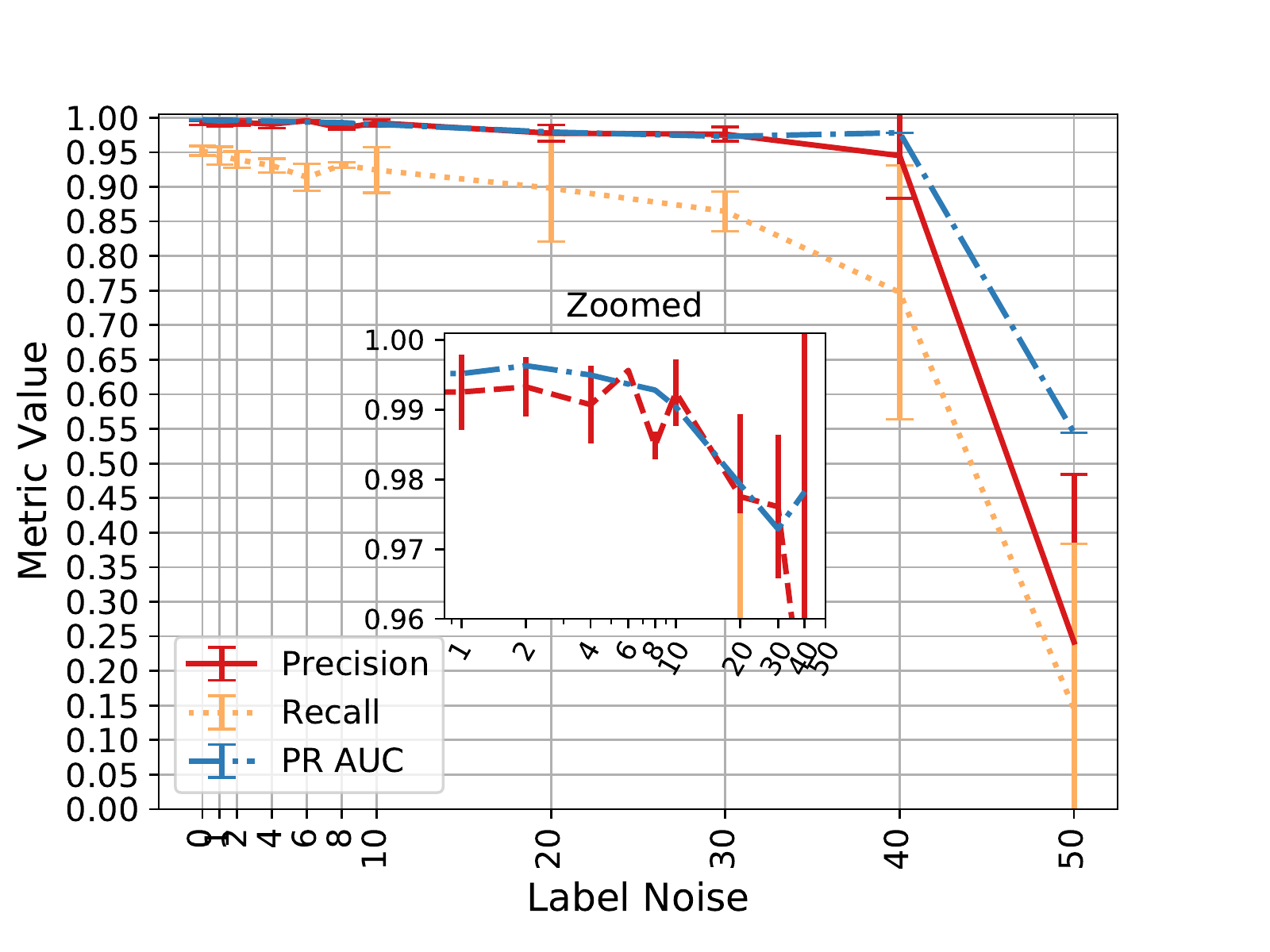}}
	\hskip -.25in
	\subfigure[Average scores at threshold=0.5.]{\label{fig:koi-noise2}\includegraphics[width=.35\textwidth]{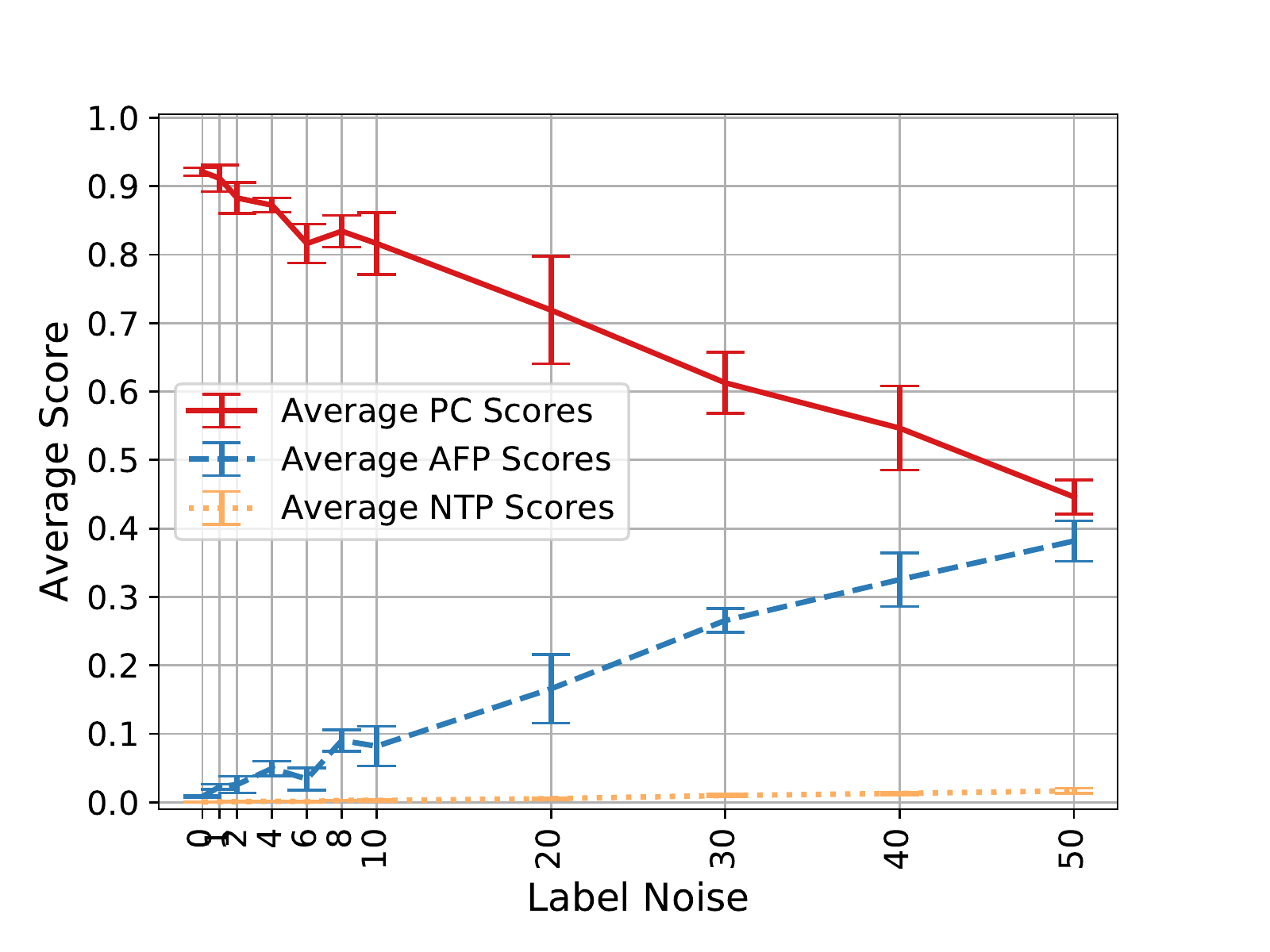}}
	\hskip -.25in
	\subfigure[KOI label noise at threshold=0.99.]{\label{fig:koi-noise-0.99}\includegraphics[width=.35\textwidth]{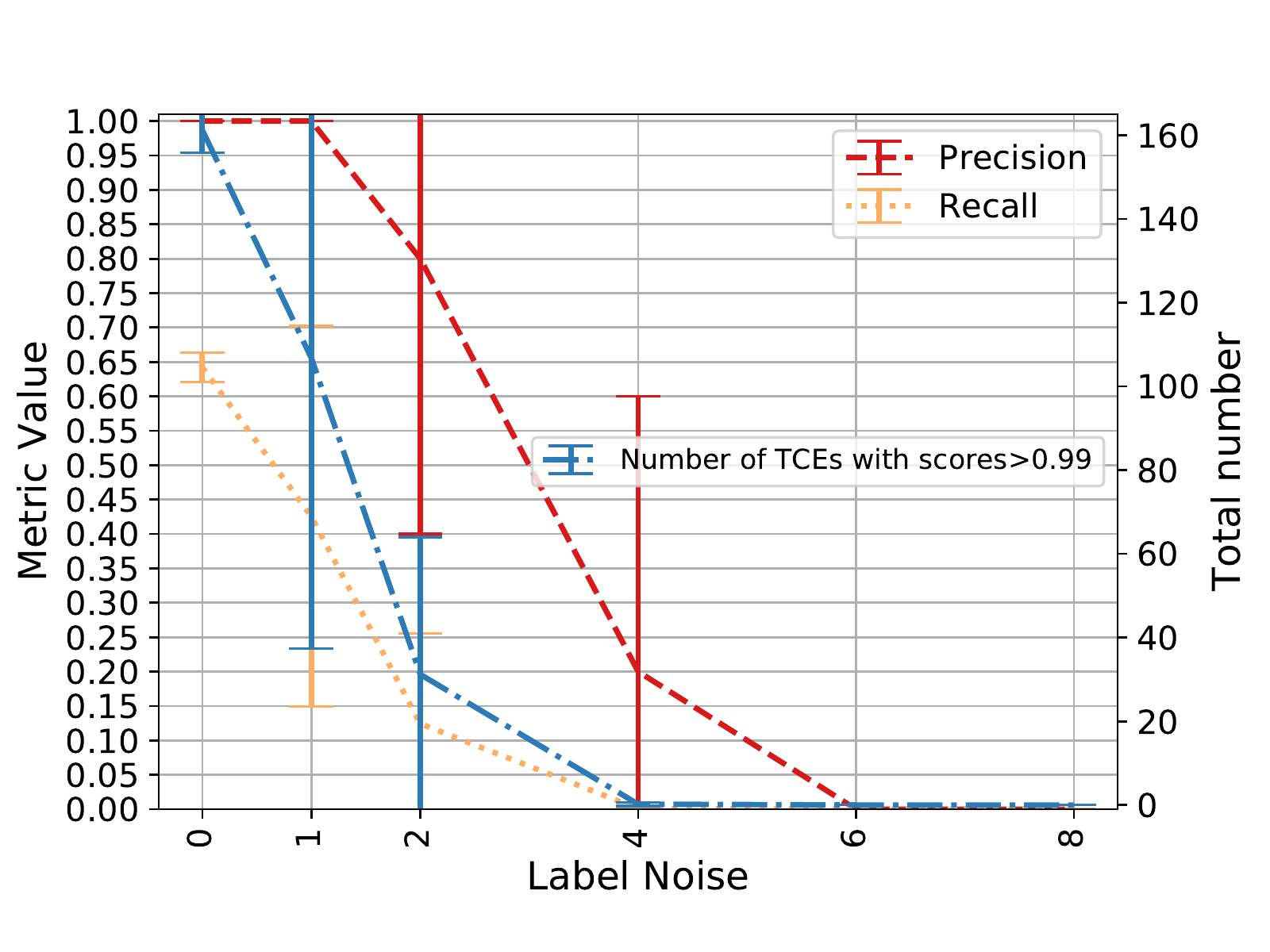}}
	\caption{Randomly switching labels of $q\%$ KOIs in the training set to train the model. The plots show the mean value $\pm$ standard deviation of different test set metrics over the 5 runs for each value of $q$. The right y-axis in Figure c shows the total number of TCEs in the test set with scores $>0.99$.} 
\label{fig:koi-training-noise-analysis}
\end{figure*}

\begin{figure*}[htb!]
	\centering
	\subfigure[Performance at threshold=0.5.  ]{\label{fig:tce-noise1}\includegraphics[width=.35\textwidth]{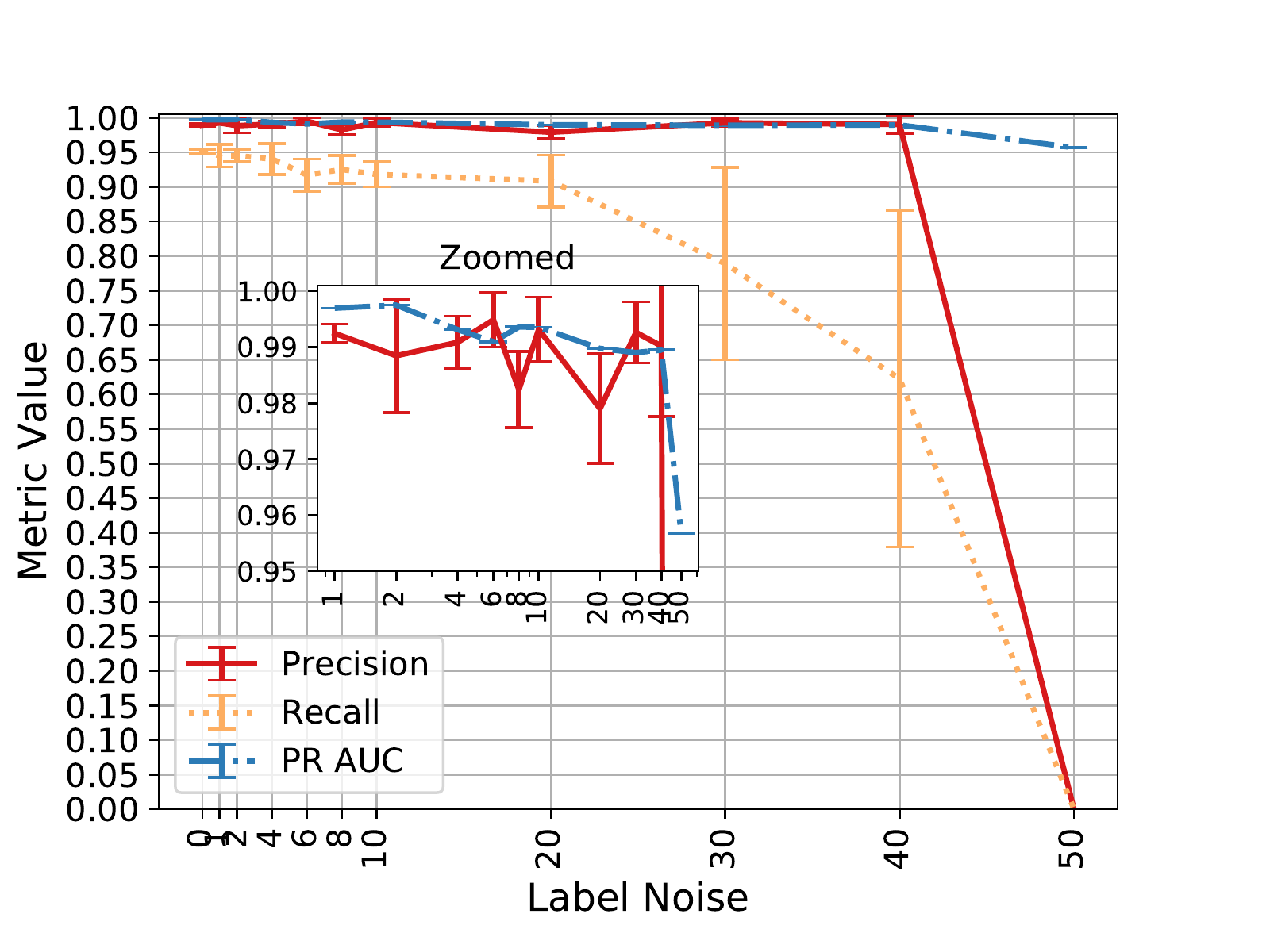}}
	\hskip -.25in
	\subfigure[Average scores at threshold=0.5.  ]{\label{fig:tce-noise2}\includegraphics[width=.35\textwidth]{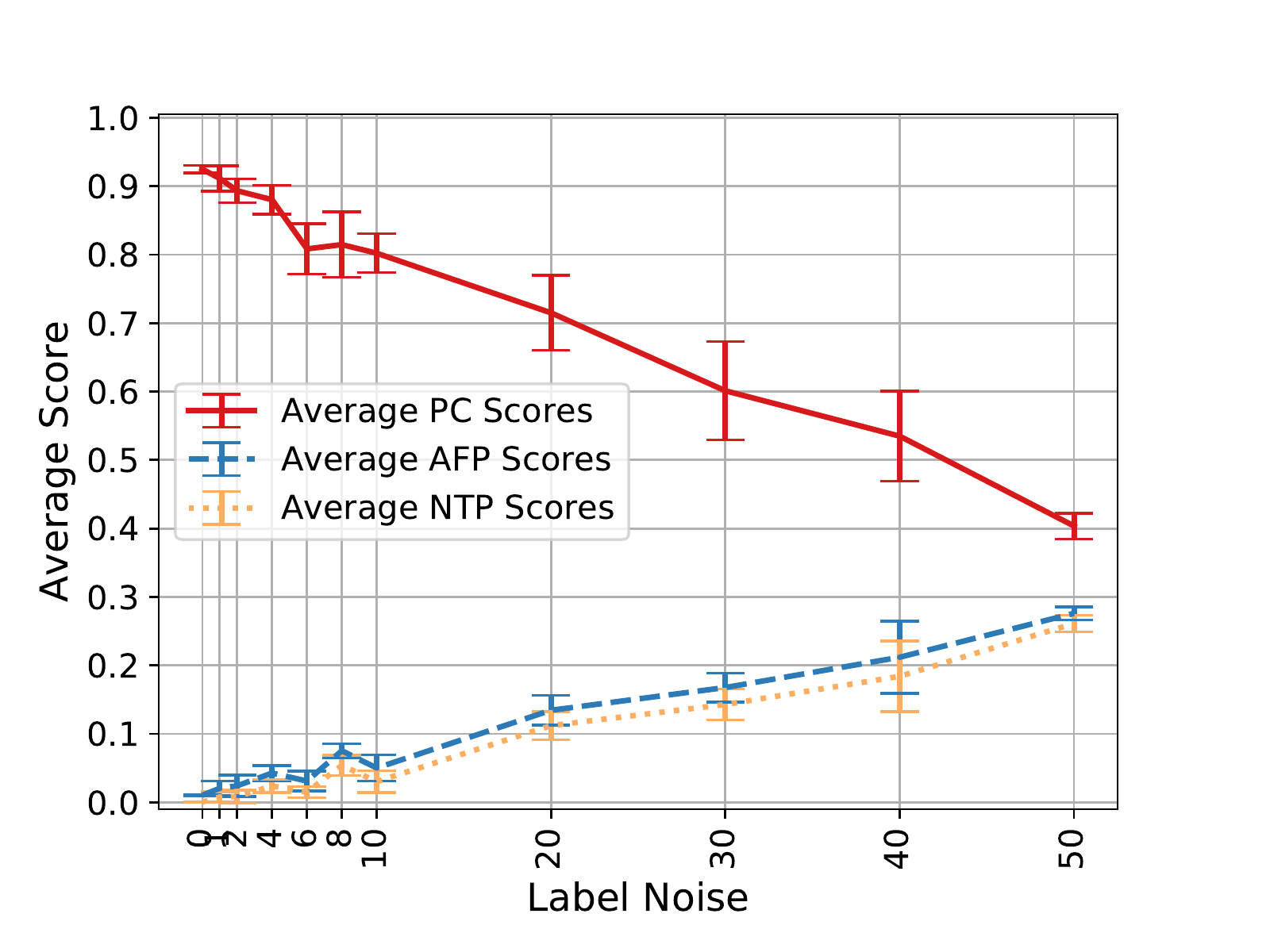}}
	\hskip -.25in
	\subfigure[TCE label noise at threshold=0.99]{\label{fig:tce-noise-0.99}\includegraphics[width=.35\textwidth]{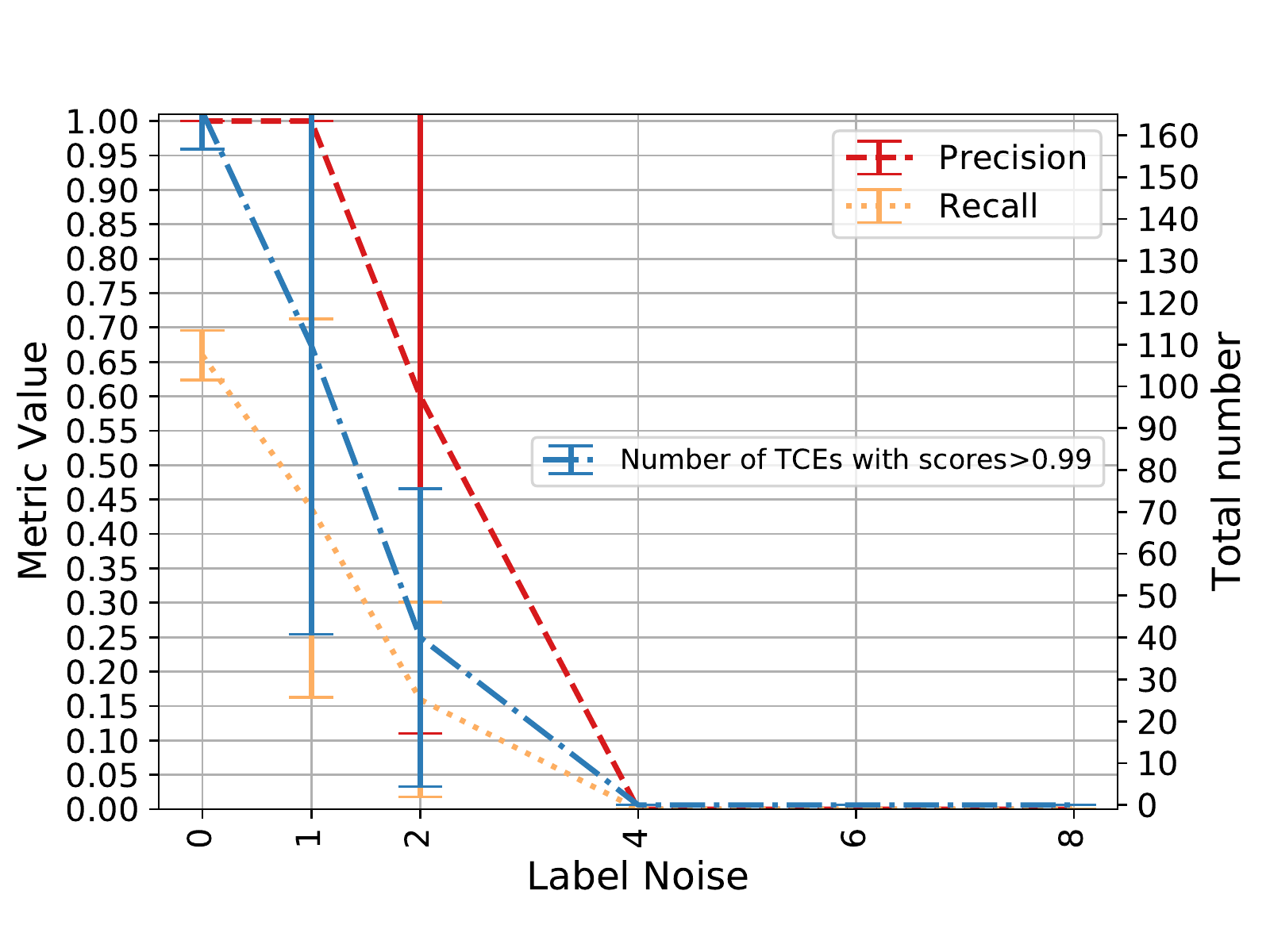}}
	\caption{Randomly switching labels of $q\%$ TCEs in the training set to train the model. The plots show the mean value $\pm$ standard deviation of different test set metrics over the 5 runs for each value of $q$. The right y-axis in Figure c) shows the total number of TCEs in the test set with scores $>0.99$.} 
\label{fig:tce-training-noise-analysis}
\end{figure*}

Second, we would like to test the effect of label noise on the performance of \ExoMiner. Similar to the previous experiment, we split the data into 80\% training, 10\% validation, and 10\% test sets. We then artificially inject label noise into both training and validation sets by switching the labels of TCEs, train the model using this noisy data, and report its performance on the original held out fixed test set. For each noise level setting, 5 runs are performed to curb stochastic effects due to the random selection of a subset of TCEs whose label is flipped.We run the following two experiments: 
\begin{itemize}
\item{\textbf{KOI label noise:}} Given that distinguishing PCs from AFPs is a more difficult task, we only flip the label of $q\%$ of AFPs to PCs and $q\%$ of PCs to AFPs in this experiment (no added label noise in NTP category). Figure~\ref{fig:koi-noise1} shows the performance of \ExoMiner\ for varying values of $q$ from $1\%$ to $50\%$. Even at $q=40\%$, \ExoMiner\ is able to perform well on the test set. This result is consistent with the previous research~\cite[e.g.,][]{Rolnick-DNNlabelnoice-2017}. However, \ExoMiner's confidence on its assigned class label for PCs decreases with increasing label noise, as shown in Figure~\ref{fig:koi-noise2}. As expected, the distribution of scores for PCs shifts towards lower values, while  the opposite happens for AFPs. At $q=50\%$, in which $50\%$ of PCs and AFPs have their labels flipped, \ExoMiner\ performs poorly and the average PC and AFP scores get very close to each other, even though the average PC score is still higher than the average AFP score. The reason that \ExoMiner\ can still have higher scores for PCs is that there are still FPs in the NTP category whose labels are not flipped. Using these FPs, the model is able to understand similar cases and perform better than random guessing. Similar to the experiment for training set size, we plot the performance of \ExoMiner\ for validation threshold=$0.99$. For KOI label noise as small as $4\%$, no KOI\footnote{To be precise there are $0.2$ KOIs, i.e., only one KOI out of five runs.} gets a score $>0.99$ and there are only about 30 TCEs passing the validation score at KOI label noise $2\%$. The total number of validated TCEs for $q=0\%$ is around 160. 
\item{\textbf{TCE label noise:}} In this experiment, we flip the labels of $q\%$ of PCs to non-PCs and $\frac{q}{2}\%$ of AFPs and $\frac{q}{2}\%$ of NTPs to PCs in both training and validation sets. This leads to $q\%$ label noise in PC class and $q\%$ label noise in FP class (AFPs+NTPs). Figure~\ref{fig:tce-noise1} shows the performance of \ExoMiner\ for this experiment for varying values of $q$ from $1\%$ to $50\%$. Because the total number of TCEs in the NTP class is much more than the other two classes, the amount of label noise is effectively larger in this experiment compared to the previous setting. Nonetheless, \ExoMiner\ performs better compared to the previous setting. This is because AFPs are more difficult to classify than NTPs. Similar to the previous setup, the model is highly robust up to $40\%$ label noise but confidence in its label decreases more rapidly when compared to KOI label noise (Figure~\ref{fig:tce-noise2}. The reason for this behavior is that a large number of NTPs are labeled as PCs. We plot in Figure~\ref{fig:tce-noise-0.99} the effect of this label noise on the validation threshold=$0.99$, which shows similar behavior to the KOI label noise.    
\end{itemize}

Note that the results of label noise experiments can also be justified from the bias/variance trade-off, similar to the training set size experiment. 

The above experiments imply that if the size of training set is not large enough or if there is a considerable amount of label noise in the training set, \ExoMiner\ is not able to classify TCEs with high scores. This is a very important property when we use \ExoMiner\ to validate new exoplanets. 

\subsection{Analysis of Misclassified Cases}
\label{sec:misclassified-cases}
In this section, we examine TCEs that were misclassified in order to better understand the behavior of \ExoMiner. 

There is a total of 60 CPs in all ten folds (six PCs in average per fold) misclassified by \ExoMiner. Most of these misclassified CPs are cases for which one diagnostic test fails, as we detail below. 
\begin{itemize}
\item{Centroid test:} One subset of CPs are misclassified by \ExoMiner\ solely because there is some degree of centroid shift, based on the DV summary report. Note that the centroid motion test may fail due to aperture crowding, imperfect background removal, or saturated stars~\citep{Twicken_2018_DV} that are unknown to the model at the time of training or classification. Thus, the model does not have access to such information to correctly classify TCEs that have a clear but unreliable centroid shift. Note that \kepler\ magnitude ($K_p$) can be used to recognize the saturated target stars for which the centroid shift test is invalid. An example of a highly saturated target star that hosts misclassified planets is Kepler-444. \ExoMiner\ does not currently use $K_p$ as an input feature; however, $K_p$ will be added in the next version of \ExoMiner\ to improve the model's performance on saturated target stars. 

\item{Odd \& even test}: There are CPs that have a considerable odd and even transit depth difference when looking at the absolute depth values of odd and even views. However, to make sure that the transit depth difference between odd and even views is statistically significant and not due to chance, one must use a statistical test that considers the variability of the depth value of different transits in the odd and even views, and also the total number of transits for each view; this can be described by a confidence interval for the sample mean of the transit depth for both views.  
Given that the information related to the number of transits per view and the variability of the depth for each view is not available to \ExoMiner, the model is not able to understand the reliability of the depth difference to correctly classify such cases. Examples of misclassified CPs in this category are KICs $8891684.1$ and $10028792.2$.  
Adding information such as the number of transits and a measure of the uncertainty for the transit depth estimate of each view to the odd \& even branch is the next step for improving the performance of \ExoMiner.

\item{Secondary test:} There are CPs that have a significant secondary transit event. These can be further categorized into three groups: 
\begin{enumerate}
   \item Some of these TCEs are certified as false positives in the CFP list~\citep{Bryson-2015-certifiedlist}. Examples of such TCEs are KICs $11517719.1$ and $7532973.1$. 
   \item Some other TCEs in this category, e.g., KICs $10904857.1$, $3323887.1$, and $11773022.1$, have a clear secondary transit event with statistically significant geometric albedo and planet effective temperature statistics. Thus, the information provided in the DV summary report leads to an FP classification for these TCEs by \ExoMiner. TCE KIC $11773022.1$ (Kepler-51~d) failed the secondary test because the weak secondary detector triggered on transits of another planet in the system. For KIC $10904857.1$, the odd and even transit depth comparison and the centroid motion tests are also flagged by DV. Thus, the non-PC classification of this TCE by \ExoMiner\ is well justified. FPWG comment in the CFP list for this TCE reads as follows: ``Clearly there is a secondary, but is it due to a planet? 250 ppm. Albedo is right near 1.0 given current stellar parameters. Logg is likely underestimated, so star is likely bigger, pushing it further into possible planet territory."

   \item A third group of TCEs in this category, e.g., hot Jupiters such as TCE KIC $4570949.1$, have a clear secondary transit event, and their geometric albedo and planet effective temperature comparison statistics are not significant enough to yield a false positive disposition. It seems that \ExoMiner\ failed to capture the relationship between the secondary transit events and the albedo and planet effective temperature comparison statistics. It is not clear why \ExoMiner\ is not able to understand such cases. We will study ways to improve this behavior of \ExoMiner\ with respect to the secondary test in our future work.
\end{enumerate}
\end{itemize}

There is a total of 74 misclassified FPs (7.4 FPs per fold in average) out of $28318$ AFPs+NTPs in the full dataset that can be categorized as follows:
\begin{itemize}
\item{Ephemeris Match Contamination:} There is a total of sixteen TCEs with their ``Ephemeris Match Indicates Contamination'' flag on. This flag means that a transit signal shares the same period and epoch as another object and is the result of flux contamination in the aperture or electronic crosstalk. For eleven out of these sixteen TCEs, the only information used to vet them as false positives is the ``Ephemeris Match Indicates Contamination'' flag, which is not available to \ExoMiner. Examples of TCEs in this category are KICs $9849884.1$, $10294509.1$, and $11251058.1$. 

\item{Centroid test:} A good portion of false positive cases have some centroid shift that \ExoMiner\ fails to capture. It is not very clear why \ExoMiner\ fails to correctly classify these TCEs. One possible explanation could be that there are CPs with centroid shift (as discussed earlier with regards to misclassified CPs) in the training set that prevent the model from capturing the correct patterns in the data. Examples of TCEs in this category are KICs $6964159.1$ and $5992270.1$. 

\item{V-shape transits:} For some TCEs, the only useful information available in the DV reports for correctly classifying them is the V-shape form of the transit signal. Given that there are also exoplanets with V-shape transit signatures with high impact parameters, it is not clear how \ExoMiner\ should learn to classify such FP cases. Examples of TCE cases in this category are KICs $6696462.1$ and $7661065.1$. 

\item{Odd \& even test:} There are FPs with statistically significant odd and even transit depth difference that \ExoMiner\ fails to classify correctly. Similar to the centroid motion test, the model is provided with conflicting information during training; i.e., there are CPs with a clear odd and even transit depth difference if one does not take into account the uncertainties for the transit depth estimates. We believe the model would be able to correctly classify cases like this if we quantified the uncertainty regarding the depth values as additional features (as discussed earlier in this section). An example of a TCE in this category is KIC 8937762.1.  
\end{itemize}

\section{Explainability of ExoMiner}
\label{sec:exominer-explainability}
In this section, we report the development of an explainability technique for analyzing the classifications made by \ExoMiner. In the past few decades, advances in computing power and improvements in ML methods have allowed for the development of highly accurate data-driven models; however, increasing a model’s performance often comes at the cost of increasing its complexity. Thus, explainability techniques are of broad and current interest in the field of ML because models that are deployed for any given application should also have human interpretable outputs to aid in the decision-making process. 

In the case of exoplanet detection, a robust model will not only be able to accurately distinguish between PC and non-PC TCEs, but can also provide an explanation for its classifications. For example, an FP classification could be complemented with an explanation that the model identified the TCE as an EB based on its transit-view of odd \& even test. Our goal is to develop a post-hoc explainability technique for labeling FP classes without explicitly training the model on these labels. Instead of searching for more granular insights about which specific region in a time series is most important to the model\textsc{\char13}s classifications as was studied by~\citep{shallue_2018}, we propose the development of a branch-occlusion sensitivity technique to identify, more broadly, which diagnostic test has the greatest contribution to the model\textsc{\char13}s classifications of each TCE. The multi-branch DNN design of \ExoMiner, inspired by the DV report, is ideally suited for performing such explainability tests.

In our preliminary study of the explainability of \ExoMiner, we devise a simple branch-occlusion explainability test that yields false positive flags by identifying the branch of the model that has the most significant contribution to the predicted score of each TCE. To determine which branch of the model makes the greatest contribution to a TCE\textsc{\char13}s score, we mask the inputs of each time series branch by setting the input data to that branch to zero. The DNN generates six new classification scores from sequentially masking each of the six branches (full-orbit flux, transit-view flux, full-orbit centroid, transit-view centroid, transit-view odd \& even, transit-view secondary). Note that we excluded stellar parameters and DV diagnostic parameters from this preliminary explainability study because these scalar parameters only provide insight when combined with other branches, a combinatorial effect we did not consider in this paper. For each TCE, we identify the branch whose exclusion from the model changes the model's disposition score the most in a positive or negative way. Thus, the input data to this maximally contributing branch contains the information that most influenced the model to label the TCE as an FP or PC, depending on the class label predicted by the model. 

To evaluate the effectiveness of this branch-occlusion explainability framework, we split the data into 80\% training, 10\% validation, and 10\% test set. We train a model using these training and validation sets and applied it on the test set using our explainability test. This provided us with the following insights about \ExoMiner:

\textbf{Classification of PCs:} In this analysis, we were interested in understanding how \ExoMiner\ classifies TCEs as PCs. After examining the contribution of different branches, we learned that the full-orbit flux view has a significant positive contribution for PC classifications, meaning that without this flux information, most PCs cannot be correctly classified. This makes sense because if all diagnostic tests (the centroid tests, odd \& even test, secondary test, and the shape of the transit in transit-view of flux data) are unremarkable, their contributions to the disposition score are minimal. We are able to come up with a simple rule for explaining how \ExoMiner\ classifies most of the PCs: if the full-orbit branch is the maximally contributing branch and its contribution toward the PC category is greater than 0.1, \ExoMiner\ likely classifies that TCE as a PC. To be more precise, 90.83\% of TCEs whose maximally contributing \ExoMiner\ branch is the full-orbit flux view, with a contribution greater than $0.1$, are indeed PCs in the test set (similar statistics in other sets). On the other hand, 99.12\% of PCs in the test set satisfy this rule, i.e., for 99.12\% of PCs, the full-orbit flux view is the maximally contributing branch with a contribution greater than 0.1 toward the PC class. In other words, this simple rule for explaining how \ExoMiner\ classifies TCEs as PCs has a precision of 0.991 and a recall of 0.908.

\textbf{Classification of AFPs:} In this analysis, we were interested in understanding whether \ExoMiner\ utilizes the correct type of diagnostic test in order to vet an AFP as non-PC. Based on domain knowledge of the diagnostic information contained in each branch\textsc{\char13}s input data, we map the different false positive classes to the branches whose inputs would be most crucial for manually diagnosing the respective false positive type. Specifically, TCEs with their greatest contribution from the transit-view secondary eclipse branch or the transit-view odd \& even flux branch are labeled as EB FPs, and TCEs with their greatest contribution from the full-orbit centroid branch or transit-view centroid branch are labeled as background object FPs. Given that we did not have access to gold standard labels for these subcategories, we utilized the flags,  i.e., `Eclipsing Binary High Level Flag' and `Offset High Level Flag', provided in the CFP list~\citep{Bryson-2015-certifiedlist} as gold standard labels. Using these noisy labels, we found that the FP labels generated by applying the branch-occlusion test to \ExoMiner\ match those provided by the CFP list 78.92\% of the time. We believe that when we perform a more careful explainability study of \ExoMiner, we will improve our ability to flag the different TCE false positive types. 

\textbf{Example TCEs:} Table~\ref{table: example-TCEs-Explainability} shows a few TCEs for which our explainability analysis provides further insights and demonstrates how useful explainability can be in practice for understanding how the model works and finding ways to improve it. The positive/negative contribution of a branch implies that removing that branch increases/decreases the score of that TCE, which makes that TCE more/less likely to be a PC. 

\begin{table*}[htb]
 \centering
\caption{Contribution of different branches in classifying interesting TCEs. Note that the contribution is calculated by subtracting \ExoMiner's original score without occluding any branches from the score when a given branch is occluded. Thus, a negative value implies that the branch is contributing towards a PC label, and a positive value implies that the branch is contributing towards a non-PC label.}
\label{table: example-TCEs-Explainability}
\begin{threeparttable}
\begin{tabularx}{\linewidth}{@{}Y@{}}
\begin{tabular}{cccccccccc}
\toprule
TCE KIC & Period (days) & Original Label & \ExoMiner\ Score & $B_1$ & $B_2$ & $B_3$ & $B_4$ & $B_5$ & $B_6$    \\
 \midrule
6543893.1\tnote{1} & 10.30 & PC & 0.498 & -0.440 & -0.400 & -0.006 & \textbf{0.270} & 0.124 & 0.120 \\

9141355.1\tnote{2} & 469.621 & PC & 0.425 & \textbf{-0.188} & \textbf{-0.325} & -0.034 & -0.027 & \textbf{0.101} &	0.016 \\ 

11517719.1\tnote{3} & 2.50 & PC & 0.353 & -0.340 & -0.168 & 0.001 & 0.034 & 0.003 & \textbf{0.245}\\ 
 \midrule
9996632.1\tnote{1} & 8.44 & AFP & 0.017 & -0.014 & -0.012 & 0.030 & \textbf{0.971} & 0.004 & 0.030\\

8937762.1\tnote{2} & 7.53 & AFP & 0.730 & \textbf{-0.606} & \textbf{-0.298} & -0.223 & 0.061 & \textbf{0.256} & -0.034\\

6500206.1\tnote{3} & 13.375 & AFP & 0.154 & -0.135 & -0.023 & -0.079 & -0.047 & 0.033 & \textbf{0.838}\\
\midrule

 6666233.2\tnote{4} & 1.0248 & NTP & 0.066 & \textbf{0.578} & -0.027 & -0.001 & -0.010 & 0.014 & 0.023 \\
 
 10186945.2\tnote{4} & 0.794 & NTP & 0.095 & \textbf{0.549} & -0.019 & -0.019 & -0.008 & 0.007 & 0.005\\
 
 8052016.4\tnote{5} & 33.1842 & NTP & 0.004 & -0.003 & \textbf{0.165} & -0.002 & -0.000 & 0.075 & 0.008 \\
 
 8043714.2\tnote{6} & 3.407 & NTP & 0.013 & -0.006 & -0.000 & 	0.004 & 0.000 & \textbf{0.696} & -0.003\\
 
 9705459.2\tnote{6} & 1.243 & NTP & 0.000 & 0.000 & 0.000 & 0.000 &	0.000 & \textbf{0.635} & 0.000  \\

5446285.2\tnote{7} & 416.653 & NTP & 0.003 & -0.003 & -0.003 & 0.000& 0.000 & 0.002 & \textbf{0.803}\\

\bottomrule
\end{tabular}
\end{tabularx}
\begin{tablenotes}[para,flushleft]\scriptsize
  \item[] $B_1$: Full-orbit flux, $B_2$: Transit-view flux, $B_3$: Full-orbit centroid, $B_4$: Transit-view centroid, $B_5$: Transit-view odd-even, $B_6$: Transit-view secondary.
  \item [1] The centroid branch is one of the main contributors to the final score.
  \item [2] The odd and even transit depth difference is one of the main contributors to the final score.
  \item [3] The secondary branch is one of the main contributors to the final score. 
  \item [4] The secondary transit of this TCE should refer to the first TCE. It is not corrected because the phase information for the secondary of the first TCE is inaccurate.
  \item [5] The transit-view of flux reveals that this is not a real transit. 
  \item [6] After removing the primary, the pipeline folded the secondaries onto the data gaps left behind by the primary event. Thus, the period for the second TCE in each case was one half that of the primary that resulted in an odd and even depth difference for the second TCE.
  \item [7] Even though this is an NTP, there is a clear secondary event helping the model to classify this as a non-PC. 
\end{tablenotes}
\end{threeparttable}
\end{table*}

For the PC category, the full-orbit flux branch is almost always (99.1\% of the time) the maximally contributing branch towards a PC classification (negative values). However, when one or more other branches have a positive non-negligible contribution, that leads to a misclassification. We provide three examples in Table~\ref{table: example-TCEs-Explainability} for which other diagnostic test branches, besides the full-orbit branch, are the maximally contributing branch. TCEs KIC~6543893.1,  KIC~9141355.1, and KIC~11517719.1 are misclassified due to centroid shift, odd-even depth difference, and existence of a weak secondary event, respectively. For each of these TCEs, the related branch is correctly used to reduce the score of \ExoMiner. Thus, \ExoMiner\ clearly utilized the correct diagnostic test for each classification even though it misclassified those TCEs.  

To correctly classify an AFP, however, the contribution of one of the diagnostic tests should be large enough to compensate for the contribution of the full-orbit flux branch. When that fails, \ExoMiner\ is not able to classify the TCE correctly. An example of an AFP that is misclassified by \ExoMiner\ is TCE KIC~8937762.1, for which the contribution of the transit-view odd \& even branch is not large enough to compensate for the negative contribution value of the full-orbit and transit-view flux branches. TCEs KIC~9996632.1 and KIC~6500206, for which the transit-view centroid and weak secondary branches, respectively, are the main contributing branches, are correctly classified by \ExoMiner. 

\begin{figure}[htb!]
	\centering
	\subfigure[Secondary event for KIC TCE $6666233.1$]{\label{fig:wrong-phase-secondarya}\includegraphics[width=\columnwidth]{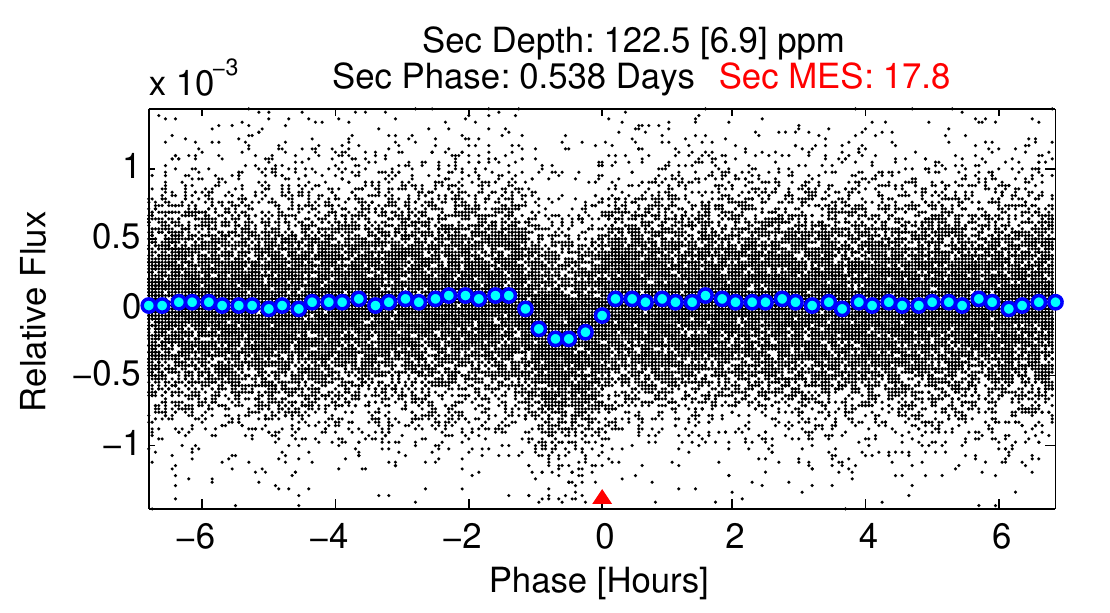}}
	\subfigure[Secondary event for KIC TCE $10186945.1$]{\label{fig:wrong-phase-secondaryb}\includegraphics[width=\columnwidth]{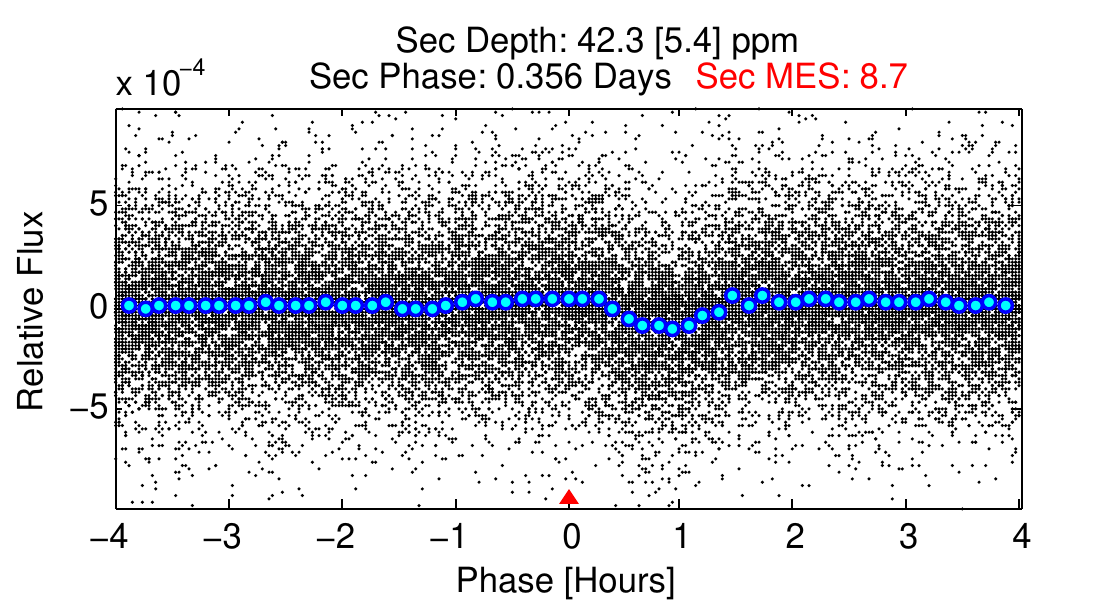}}
	\caption{The phase value pointing to the secondary event is inaccurate, which prevents us from correcting the secondary of the second TCE of that EB system (see Section~\ref{sec:data-driven}).} 
\label{fig:wrong-phase-secondary}
\end{figure}

\begin{figure}[htb!]
	\centering
	\subfigure{\label{fig:b}\includegraphics[width=\columnwidth]{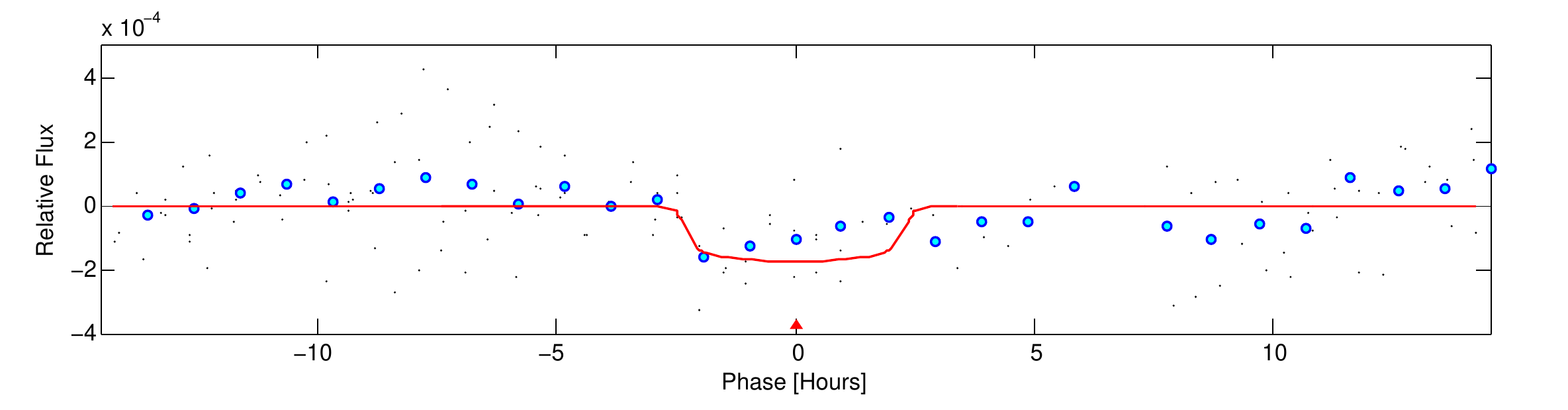}}
	\caption{KIC TCE~8052016.4 is an NTP and is correctly classified based on its unusual transit-view flux shape. Note that the blue dots are the real flux data, and the red solid line is the transit fit generated by the pipeline.} 
\label{fig:weird-transit-view}
\end{figure}

The most interesting insights provided by this explainability analysis are those related to the NTP class. For two TCEs, i.e., KIC~6666233.2 and KIC~10186945.2, which are labeled as NTP, the branch contributing maximally toward a non-PC classification (i.e., positive value) is the full-orbit view. Both of these TCEs are the secondary eclipse of an EB; thus, their correct label is AFP. Our preprocessing failed to correct the secondary event for these TCEs because the phase generated by the pipeline for the secondary transit of the primary event (i.e., TCEs KIC~6666233.1 and KIC~10186945.1) was inaccurate (Figure~\ref{fig:wrong-phase-secondary}). Thus, there was no transit in the secondary test for \ExoMiner\ to capture and utilize to generate a correct classification. However, because we do not remove other TCEs when processing our flux data for each TCE, the secondary transit (i.e., transit corresponding to TCEs KIC~6666233.1 and KIC~10186945.1) was visible in the full-orbit view of KIC~6666233.2 and KIC~10186945.2, and the model could use them in order to classify these TCEs correctly. 

Another interesting NTP is TCE KIC~8052016.4. The branch contributing the most to \ExoMiner's score for this TCE is the transit-view flux branch. A close look at the transit-view of this TCE (Figure~\ref{fig:weird-transit-view}) reveals that the model captured the unusual transit shape of this TCE and labeled it correctly. Other interesting scenarios are provided by TCEs KIC~8043714.2 and KIC~9705459.2, which are actually part of an EB but are labeled as NTP. In both cases, the primary events were completely removed by the pipeline and the second TCE was triggered by the remaining secondaries. However, the pipeline folded the secondaries onto the data gaps left behind by the primary event, so the period for the second TCE in each case was one half that of the primary. Because of this, the depth of the odd and even views are different for the subsequent TCE since the transits in one view are residuals of the original TCE while transits in the other view represent another transit. The last example is TCE KIC~5446285.2, which is correctly classified using the secondary event branch. The target star associated with this TCE is Kepler-88 which is a Transit Timing Variation (TTV) system and hosts multiple exoplanets. The TTVs for Kepler-88~b are sufficiently large ($\pm$0.5~days) that the pipeline was not able to lock on to the nominal orbital period. At the detected period, other transits of the planet are actually identified as the secondary for TCE KIC~5446285.2 leading to its correct classification.

The results of this preliminary study indicate that identifying the maximally contributing branch to each TCE\textsc{\char13}s predicted score holds a high degree of explainability power for an extremely simple explainability technique. To develop higher fidelity explainability tests that can be integrated into the \ExoMiner\ classification pipeline, several improvements to this technique will be explored in future work. The model architecture diagram shown in Figure~\ref{fig:dnn-architecture} shows that the stellar parameter scalars branch and the DV diagnostic scalars branch exist at a different hierarchical level than the convolutional branches. Rather than setting the inputs of each convolutional branch to zero, if we instead set each branch\textsc{\char13}s contribution to the fully connected block to zero, the FP flags we obtain may be more accurate. In particular, this is important for the transit-view odd \& even flux branch where an input of all zeroes has a different meaning due to the subtraction layer. Achieving more accurate flags of eclipsing binaries through the transit-view odd \& even flux branch may be possible by implementing noise sensitivity mapping, as discussed in Section~\ref{sec:explainability}. More efficient gradient-based methods like Grad-CAM~\citep{Selvaraju_2019_explainability} will also be explored as components of a more robust explainability testing for \ExoMiner. Refining these explainability tests will not only provide more information about the model’s classifications, but will also provide insight into how the model can be improved in future iterations. As \ExoMiner\ is applied to TESS data, these explainability tests will be helpful in providing additional confidence in the dispositions and pointing out the limitations of the model for classifying new signals.

\section{ExoMiner for exoplanet validation}
\label{sec:exominer_validation}

\subsection{Classification with Rejection}
\label{sec:rejection}
\begin{figure}[htb!]
\begin{center}
\includegraphics[width=85mm]{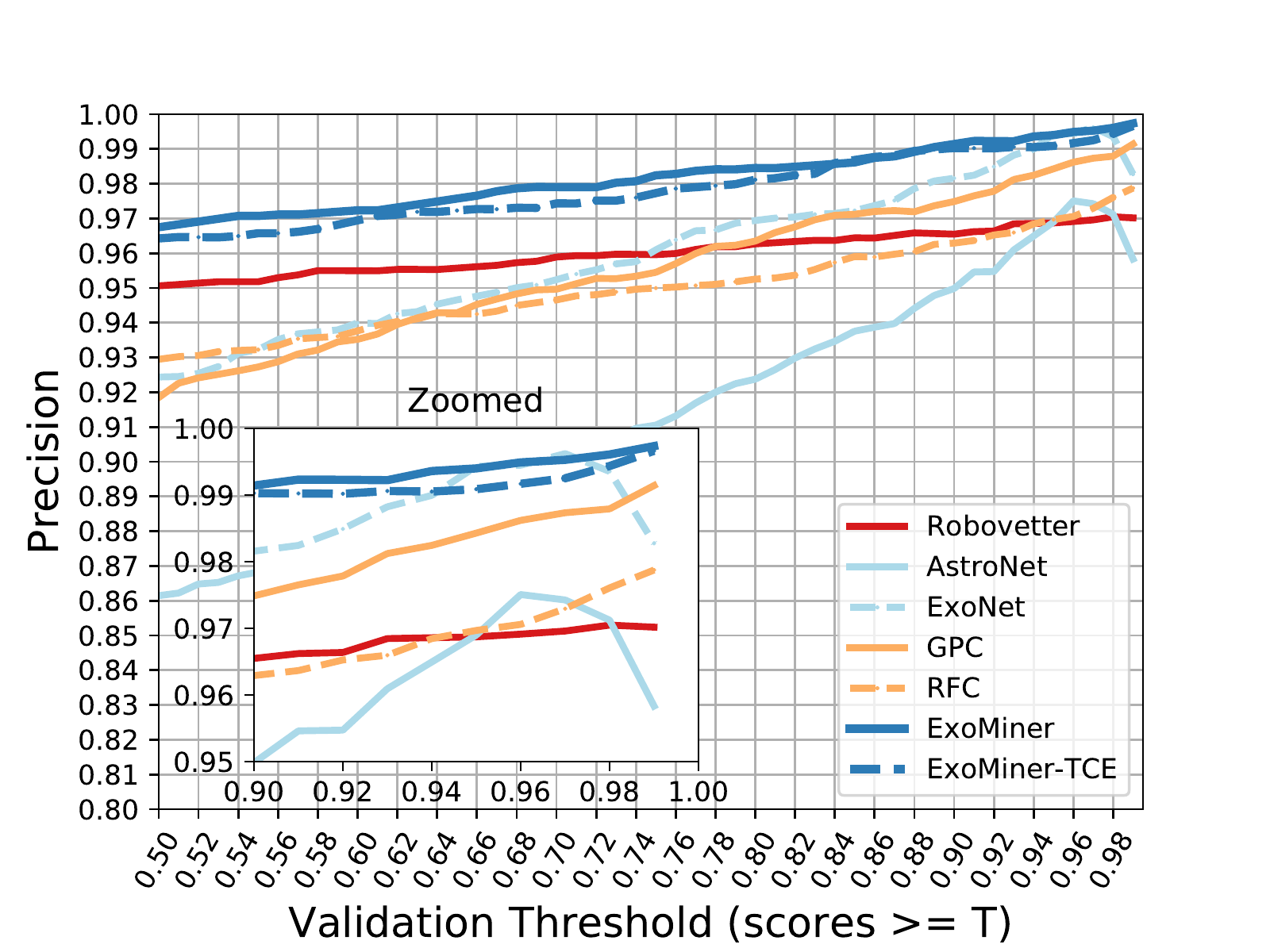}
\caption{Precision of different classifiers when a threshold $T$ is used to reduce the false positive rate.}
\label{fig:rejectionplots}
\end{center}
\end{figure}

For applications where the cost of misclassification is high (e.g.,~validating new exoplanets), we need ways to reduce the error rate of the classifier. This is a relatively old concept in statistics~\citep{Chow-1970-Rejection} and a standard one in ML~\citep{Cortes-2016-rejection, Nadeem-2010-Rejction}. It is called classification with rejection. One classification with rejection approach widely used in ML is based on failing to classify examples when the model is not very confident of its label. The idea is that if the classifier cannot classify a case with high confidence, then it is not sure about its label. When the classifier predicts the label of an example with a very high score, it means that the model has a good understanding of the feature space in which that example lies (think about it in terms of $p(y|X)$). As we reported in Section~\ref{sec:training_set_sensitivity}, by reducing the size of the training set or adding label noise, \ExoMiner\ becomes less confident about its disposition. This verifies that by setting a confidence-based rejection option, we control the risk of misclassification. To study confidence-based rejection, we plot in Figure~\ref{fig:rejectionplots} the precision of different classifiers when different threshold values for rejection are used to reduce the rate of false positives. As can be seen, all methods perform reasonably well for large thresholds. Note that the precision of \ExoMiner\ at $T=0.99$ is around 99.7\%, much higher than the required precision value of 99\% for that threshold. The precision of \ExoMiner\ even for $T=0.9$ is around 99\%.  

It is important that one uses a model-specific threshold $T$ for classification with rejection because the precision does not necessarily increase with larger thresholds, e.g., check \AstroNet\ and \ExoNet\ for $T$ $>0.97$. However, given that $T=0.99$ has been accepted to use for validating new exoplanets~\citep{Morton-2016-vespa,armstrong-2020-exoplanet} and \ExoMiner\ performs very well for that threshold, in the remainder of this paper, we focus on using this threshold for different analyses (in addition to the standard classification threshold of 0.5). 

\begin{figure}[htb!]
	\centering
	\subfigure{\label{fig:pp-pp}\includegraphics[width=\columnwidth]{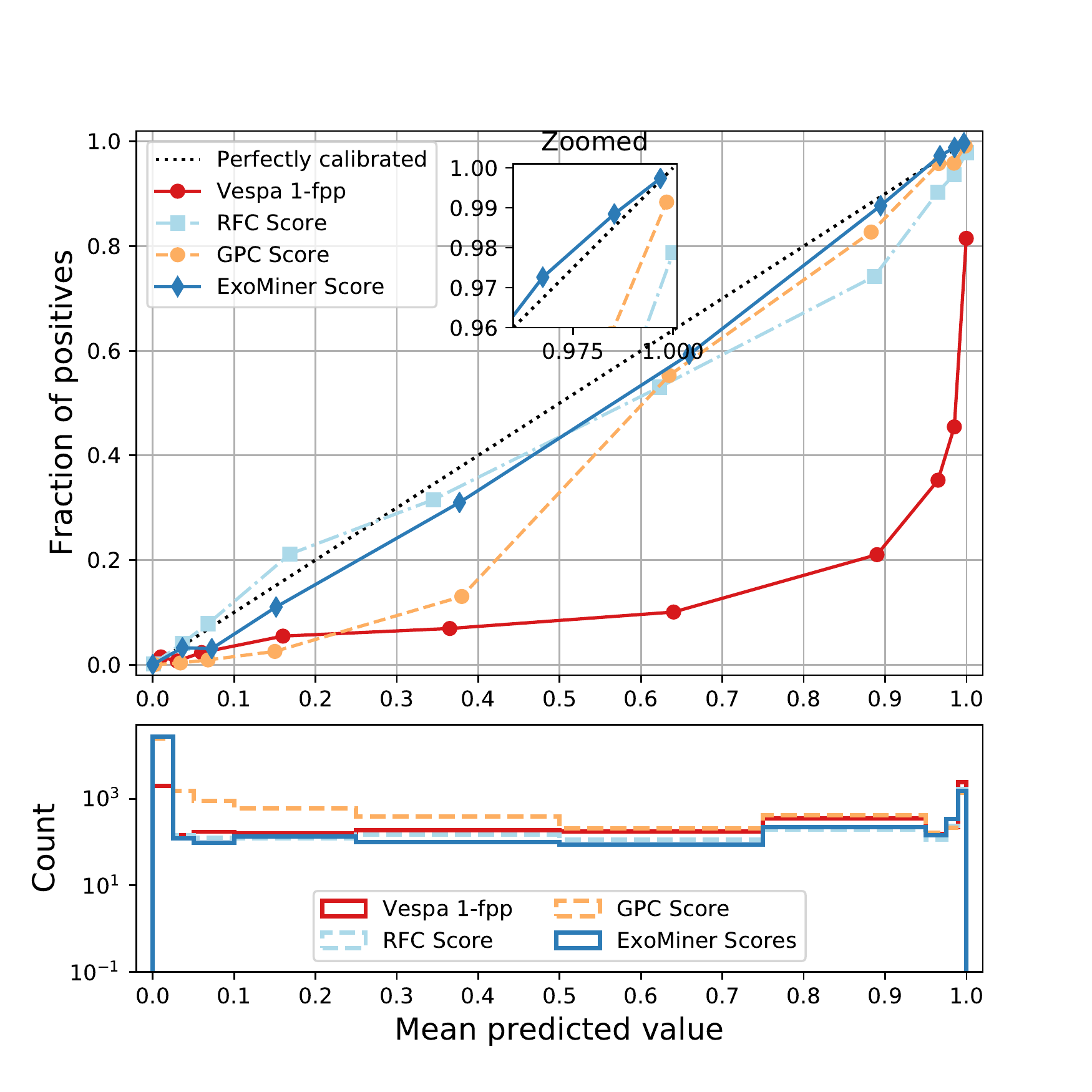}}	
\caption{Reliability plots of \ExoMiner, \GPC, \RFC, and \vespa.}
\label{fig:reliabilitycurve}
\end{figure}

\newpage
\subsection{Classifier Calibration}
\label{sec:calibrated_classifier}
In order to make a decision based on the output of a classifier (e.g., to validate new exoplanets), we should be able to interpret the scores generated by that classifier. If a classifier generates probability scores, we can estimate the error rate for different probability scores, i.e., for a given probability score value of $s$, the classifier correctly classifies $100\times s\%$ of examples. Such classifiers are called calibrated. Most discriminative models for machine classification, however, do not generate calibrated scores even when their generated scores are in the $[0,1]$ range. Such classifiers require calibration before their scores can be used for decision making, especially when a decision must be taken for any scenario (e.g., consider a self driving car making a decision to stop or keep moving). Because of this,~\citet{armstrong-2020-exoplanet} calibrated their classifiers prior to validating new exoplanets using isotonic regression~\citep{Niculescu-2015-isotonic}, one of the two most common calibration methods~\citep{Platt-1999-plattscaling, Niculescu-2015-isotonic}.

\begin{table}[!htbp]
 \centering
\caption{The level of calibration for different methods. Lower values are better.}
\label{table:ECE_MCE}
\begin{threeparttable}
\begin{tabularx}{\linewidth}{@{}Y@{}}
\begin{tabular}{ccc}
\toprule
Classifier & ECE & MCE  \\
 \midrule
\vespa (1-fpp) & 0.246 & 0.68\\

\GPC & 0.013 & 0.249 \\

\RFC & 0.009 & 0.145 \\

\ExoMiner\ & 0.001 & 0.067 \\

\bottomrule
\end{tabular}
\end{tabularx}
\end{threeparttable}
\end{table}

To study the calibration of a classifier, people use calibration plots (reliability curves) that show the error rate of the classifier for different probability score values. A classifier is perfectly-calibrated if the error rate is equal to the score it generates. Figure~\ref{fig:reliabilitycurve} shows the calibration plots of uncalibrated \ExoMiner, \vespa\ 1-fpp, and two calibrated classifiers\footnote{Note that \GPC\ is calibrated by design and RFC was calibrated in post-processing.} from~\citet{armstrong-2020-exoplanet}, \GPC\ and \RFC, for the full dataset (using CV). Note that \vespa\ fpp is only available for KOIs, so its reliability plot only uses a subset of the data used for the other three classifiers. The results and conclusion, with regard to the level of calibration, do not change if we use only KOIs for all models. As can be seen, \ExoMiner\ appears well calibrated already. To measure the level of calibration of these classifiers, we use Expected Calibration Error (ECE) and Maximum Calibration Error (MCE)~\citep{Guo-calibrationmetrocs-2017}. ECE and MCE are, respectively, the weighted average and maximum of the absolute difference between the classifier's average accuracy and score for each score bin. Table~\ref{table:ECE_MCE} shows the level of calibration for different classifiers used for validation of new exoplanets. As can be seen, the out-of-box \ExoMiner\ is well calibrated and can be used without further calibration. It has been previously shown that the scores generated by certain classifiers such as neural networks are well-calibrated~\cite{Niculescu-Mizil-calibratedclassifiers-2005}. Nonetheless, we tried both isotonic regression~\citep{Niculescu-2015-isotonic} and Platt scaling~\citep{Platt-1999-plattscaling} to further improve the calibration of \ExoMiner. These calibration techniques did not work well because the bins in the middle of the reliability plots do not have enough points per fold, leading to poor post-processing calibration. To overcome this problem,~\citet{armstrong-2020-exoplanet} generated synthetic examples by interpolating between members of each class. Generating such synthetic datasets for diagnostic test time series is not easy and does not guarantee better results. As such and because the level of calibration of \ExoMiner\ is already high, we use the scores generated by the classifier directly for planet validation. 

We would like to emphasize that the calibration of methods for scores $>0.99$ is of particular interest for validating new exoplanets. As can be seen in the zoomed area of Figure~\ref{fig:reliabilitycurve}, only  \ExoMiner\ is an underestimator classifier. A classifier is an underestimator when the fraction of positive examples for a given probability score is higher than the score assignment; i.e., the mean predicted value assigned by such a classifier is lower than its precision. This is an useful property for validating exoplanets because a classifier that is an underestimator in the high score range (close to 1.0) makes fewer errors when validating new exoplanets. \GPC, \RFC, and \vespa\ are all overestimator classifiers in that range, meaning that there is more than 1\% error when validating exoplanets using a threshold value of $0.99$. Note that these results are consistent with Figure~\ref{fig:rejectionplots} and the classification with rejection discussion.

\begingroup
\setlength{\tabcolsep}{2.5pt} 
\begin{table}[htb]
\centering
\caption{Scores of different classifiers trained in this work after applying priors. This table describes the available columns. The full table is available online.}
\label{table:classification_results_prior}
\begin{threeparttable}
\begin{tabularx}{\linewidth}{@{}Y@{}}
\begin{tabular}{ll}
\toprule
Column & Description \\
\midrule
target\_id & KIC ID\\
tce\_plnt\_num & TCE planet number\\
tce\_period & TCE period\\
mes & TCE MES\\
original\_label & NTP, AFP, or PC\\
binary\_label & 0 for NTP or AFP, 1 for PC\\
astronet\_prior & \AstroNet\ score with priors \\
exonet\_prior & \ExoNet\ score with priors\\
exominer\_prior & \ExoMiner\ score with priors \\
exominer\_tce\_prior & \ExoMiner-TCE score with priors \\ 
\bottomrule
\end{tabular}
\end{tabularx}
\end{threeparttable}
\end{table}
\endgroup

\begin{table*}[!htbp]
\small
 \centering
\caption{The effect of prior on the precision \& recall values of different models. The boost in performance due to prior diminishes as the model uses more complete set of diagnostic tests.}
\label{table:prior_probabilities}
\begin{threeparttable}
\begin{tabularx}{\linewidth}{@{}Y@{}}
\begin{tabular}{c|cccc }
\toprule
 & \multicolumn{2}{c}{Threshold=0.5} & \multicolumn{2}{c}{Threshold=0.99} \\
Model/Metric & Original & Using priors & Original & Using priors \\
\midrule
  \AstroNet & 0.861 \& 0.885 & 0.920 \& 0.920  &  0.958 \& 0.010 & 1.000 \& 0.102 \\

 \ExoNet & 0.925 \& 0.864 & 0.933 \& 0.902 & 0.983 \& 0.025 & 0.993 \& 0.123 \\
 
 \GPC & 0.921 \& 0.964 & 0.910 \& 0.977 & 0.991 \& 0.605 & 0.993 \& 0.663 \\
 
 \RFC\tnote & 0.929 \& 0.955 & 0.942 \& 0.964 & 0.979 \& 0.724 & 0.986 \& 0.791\\

 \ExoMiner  &  0.968 \& 0.974 & 0.971 \& 0.978 & 0.997 \& 0.656 & 0.997 \& 0.752\\

 \ExoMiner-TCE  & 0.965 \& 0.977 & 0.968 \& 0.983 & 0.997 \& 0.775 & 0.996 \& 0.823 \\
\bottomrule
\end{tabular}
\end{tabularx}
\end{threeparttable}
\end{table*}

\subsection{Utilizing prior information}
\label{sec:scenarior-priors}
As we mentioned in Section~\ref{sec:misclassified-cases}, \ExoMiner\ does not perform well under some specific data scenarios. In order to address that, we utilize prior probabilities for different scenarios as used in~\citet{Morton_2011_Vespa, Morton_2012_vespa, armstrong-2020-exoplanet}. More specifically, ~\citet{armstrong-2020-exoplanet} utilized prior probabilities for different scenarios (EB, Hierachical EB,  BEB, hierarchical transiting planet, background transiting planet, known other stellar source, non-astrophysical/systematic, planet) and combined them with their calibrated classifier scores to improve the performance of the classifier.  We take the same approach to combine the prior probability calculated in~\citet{armstrong-2020-exoplanet} with the scores of \ExoMiner, with a somewhat different justification, as follows. There are two sets of probability scores computed from two different sources of information: (1) $p(y=1|M, X, x^*)$ that learner $M$ (in our case \ExoMiner) generates for signal $x^*$ using the information from the training set $X$ and (2) $P(y=1|I)$ where $I$ represents target star's individual characteristics (stellar type, local environment, spacecraft status at the transit time, etc.). In order to combine these two scores, assuming that (1) I and (2) $x^*$, $M$, $X$ are conditionally independent (i.e., $p(x*,M,X,I|y)=p(x*,M,X|y)\times p(I|y)$), we can use the Bayes theorem to get the following equation that is the identical equation as the one reported in~\citet{armstrong-2020-exoplanet}:
\begin{equation}
\label{eq:prior_information}
p(y=1|M, X, x^*, I)=\frac{p(y=1|M, X, x^*)p(y=1|I)}{\sum_y p(y|M, X, x^*) p(y|I)}
\end{equation}
The results of applying this technique to different classifiers are reported in Table~\ref{table:classification_results_prior}, with the performance metrics reported in Table~\ref{table:prior_probabilities}. As can be seen, the performance of all classifiers improved by applying the prior information $I$. However, the use of this prior information is more helpful to those classifiers, e.g., \AstroNet\ and \ExoNet, that do not use the comprehensive diagnostic test data provided by Kepler pipeline. The performance of the classifiers that use a more comprehensive set of diagnostic tests, such as \ExoMiner\ and \GPC, is not improved as much. Interestingly, the application of the prior information $I$ to these classifiers often makes the classifiers more confident about the labels. This is because if either $p(y=1|M, X, x^*)$ or $p(y=1|I)$ are close to $1$, $p(y=1|M, X, x^*, I)$ becomes close to 1\footnote{This is because if either of these probabilities is close to 1.0, numerator and denominator in Equation~\ref{eq:prior_information} become equal.}. Thus, there will be more TCEs with scores $>0.99$. This is why the recall values for all classifiers at threshold $0.99$ increase. \ExoMiner\ is a conservative classifier and applying prior information $I$ leads to a large recall improvement at the threshold $0.99$. 

We also checked the performance of \ExoMiner\ updated scores using the scenario priors for changes in training set size and label noise, as we have done in Section~\ref{sec:training_set_sensitivity}. Even though we are not providing the plots here for conciseness, using the scenario priors improves the performance of \ExoMiner, more significantly when the size of the training set is small or there is more label noise. Given that using the scenario priors improves the performance of \ExoMiner, we use these updated \ExoMiner\ scores for the validation of new exoplanets.

\subsection{Dependence on candidate parameters}
\label{sec:dependence_parameters}
In this section, we study \ExoMiner\ dependence on candidate parameters using the Cumulative KOI catalog. Given that we use the updated \ExoMiner\ scores using scenario priors, as described in Section~\ref{sec:scenarior-priors}, for validation of new exoplanets, we will focus here on these updated scores.  Table~\ref{table:p_rad} shows \ExoMiner\ scores for different values of KOI radius and multiplicity. As can be seen, \ExoMiner\ generally assigns higher scores to KOIs in multiple systems compared to single systems even though it does not have access to the information related to the number of detected transits for each target star. Moreover, \ExoMiner\ assigned lower scores to larger KOIs.

\begin{figure*}[htb!]
	\centering
	\subfigure[Precision for orbital period versus planet radius ]{\label{fig:precision_period_radius}\includegraphics[width=.45\textwidth]{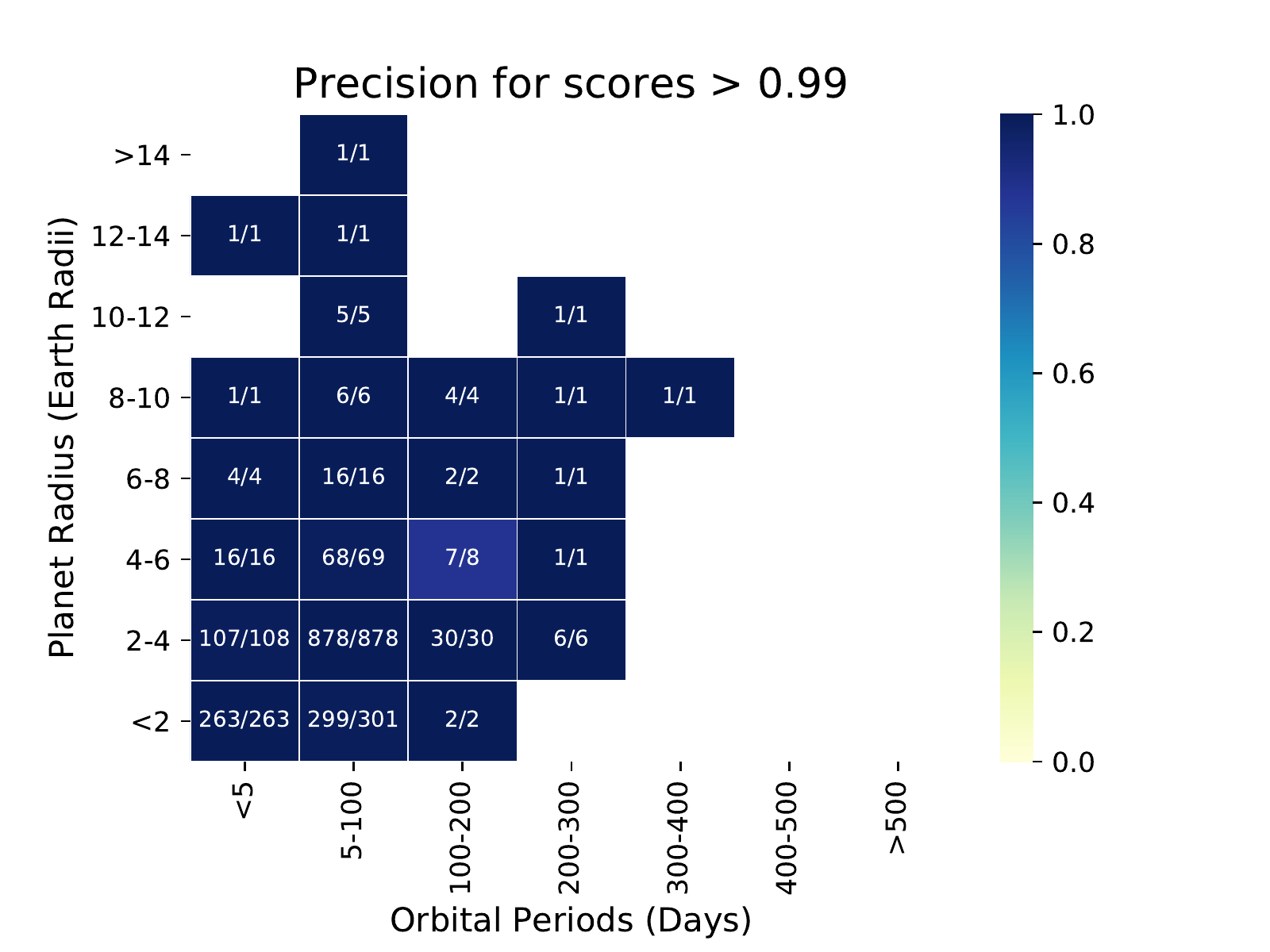}}	
	\hskip -.2mm
	\subfigure[Recall for orbital period versus planet radius.]{\label{fig:recall_period_radius}\includegraphics[width=.45\textwidth]{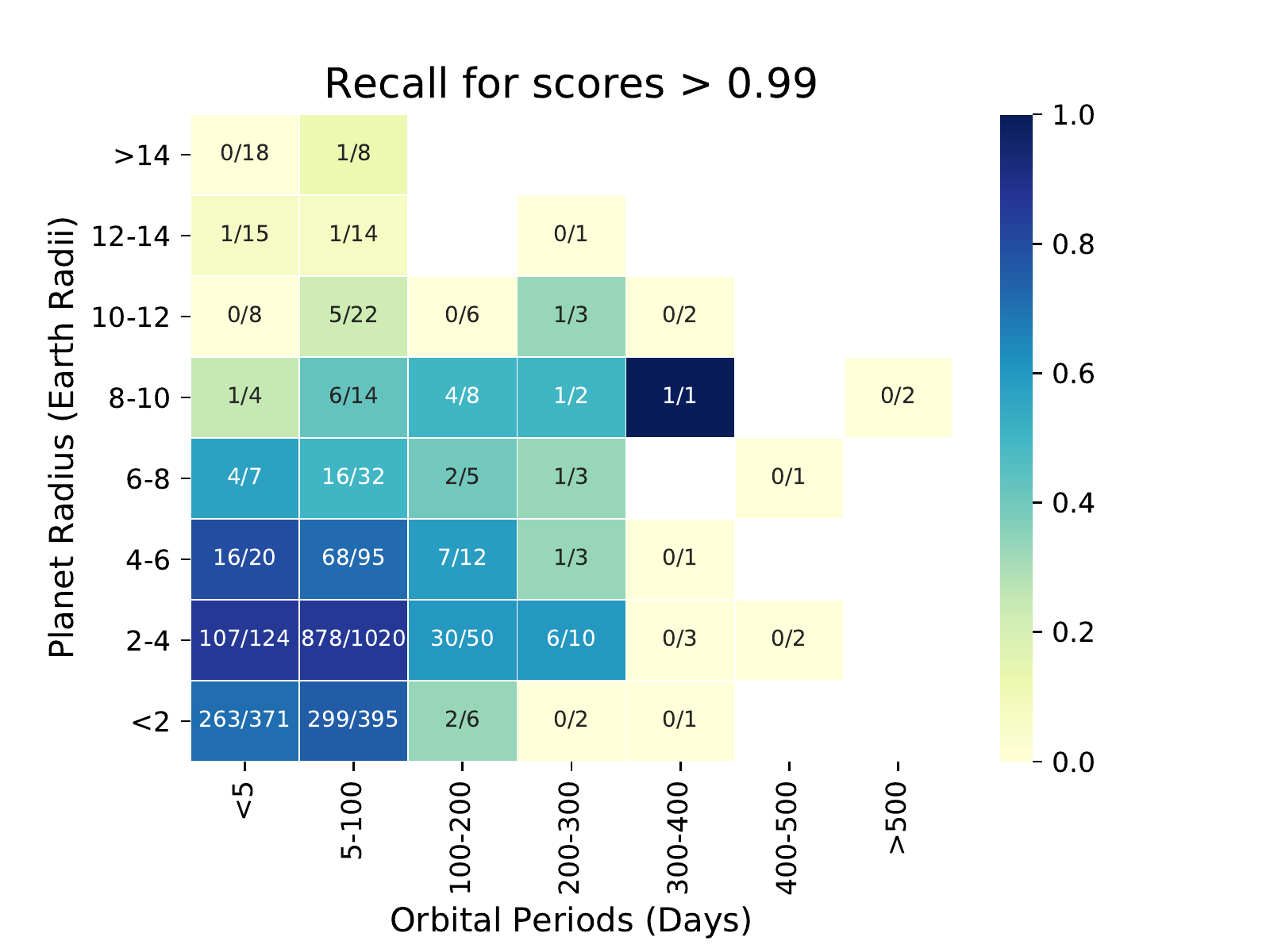}}	
	\vskip -2mm
	\subfigure[Precision for orbital period versus MES. ]{\label{fig:precision_period_MES}\includegraphics[width=.45\textwidth]{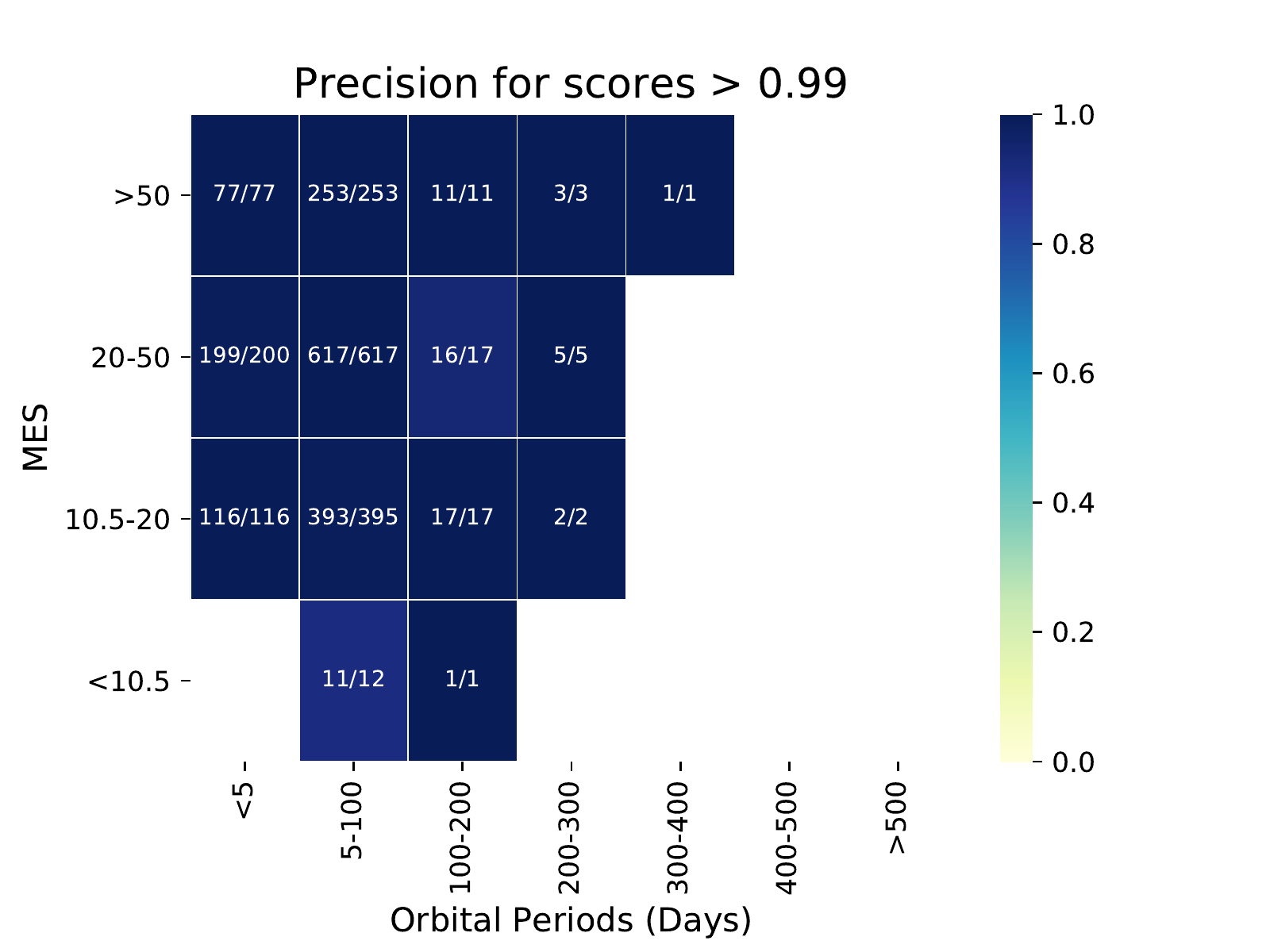}}	
	\hskip -.2mm
	\subfigure[Recall for orbital period versus MES.]{\label{fig:recall_period_MES}\includegraphics[width=.45\textwidth]{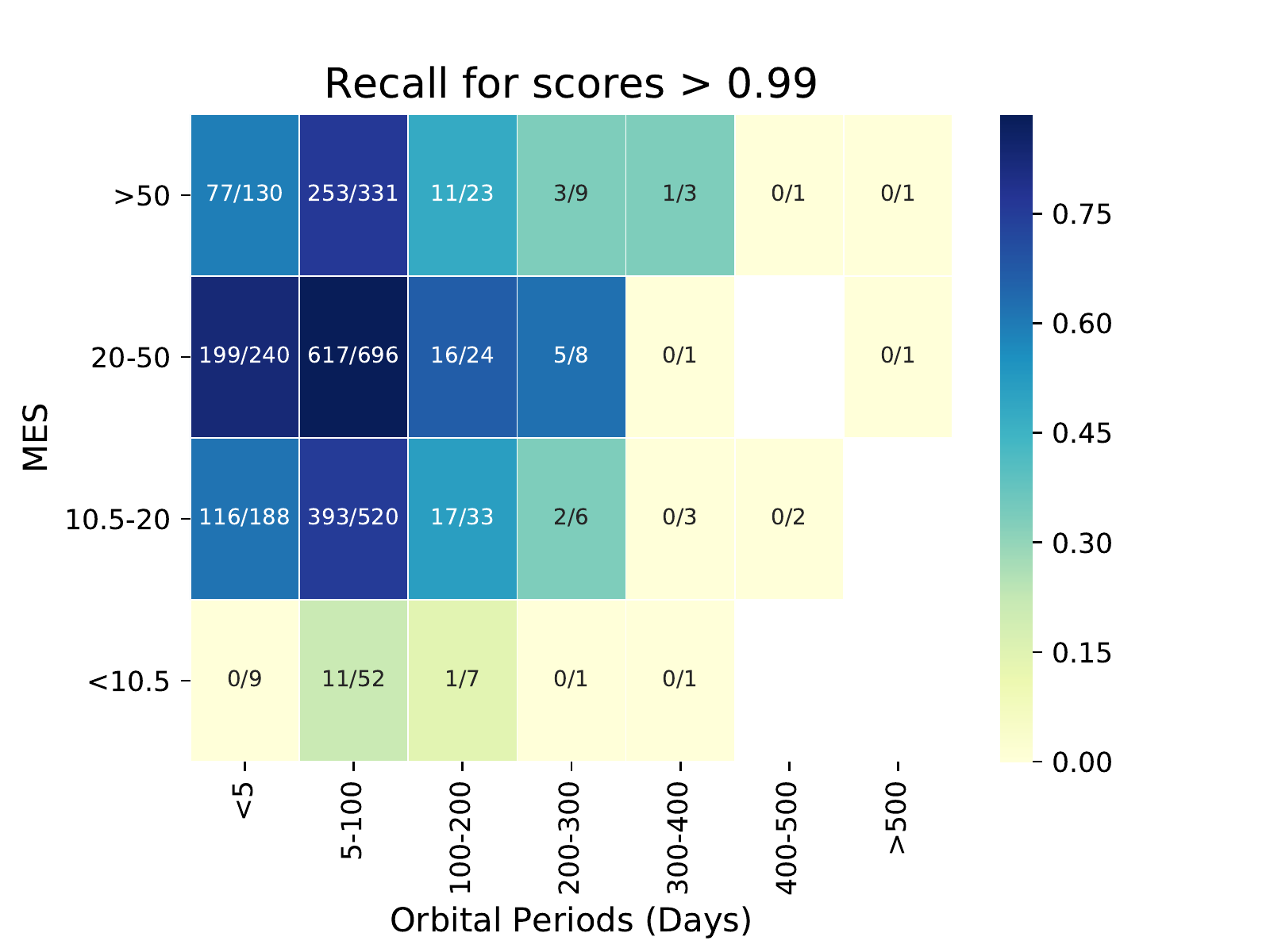}}	
	\vskip -2mm
	\subfigure[Precision for orbital period versus galactic latitude.]{\label{fig:precision_period_position}\includegraphics[width=.45\textwidth]{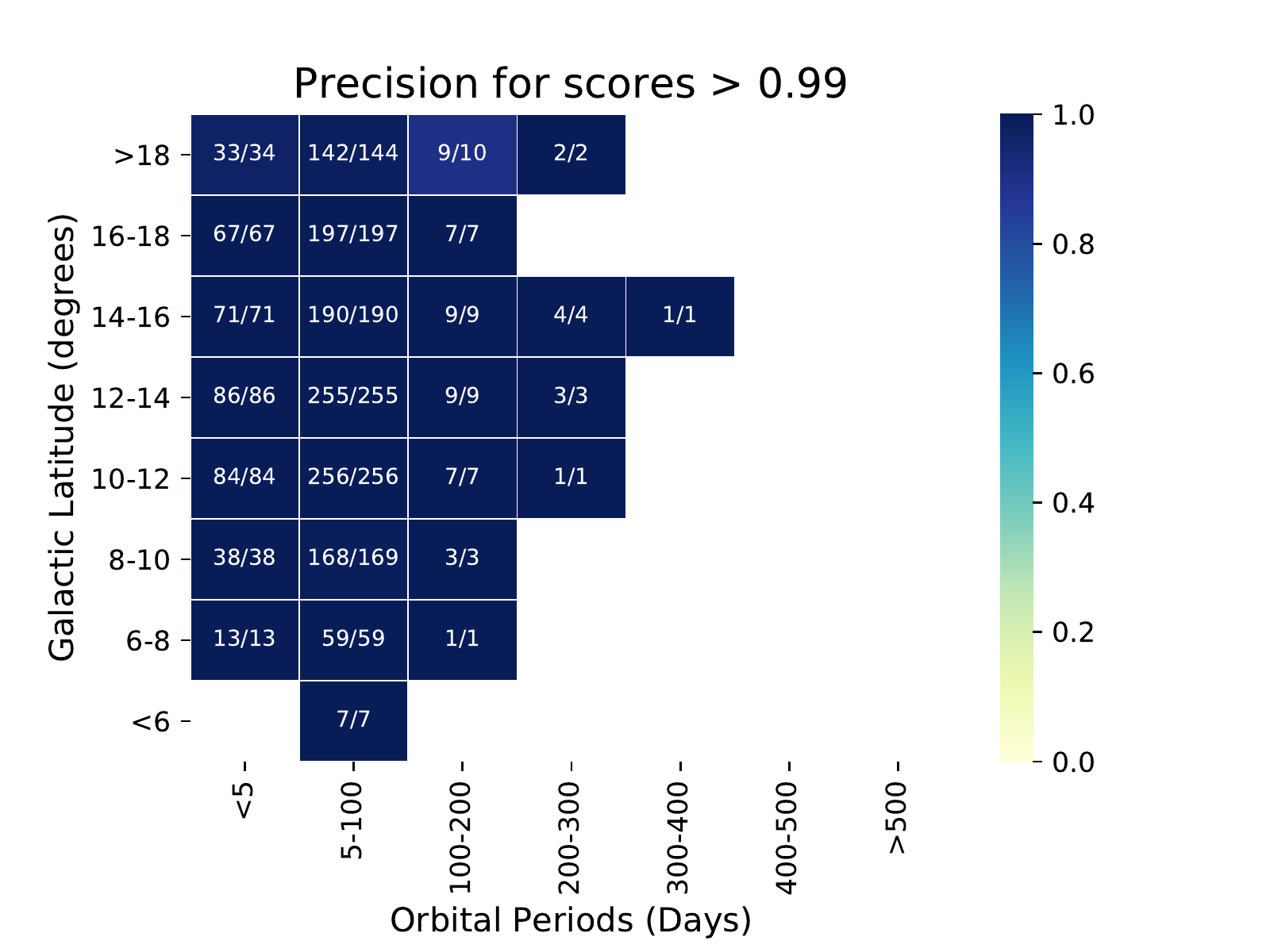}}	
	\hskip -.2mm
	\subfigure[Recall for orbital period versus galactic latitude.]{\label{fig:recall_period_position}\includegraphics[width=.45\textwidth]{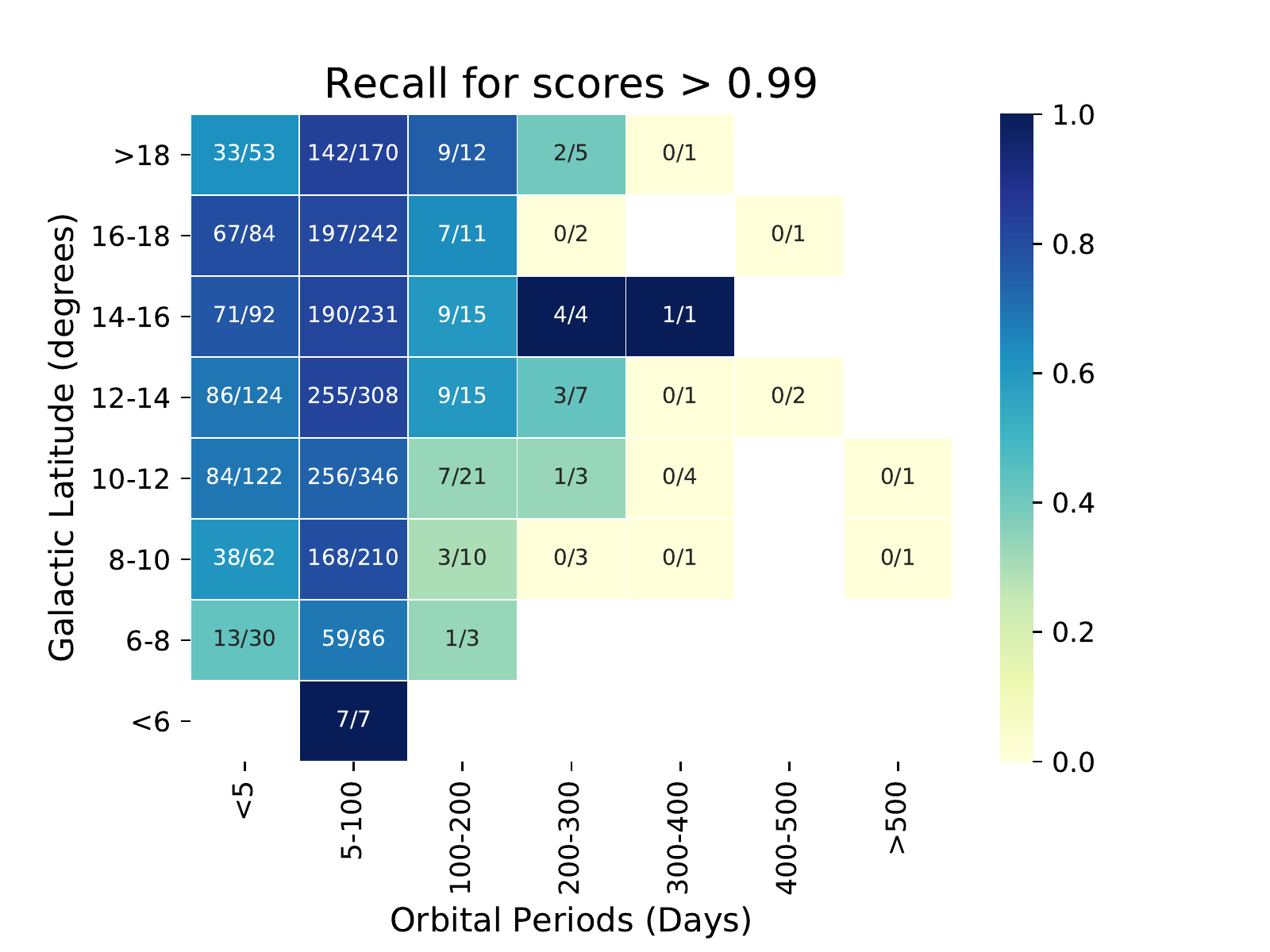}}	
\caption{Precision and recall for different parameters when threshold 0.99 is used for validation.}
\label{fig:performance_params}
\end{figure*}

\begin{table}[htb]
 \centering
\caption{\ExoMiner\ Scores by KOI radius and multiplicity.}
\label{table:p_rad}
\begin{threeparttable}
\begin{tabularx}{\linewidth}{@{}Y@{}}
\begin{tabular}{cccc}
\toprule
Selection & Count & Mean & Median  \\
 \midrule
All & 6464 & 0.379 & 0.002\\
KOI in multiple systems & 1313 & 0.857 & 0.996\\
KOI singles  & 5151 & 0.257 & 0.000\\
 $R_p<2R_\oplus$ & 2044 & 0.402 & 0.000\\
 $2R_\oplus<R_p<4R_\oplus$ & 1902 & 0.661 & 0.993\\
 $4R_\oplus<R_p<10R_\oplus$ & 682 & 0.351 & 0.001\\
 $10R_\oplus<R_p<15R_\oplus$ & 263 & 0.336 & 0.011\\
 $R_p>15R_\oplus$ & 1487 & 0.028 & 0.000\\
\bottomrule
\end{tabular}
\end{tabularx}
\end{threeparttable}
\end{table}
To study the performance with regards to candidate parameters above the validation threshold, we plotted the precision and recall of \ExoMiner\ for scores $>0.99$ for different pairs of candidate parameters in Figure~\ref{fig:performance_params}. \ExoMiner\ is reliable in different areas of candidate parameters for planet validation. At the validation threshold of 0.99, \ExoMiner\ is very conservative in labeling KOIs as planets so it has high precision and low recall. Also, note that there are regions of parameter space for which there are not enough KOIs. Because there are not enough examples in those area in the training set, the model is not confident about the scores even if it assigns the cases correctly to the class. This shows a similar effect to the training set size experiment we have done in Section~\ref{sec:training_set_sensitivity}.  

In the low-MES region (MES $<10.5$), there is only one FP, KIC TCE 10187017.6, out of a total of thirteen TCEs that have scores $>0.99$. Interestingly, this is the 6th TCE of a system for which the first five TCEs are all confirmed exoplanets. Thus, we believe this could be planet number six with a period that fits between planet e and planet f of KOI-82. This object is marked as a centroid offset false positive in the certified false positive table, but the TCERT report does not show clear evidence for this offset, and upon review this object's original vetter for the false positive working group agrees that the background false positive classification is an error due to not accounting for saturation (Steve Bryson, personal communication).

\begin{figure}[htb!]
	\centering
	\subfigure[Log orbital period versus log planet radius. ]{\label{fig:PC_prob_period_radius}\includegraphics[width=\columnwidth]{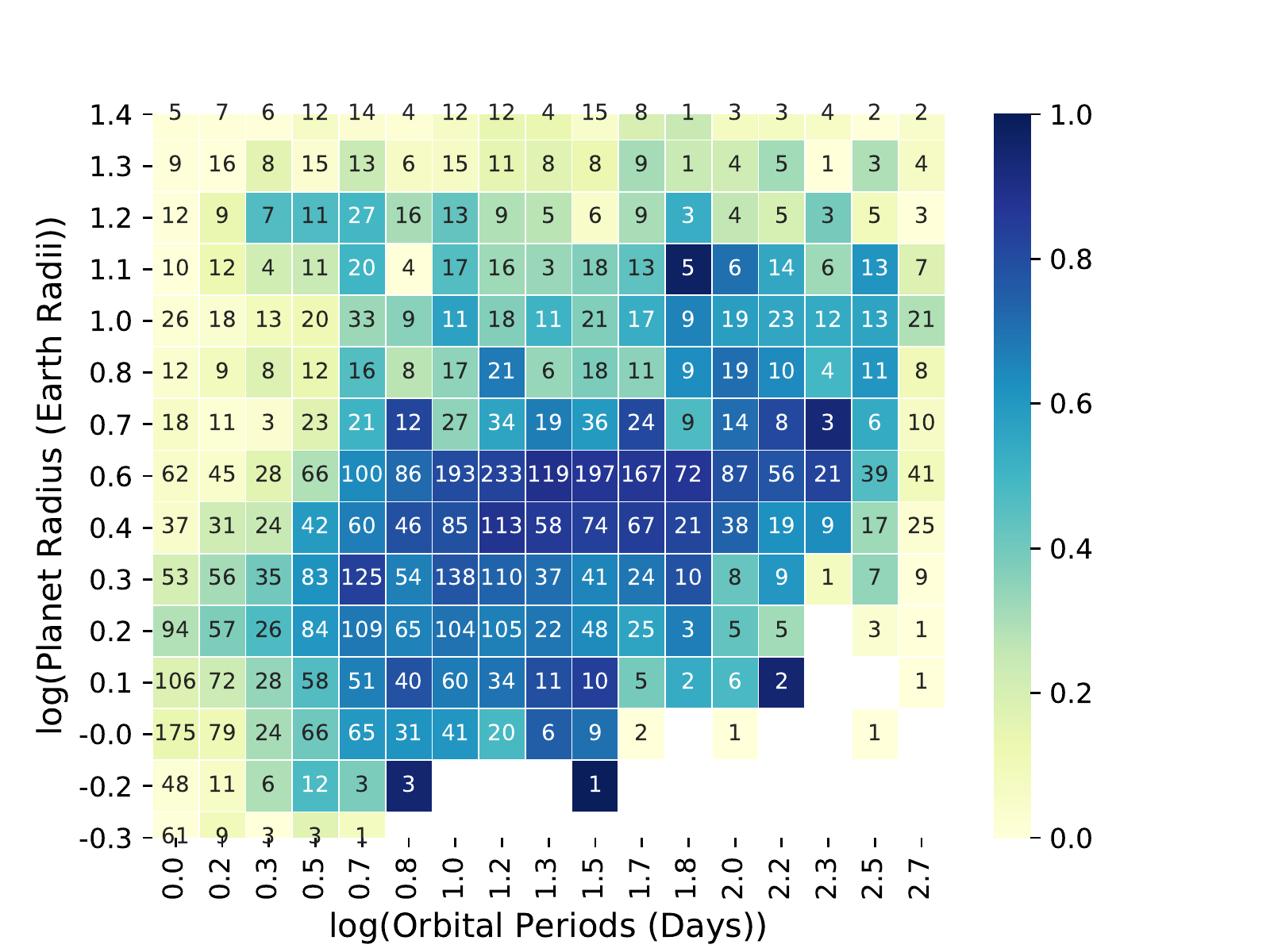}}	
\vskip -0.1mm
	\subfigure[Log orbital period versus log MES. ]{\label{fig:PC_prob_period_mes}\includegraphics[width=\columnwidth]{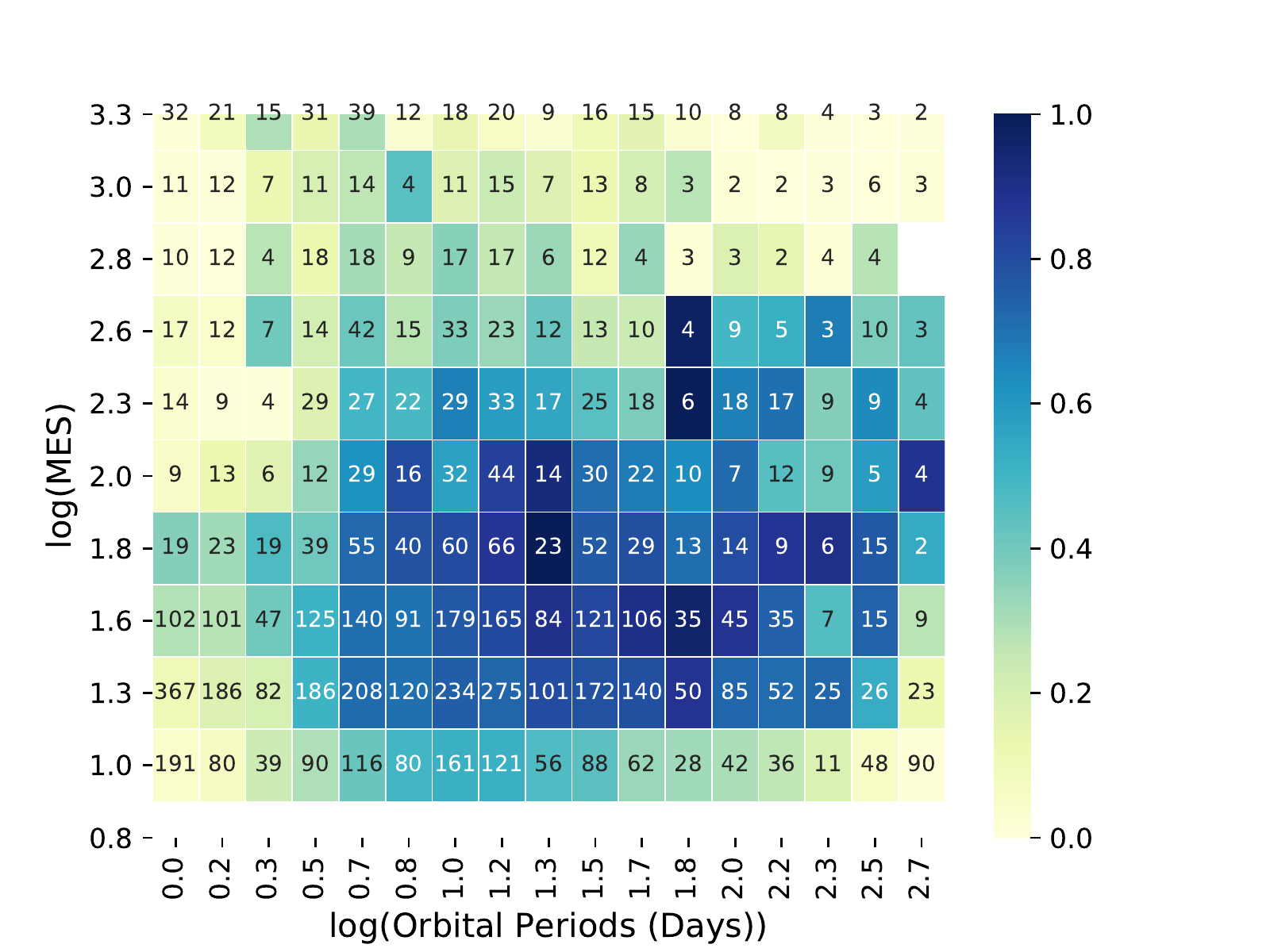}}	
\vskip -0.1mm
	\subfigure[Log orbital period versus galactic latitude. ]{\label{fig:PC_prob_period_position}\includegraphics[width=\columnwidth]{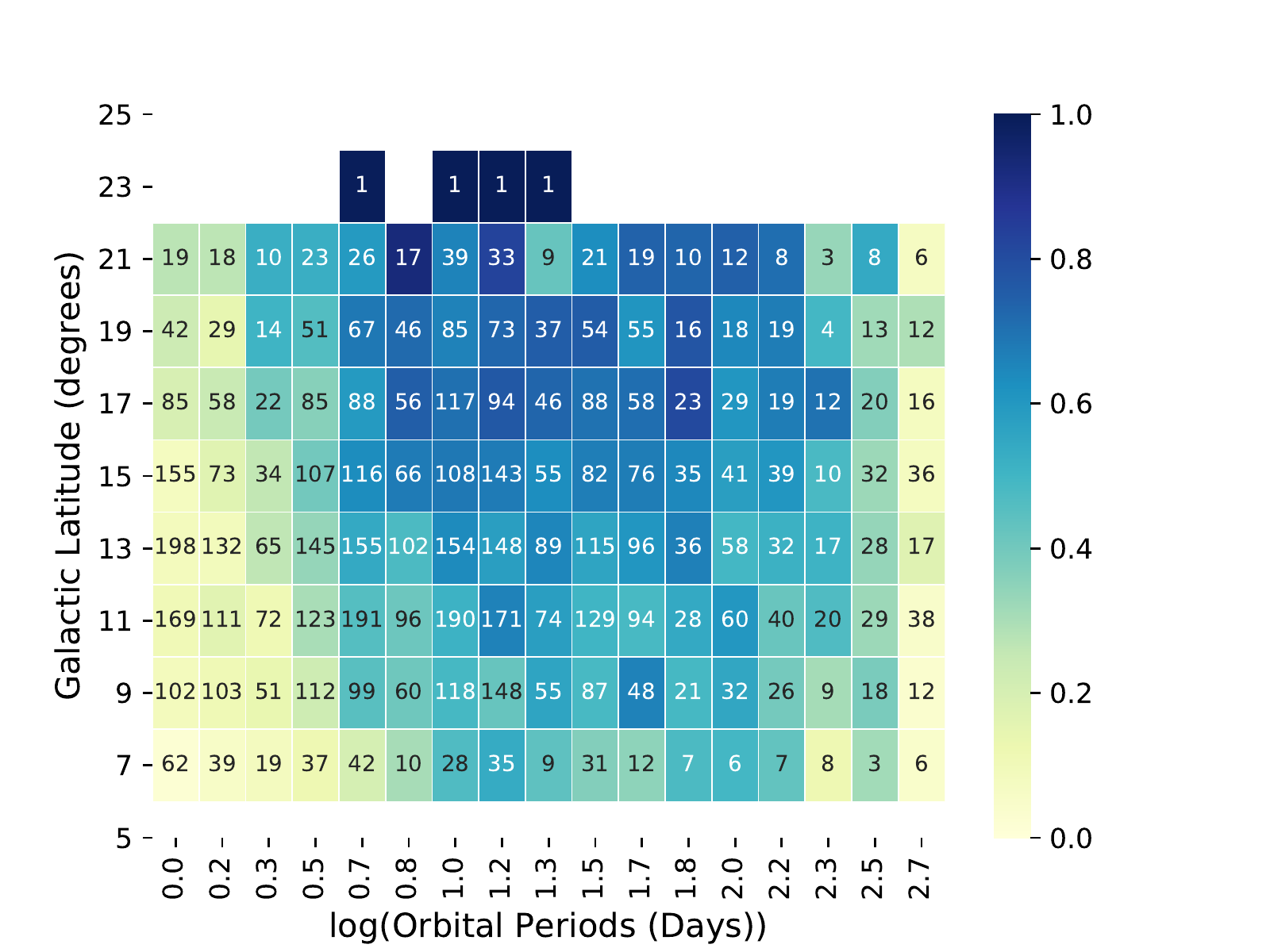}}	
\caption{Mean score of KOIs for different parameters. The number in each bin shows the total number of KOIs in that bin. There are no KOIs in white bins.}
\label{fig:pc_prob_params}
\end{figure}

Figure~\ref{fig:pc_prob_params} shows the  distribution of scores for planet radius, MES, and galactic latitude of KOI versus KOI orbital period. As can be seen, short- and long-period KOIs, KOIs in the low-MES region, and KOIs at lower galactic latitudes are all assigned low scores.  

\section{Planet Validation}
\label{sec:planetvalidation}
\subsection{Setup}
\label{sec:validation-setup}
Similar to existing planet validation approaches, we use the rejection threshold of $>0.99$ to validate new exoplanets. To make the validation process more reliable, we need to identify where \ExoMiner\ does not perform well in order to design vetoing criteria to improve the validation process even further.

As we detailed in Section~\ref{sec:misclassified-cases}, \ExoMiner\ classifies some FPs into PC category for different reasons: lack of enough information such as that given by the contamination flag, or poor interpretation of diagnostic tests such as the centroid test. As we discussed in Section~\ref{sec:scenarior-priors}, utilizing scenario priors addressed some of these misclassifications and improved the performance. Nonetheless, there are still cases that are not resolved by the application of scenario priors, even though \ExoMiner\ is making less than one error out of every 100 TCEs that exceed the rejection threshold of 0.99. To be precise, \ExoMiner\ classifies 1727 TCEs with score $>0.99$, of which there are five FPs\footnote{These numbers are for the adjusted \ExoMiner\ scores using scenario priors. The total number of TCEs with original scores $>0.99$ is 1507, of which there are four FPs}, leading to a precision of 0.997, better than the required precision of $0.99$ for the threshold of $0.99$. To further improve the precision of \ExoMiner\ in validating new exoplanets, we ensure that the KOI is dispositioned as a `candidate' in the Cumulative KOI catalog \citep[similar to][]{Morton-2016-vespa} and that no disposition flag is on in the Cumulative KOI catalog. These vetoes remove those five FPs from the validated list. Regarding low-MES region, even though the model performs well in this region\footnote{There was only one FP in the low-MES region whose correct disposition might be planet.} (see Figure~\ref{fig:precision_period_MES}), we take the more conservative approach and do not validate any KOI with MES $<10.5$. 

\begin{figure}[htb!]
	\centering
	\subfigure{\label{sfig:pp-CP}\includegraphics[width=92mm]{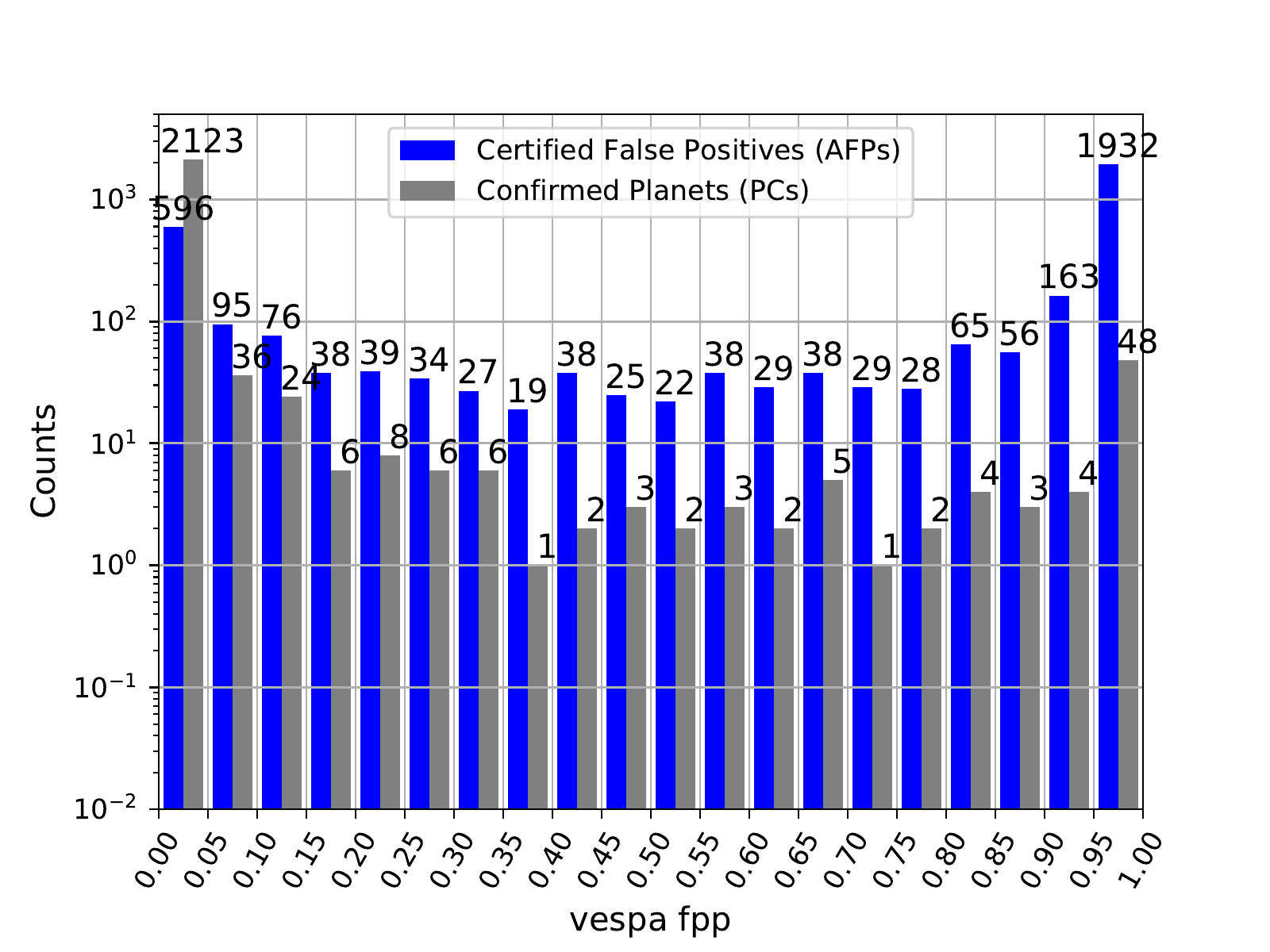}}	
\caption{Distribution of CPs and CFPs as function of \vespa\ fpp score. There are 596 CFPs out of a total of 2719 CPs+CFPs with \vespa\ fpp $\leq0.05$. This implies a misclassification rate of 21.9\% if we use \vespa\ fpp $\leq0.05$ to validate new exoplanets. Not shown in this picture: there are 403 CFPs with \vespa\ fpp $\leq0.01$ out of 2372 CPs+CFPs (a misclassification rate of 17.0\%). These results demonstrate that \vespa\ fpp is not effective for screening against AFPs at the 0.01 confidence level in the absence of other constraints (see Section~\ref{sec:planetvalidation}).}
\label{fig:vespa}
\end{figure}

Unlike~\citet{armstrong-2020-exoplanet}, we do not recommend using \vespa\ fpp scores to veto the validated exoplanets for multiple reasons:
\begin{itemize}
\item \vespa\ is poorly calibrated, as shown in Figure~\ref{fig:reliabilitycurve}. As an overestimator classifier, its scores cannot be used for validation because the error for any rejection threshold is higher than that threshold, as seen in in Figures~\ref{fig:reliabilitycurve} and~\ref{fig:vespa}. As an example, \vespa\ gives scores below $0.01$ to 2409 TCEs, of which 446 TCEs are FPs, leading to a precision of 0.815, which is much lower than the required value of 0.99 for the applied threshold. As we mentioned in Section~\ref{sec:planetvalidation},~\citet{Morton-2016-vespa} used other criteria in order to improve the vetting. This led to a highly accurate approach for exoplanet validation in terms of precision, but very low recall. 
\item Inspired by~\citet{armstrong-2020-exoplanet}, we use the scenario priors $I$ as complementary sources of information. These scenario priors are also used in calculating \vespa\ fpp scores. Thus, we are using part of the \vespa\ model. By only using scenario priors, we ignore the likelihood used to calculate \vespa\ fpp. We discussed in Section~\ref{sec:automatic-classification} the problems with the generative approach and why a discriminative approach is preferred for classification problems. 
\item The \vespa\ model is highly dependent on the accuracy of stellar parameters. Given the uncertainty about the values of stellar parameters, \vespa\ fpp is subject to continuous change. 
\item As discussed in Section~\ref{sec:comparison-vespa}, when there is a disagreement between \ExoMiner\ and \vespa\ scores for validating exoplanets at a threshold of 0.99, \ExoMiner\ is almost always correct. Thus, the advantage of vetoing based on \vespa\ fpp is not clear. 
\end{itemize}

We also do not suggest using an outlier detection test to veto validated exoplanets for two main reasons: 
\begin{itemize}
\item Unsupervised outlier detection is a subjective task where the results vary from one algorithm to another or by changing the hyper-parameters of the same algorithm; i.e., one can always come up with an outlier detection algorithm such that some of these validated exoplanets are outliers.
\item By using the rejection threshold, we already identified TCEs for which the model has not seen similar cases, as discussed in Section~\ref{sec:rejection}. 
\end{itemize}
Nonetheless, as reported in Section~\ref{sec:validating_new_exoplanet}, all validated exoplanets in this work pass the outlier tests of~\citet{armstrong-2020-exoplanet}.

\begingroup
\setlength{\tabcolsep}{2.5pt} 
\begin{table}[htb]
 \centering
\caption{Original \ExoMiner\ score and \ExoMiner\ score after applying priors for the unlabeled KOIs. This table describes the available columns. The full table is available online.}
\label{table:classification_results_unlabeled}
\begin{threeparttable}
\begin{tabularx}{\linewidth}{@{}Y@{}}
\begin{tabular}{ll}
\toprule
Column & Description \\
\midrule
target\_id & KIC ID\\
tce\_plnt\_num & TCE planet number\\
tce\_period & TCE period\\
mes & TCE MES\\
exominer & original \ExoMiner\ score \\
exominer\_prior & \ExoMiner\ score with priors \\ 
\bottomrule
\end{tabular}
\end{tabularx}
\end{threeparttable}
\end{table}
\endgroup

\subsection{Validated Exoplanets Using \ExoMiner}
\label{sec:validating_new_exoplanet}
To validate new exoplanets, we apply ten \ExoMiner\ models trained using 10-fold CV to those KOIs that are not part of our working set, i.e., KOIs that are not CPs, CFPs, or CFAs. We will designate this subset by unlabeled KOI set. There are a total of 1922 such KOIs in the Cumulative KOI catalog from the Q1-Q17 DR25 run. We use the average score of ten models to classify these KOIs.  Table~\ref{table:classification_results_unlabeled} shows the scores provided by \ExoMiner\ before and after applying priors for the unlabeled KOI set. Ephemerides and other relevant parameters are also presented. Given that a majority of TCEs in the training set are NTPs, and the KOIs from which we are validating do not include any NTPs, the rate of FPs in the training set is much higher than that of the unlabeled KOI set. Thus, the probability scores generated by \ExoMiner\ are more conservative in labeling these KOIs as PC and validating them. \ExoMiner\ labels 1245\footnote{this number is 1186 for original \ExoMiner\ scores without applying scenario priors.} KOIs as PC (scores $>0.5$) out of 1922 KOIs. Other existing classifiers also provide similar results. For example, for posterior GPC~\citep{armstrong-2020-exoplanet}, this number is 1126 KOIs.

Out of these 1245 positively labeled KOIs by \ExoMiner, eight KOIs are dispositioned as `False Positive' or have an FP flag on in the Cumulative KOI catalog. Given that most of these positively labeled KOIs are already dispositioned as `candidate' in the Cumulative KOI catalog, this number should not be surprising. One possible justification for why there are so many candidates left in the unlabeled set could be because vetting false positives is easier than confirming PCs. A TCE is certified as a false positive if one of the diagnostic tests indicates a FP. However, it is not enough to confirm an exoplanet if none of the diagnostic metrics is inconsistent with a true transit signal; confirming new exoplanets requires follow-up studies or statistical/ML validation. 

Out of 1922 unlabeled KOIs, \ExoMiner\ labels 368 KOIs as PC with validation score $>0.99$, of which 46 are among the previously validated exoplanets by~\citet{armstrong-2020-exoplanet} and one among the previously validated by~\cite{Jontof-Hutter-2021-validation}, leaving 321 new ones. Interestingly, out of a total of 943 unlabeled KOIs with MES $<10.5$, \ExoMiner\ gives a score $>0.99$ to twenty KOIs, even though 839 of them have a `candidate' disposition in the Cumulative KOI catalog. This implies that \ExoMiner\ is more conservative in the low-MES region, as discussed in Section~\ref{sec:dependence_parameters}. Removing these 20 KOIs with MES $<10.5$, we validate 301 new exoplanets. Out of these 301 newly validated exoplanets, 42 KOIs are in multiplanet systems. We provide the list of newly validated exoplanets in Table~\ref{table:confirmedplanetlist}. None of these newly validated exoplanets are found to be outliers based on the two algorithms in~\citet{armstrong-2020-exoplanet}. We also report those KOIs with ExoMiner scores $>0.99$ in the low-MES region in Table~\ref{table:confirmedplanetlist_lowMES}, even though we are not validating them. Nine out of twenty such KOIs are in multiplanet systems.

\begin{table*}[htb!]
 \centering
\caption{List of newly validated exoplanets with \ExoMiner\ score $>0.99$ and MES $>10.5$. The full list is available online.}
\label{table:confirmedplanetlist}
\begin{threeparttable}
\begin{tabularx}{\linewidth}{@{}Y@{}}
\begin{tabular}{lcccccccc}
\toprule
Number & TCE KIC & KOI Name & Period (days) & Radius (Re) & MES & \ExoMiner\ score  & \vespa\ Score & Kepler Name\\
\midrule
\rownumber & 11852982.1 & K00247.01 &   13.8150 &    2.2837 &   41.4900 &     0.9989 &     0.0076 &  Kepler-1712b \\
\rownumber & 7700622.1 & K00315.01 &   35.5810 &    2.4248 &   73.5900 &     0.9981 &     0.0000 &  Kepler-1716b \\
\rownumber & 7135852.1 & K00875.01 &    4.2210 &    3.8940 &  134.5000 &     0.9981 &     0.0009 &  Kepler-1729b \\
\rownumber & 6803202.1 & K00177.01 &   21.0608 &    1.9230 &   43.0500 &     0.9959 &     0.0005 &  Kepler-1711b \\
\rownumber & 2444412.1 & K00103.01 &   14.9111 &    3.1974 &  108.0000 &     0.9980 &     0.0001 &  Kepler-1710b \\
\rownumber & 7941200.1 & K00092.01 &   65.7046 &    3.1242 &   81.8900 &     0.9977 &     0.2000 &  Kepler-1709b \\
\rownumber & 8087812.1 & K04343.01 &   27.2108 &    1.2392 &   13.3900 &     0.9931 &     0.0005 &  Kepler-1952b \\
\rownumber & 7749773.1 & K02848.01 &   13.7873 &    1.4867 &   16.2100 &     0.9983 &     0.0006 &  Kepler-1882b \\
\rownumber & 11599038.1 & K01437.01 &    7.0173 &    1.8242 &   12.5500 &     0.9981 &     0.0001 &  Kepler-1751b \\
\rownumber & 10587105.3 & K00339.03 &   35.8662 &    2.7950 &   18.2200 &     0.9981 &     0.0006 &   Kepler-529d \\
\rownumber & 3645438.2 & K04385.01 &    7.2980 &    1.9421 &   12.2000 &     0.9988 &     0.0004 &   Kepler-1600c \\
\rownumber & 4665571.2 & K02393.01 &    4.6030 &    1.4594 &   14.9800 &     0.9980 &     0.0002 &  Kepler-1834b \\
\rownumber & 7364176.1 & K00373.01 &  135.1880 &    2.5685 &   62.9900 &     0.9980 &     0.8100 &  Kepler-1718b \\
\rownumber & 9823487.1 & K01489.01 &   16.0046 &    2.2268 &   27.3200 &     0.9986 &     0.0000 &  Kepler-1753b \\
\rownumber & 5374854.1 & K00645.02 &   23.7832 &    2.2960 &   28.8000 &     0.9957 &     0.0035 &   Kepler-633c \\
\rownumber & 8410415.1 & K02291.01 &   44.2992 &    2.2735 &   19.3700 &     0.9980 &     0.1200 &  Kepler-1817b \\
\rownumber & 3937519.2 & K00221.02 &    5.8966 &    1.2194 &   12.1100 &     0.9986 &     0.0017 &   Kepler-495c \\
\rownumber & 10134152.1 & K02056.01 &   39.3137 &    2.3190 &   25.3000 &     0.9984 &     0.0280 &  Kepler-1787b \\
\rownumber & 9150827.1 & K01408.01 &   14.5341 &    1.6130 &   25.6200 &     0.9986 &     0.0010 &  Kepler-1749b \\
\rownumber & 3641726.1 & K00804.01 &    9.0293 &    3.1732 &   47.3700 &     0.9975 &     0.0000 &  Kepler-1727b \\
\rownumber & 12121570.1 & K02290.01 &   91.5010 &    1.9679 &   18.6600 &     0.9925 &     0.0003 &  Kepler-1816b \\
\rownumber & 11046025.1 & K01646.01 &    3.7202 &    0.9038 &   14.0200 &     0.9972 &     0.0007 &  Kepler-1759b \\
\rownumber & 9837661.3 & K02715.03 &    5.7209 &    2.6284 &   15.0600 &     0.9974 &     0.1800 &   Kepler-1321d \\
\rownumber & 11821363.1 & K01494.01 &    8.1959 &    2.4400 &   17.2000 &     0.9980 &     0.0009 &  Kepler-1754b \\
\rownumber & 10074466.1 & K01917.01 &   35.0119 &    2.6116 &   27.8000 &     0.9977 &     0.0730 &  Kepler-1776b \\
\rownumber & 10386922.2 & K00289.01 &   26.6295 &    4.2096 &   76.4700 &     0.9974 &     0.0000 &   Kepler-511c \\
\rownumber & 7584650.1 & K02631.01 &   44.9976 &    2.2631 &   15.9600 &     0.9968 &     0.0008 &  Kepler-1863b \\
\rownumber & 7670943.1 & K00269.01 &   18.0116 &    1.7167 &   38.8800 &     0.9962 &     0.0005 &  Kepler-1713b \\
\rownumber & 5601258.1 & K02191.01 &    8.8479 &    1.5450 &   18.3500 &     0.9979 &     0.0650 &  Kepler-1805b \\
\rownumber & 7826659.1 & K02686.01 &  211.0340 &    3.1495 &   59.0000 &     0.9959 &     0.0001 &  Kepler-1868b \\
\rownumber & 7386827.1 & K01704.01 &   10.4189 &    2.4951 &   36.8400 &     0.9987 &     0.0250 &  Kepler-1765b \\
\rownumber & 9307509.1 & K02096.01 &    9.7630 &    1.7943 &   21.3900 &     0.9983 &     0.0260 &  Kepler-1794b \\
\rownumber & 6183511.1 & K02542.01 &    0.7273 &    1.2162 &   14.7400 &     0.9967 &     0.0054 &  Kepler-1855b \\
\rownumber & 7877978.1 & K02760.01 &   56.5732 &    2.2111 &   18.7700 &     0.9918 &     0.0029 &  Kepler-1874b \\
\rownumber & 5436013.1 & K02471.01 &   25.8860 &    1.4891 &   13.9800 &     0.9957 &     0.0053 &  Kepler-1841b \\
\rownumber & 11297236.1 & K01857.01 &   88.6440 &    4.0554 &   38.3600 &     0.9911 &     0.0000 &  Kepler-1771b \\
\rownumber & 12469800.1 & K02543.01 &    1.3020 &    1.7345 &   19.0400 &     0.9946 &     0.0044 &  Kepler-1856b \\
\rownumber & 9836959.1 & K01715.01 &  105.0150 &    2.2790 &   25.7600 &     0.9970 &     0.0045 &  Kepler-1766b \\
\rownumber & 3940418.1 & K00810.01 &    4.7830 &    2.9550 &   57.6400 &     0.9986 &     0.0370 &  Kepler-1728b \\
\rownumber & 5773121.1 & K04002.01 &    0.5242 &    1.5148 &   18.3500 &     0.9945 &     0.0160 &  Kepler-1932b \\
\rownumber & 5020319.1 & K00635.01 &   16.7199 &    2.8125 &   51.8700 &     0.9978 &     0.0170 &  Kepler-1725b \\
\rownumber & 5978361.1 & K00558.01 &    9.1786 &    2.8672 &   44.1000 &     0.9984 &     0.0280 &  Kepler-1722b \\
\rownumber & 3109930.1 & K01112.01 &   37.8103 &    3.8766 &   28.3800 &     0.9979 &     0.0130 &  Kepler-1736b \\
\rownumber & 9475552.2 & K02694.02 &    6.5661 &    1.8440 &   19.8300 &     0.9987 &     0.3000 &   Kepler-1315c \\
\rownumber & 10793172.2 & K02871.02 &    5.3637 &    0.9770 &   11.9800 &     0.9969 &     0.0190 &   Kepler-1693c \\
. & . &    . &    . &   . &     . &     . &         .  & .\\
. & . &    . &    . &   . &     . &     . &         . & .\\
. & . &    . &    . &   . &     . &     . &      . & .\\
\bottomrule
\end{tabular}
\end{tabularx}
\end{threeparttable}
\end{table*}

\begin{table*}[htb!]
 \centering
\caption{List of KOIs with \ExoMiner\ score $>0.99$ and MES $<10.5$.}
\label{table:confirmedplanetlist_lowMES}
\begin{threeparttable}
\begin{tabularx}{\linewidth}{@{}Y@{}}
\begin{tabular}{lccccccc}
\toprule
Number & TCE KIC & KOI Name & Period (days) & Radius (Re) & MES & \ExoMiner\ score  & \vespa\ Score \\
\midrule
\rownumberLowMES & 6871071.4 & K02220.04 &    7.6648 &    1.6288 &    9.7590 &     0.9969 &     0.0005 \\
\rownumberLowMES & 10793576.1 & K03451.01 &   10.9076 &    1.6938 &    9.2950 &     0.9971 &     0.0005 \\
\rownumberLowMES & 12204137.1 & K04590.01 &    9.5406 &    2.0832 &    9.9220 &     0.9958 &     0.0850 \\
\rownumberLowMES & 9896018.3 & K02579.03 &   10.3015 &    1.6263 &    9.2720 &     0.9927 &     0.0002 \\
\rownumberLowMES & 7289317.2 & K02450.02 &    7.1929 &    1.0897 &    9.6700 &     0.9918 &     0.0008 \\
\rownumberLowMES & 9011877.1 & K05597.01 &    7.7771 &    0.9646 &   10.1700 &     0.9940 &     0.0003 \\
\rownumberLowMES & 9457948.1 & K04694.01 &   19.2527 &    1.6884 &    9.5330 &     0.9961 &     0.0015 \\
\rownumberLowMES & 4919550.1 & K04766.01 &    5.5668 &    1.1317 &   10.3100 &     0.9980 &     0.0380 \\
\rownumberLowMES & 4164922.2 & K03864.02 &   18.2574 &    1.0389 &   10.2100 &     0.9962 &     0.0007 \\
\rownumberLowMES & 2860656.1 & K04857.01 &    4.1150 &    1.8554 &    9.0710 &     0.9944 &     0.0001 \\
\rownumberLowMES & 9489524.4 & K02029.04 &    4.7885 &    0.7428 &    9.9180 &     0.9966 &     0.0110 \\
\rownumberLowMES & 7098255.1 & K04762.01 &   11.5673 &    1.4018 &    9.7880 &     0.9919 &     0.0790 \\
\rownumberLowMES & 7692093.1 & K03337.01 &   11.1662 &    1.5631 &   10.1600 &     0.9933 &     0.0012 \\
\rownumberLowMES & 7100673.5 & K04032.05 &    7.2352 &    0.9662 &    9.0370 &     0.9903 &     0.2000 \\
\rownumberLowMES & 6131130.1 & K03307.01 &   10.5071 &    1.4087 &    9.4610 &     0.9918 &     0.0250 \\
\rownumberLowMES & 5542466.3 & K01590.03 &    4.7467 &    1.5175 &    9.5630 &     0.9927 &     0.0071 \\
\rownumberLowMES & 5903749.1 & K03029.01 &   18.9763 &    2.4880 &   10.2100 &     0.9931 &     0.0150 \\
\rownumberLowMES & 6144039.1 & K04602.01 &    2.8586 &    0.8922 &   10.2000 &     0.9916 &     0.0170 \\
\rownumberLowMES & 9002538.2 & K03196.02 &    6.8830 &    0.7757 &   10.1100 &     0.9916 &     0.4700 \\
\rownumberLowMES & 8874090.2 & K01404.02 &   18.9063 &    1.7561 &   10.1300 &     0.9941 &     0.0014 \\
\bottomrule
\end{tabular}
\end{tabularx}
\end{threeparttable}
\end{table*}

\begin{figure}[htb!]
	\centering
	\subfigure{\includegraphics[width=\columnwidth]{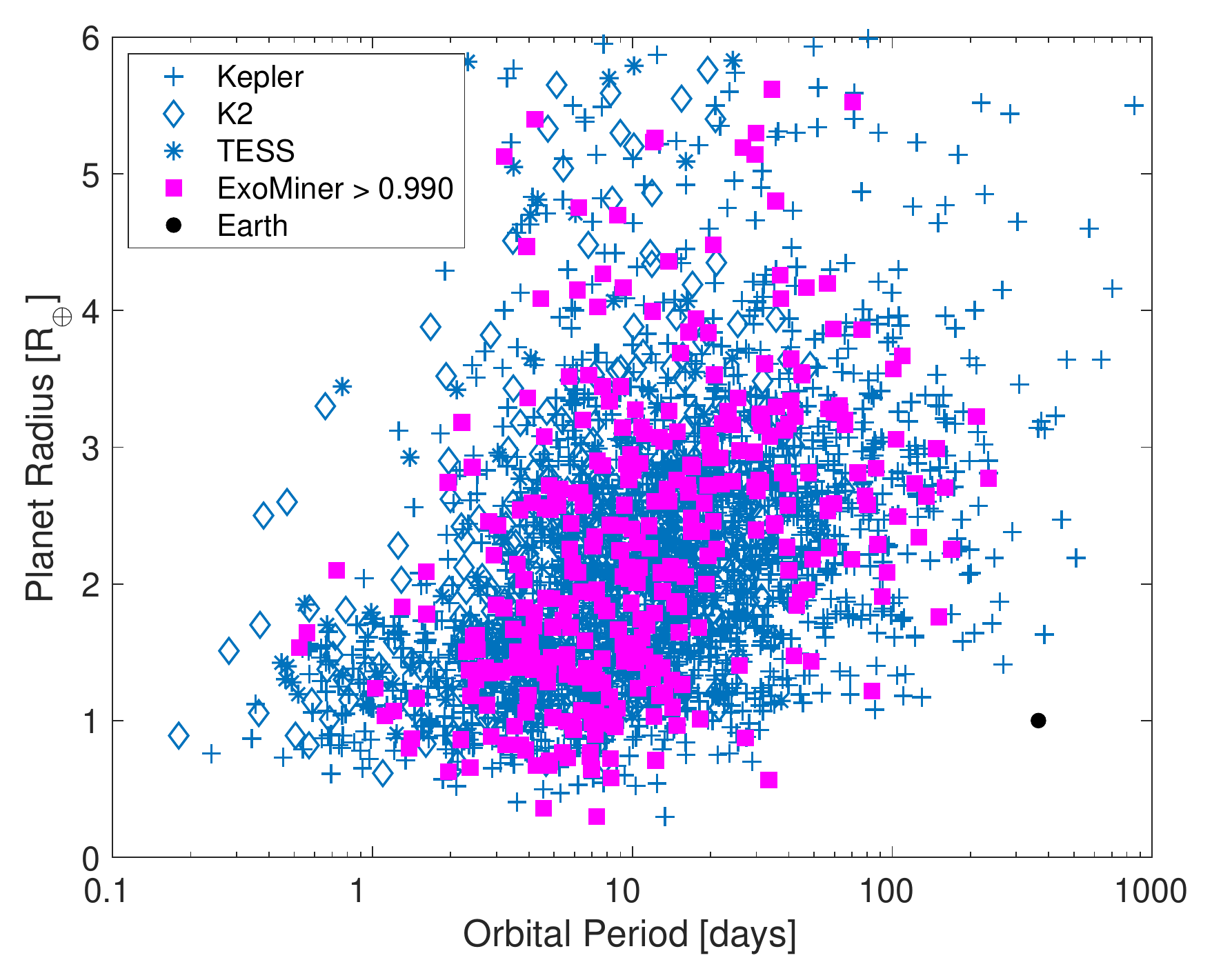}}	
\caption{Planet radius vs. orbital period for confirmed transiting planets and \ExoMiner-validated planets with a score greater than 0.99. The CPs are indicated by the discovery source with Kepler indicated by `+', K2 by a diamond `$\diamond$', and TESS by `*'. \ExoMiner-validated planets are indicated by a magenta square. For reference, Earth is indicated by a black circle.}
\label{fig:planetvsradius}
\end{figure}

\begin{figure}[htb!]
	\centering
	\subfigure{\includegraphics[width=\columnwidth]{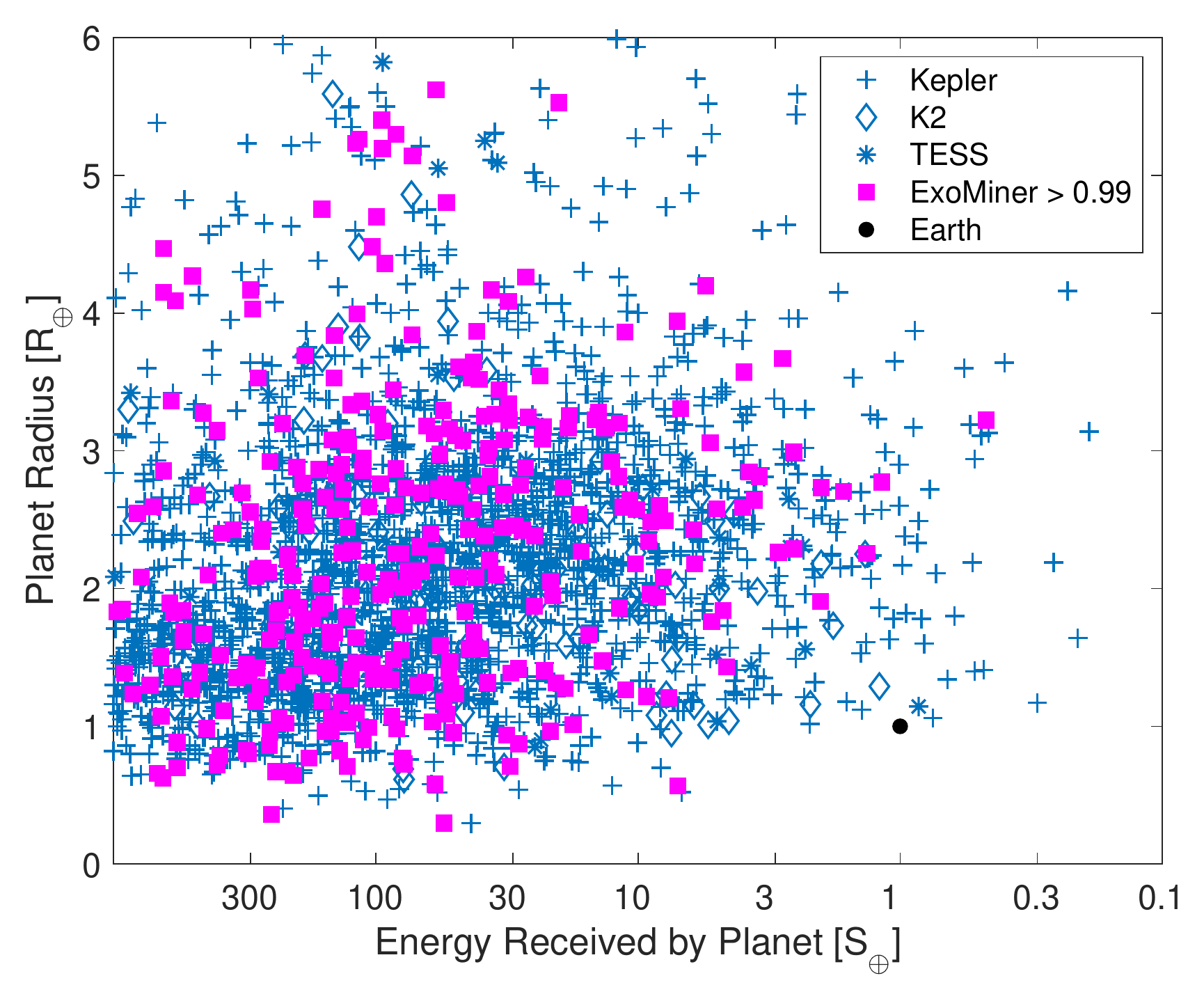}}	
\caption{Planet radius vs. energy received by planet for confirmed transiting planets and \ExoMiner-validated planets with a score greater than 0.99. Symbols are the same as in Fig.~\ref{fig:planetvsradius}.}
\label{fig:radiusvsenergy}
\end{figure}

Figure~\ref{fig:planetvsradius} displays a scatter plot of the planetary radius versus orbital period for the previously confirmed and validated exoplanets from the Kepler, K2 and TESS missions along with the 301 new  Kepler exoplanets validated by \ExoMiner. The distribution of the \ExoMiner-validated exoplanets is consistent with that of the Kepler sample, with periods ranging from $\sim$0.5 days to over 280 days, and with planetary radii as small as $0.6\,R_\oplus$ to as large as $9.5\,R_\oplus$. Figure~\ref{fig:radiusvsenergy} shows a scatter plot of the planetary radius versus the energy received by each planet in the Kepler, K2, TESS and \ExoMiner\ samples. As with the previous parametric plot, the distribution of the radius versus received energy of the \ExoMiner\ sample generally follows that of the Kepler sample.

\section{Transfer Learning to TESS data}
\label{sec:TESS}	 
Training a machine classifier to vet TESS data is challenging because there is a limited number of TCEs with gold standard dispositions available for the \TESSMission. Existing works try to overcome the challenge of limited data by using either synthetic data~\citep{Osborn-deeplearning-2020} or self-dispositions based on high level heuristics~\citep{Yu_2019} in order to train a machine classifier to vet TESS TCEs. However, it is not easy to ensure that simulated signals, including transiting planets, astrophysical signals and systematic noise and artifacts, are sufficiently similar to the flight data; when used for training, these synthetic data may harm the performance of the model. The self-disposition approach is also noisy and could lead to an ineffective model that is not able to vet TCEs accurately. Instead, we argue that if the architecture of a machine classifier is general enough for the classification of TCEs and the data are preprocessed appropriately, a trained classifier should be able to vet TCEs from different missions comparably well. The processes for manually vetting transit signals from TESS and \kepler\ are very similar; i.e., independent of the source of the TCEs, vetting a TCE requires evaluating the same diagnostic plots and metrics. If a TCE passes these tests, it is vetted as a PC no matter where the data come from. Thus, we can transfer~\citep{NG-2016} a model learned from \kepler\ data to vet TESS TCEs. 
	
We conducted a preliminary analysis of \ExoMiner's performance at classifying transit signals obtained from the ongoing \TESSMission. The Science Operations Center (SOC) pipeline, which was created to process the data from the \KeplerMission, was ported and tuned for TESS data in the Science Processing Operations Center \citep[SPOC,][]{Jenkins2016SPIE}. The SPOC pipeline is currently used to generate TCEs that can be further designated as TESS Objects of Interest~\citep[TOIs,][]{guerrero2021TOI}. Even though the population of target stars observed by TESS\footnote{TESS target stars are on average 30 to 100 times brighter than those surveyed by the \KeplerMission.} and the characteristics of the systematic noise are different, the nature of the signals captured by the two satellites is similar. In this first analysis, the same preprocessing steps were applied to the TESS data to generate the input data to feed into the model trained exclusively on data from the \KeplerMission. 
	
One specific feature of TESS data is that the pipeline is run for each individual sector, but there are also multi-sector runs that are conducted using data from multiple sectors. The single-sector runs produce TCEs in the approximately 28 day-long observation window, so the detected TCEs have orbital periods shorter than this time interval\footnote{A constraint is defined such that TESS TCEs must have at least two observed transits.}. Since there is overlap between sectors, especially at the ecliptic poles, multi-sector runs are able to produce TCEs with longer orbital periods when transits are observed in multiple sectors. Currently, the data for the first 36 sectors are publicly available\footnote{https://archive.stsci.edu/tess/bulk\_downloads/bulk\_downloads\_ffi-tp-lc-dv.html, last accessed on 3-9-2021} as well as the TCEs detected by the SPOC pipeline\footnote{https://archive.stsci.edu/tess/bulk\_downloads\_bulk\_downloads\_tce.html, last accessed on 3-9-2021}. A total of approximately 42,500 TCEs were detected, with an average of $\sim 1200$ TCEs per observed sector. Spanning from Sector 1 to Sector 26, nine multi-sector runs were conducted, yielding a total of $\sim 29,500$ TCEs. From these runs, experts vetted the detected TCEs and combine data from multiple sources to arrive at a catalog of TOIs\footnote{https://archive.stsci.edu/missions/tess/catalogs/toi/tois.csv}~\citep{guerrero2021TOI}. 

The TOI catalog made publicly available on 4 December 2020, which covered Sectors 1-30, was used to generate the data on which we evaluated \ExoMiner's performance. 
After preparing the TOI catalog (removing TOIs with missing ephemerides or observation sectors), 1167 SPOC TOIs were preprocessed using our pipeline to generate the data inputs to \ExoMiner. Given that the TOI catalog does not include the phase of the secondary event, ephemeris matching between the TOIs and the detected TCEs was performed such that our preprocessing pipeline could generate the transit-view secondary eclipse time series. In this preliminary analysis of TESS data, we trained a new model on \kepler\ Q1-Q17 DR25 TCE data using only the time series and stellar parameters as features. We call this model \ExoMiner-Basic. The classifications produced by the model were compared against the TESS Follow-up Observing Program Working Group \citep[TFOPWG --][]{collins2018tess} dispositions, which are based on additional observations and are the current gold standard dispositions. Our analysis focuses on the TOIs dispositioned as Known Planets (KPs), CPs, FPs, and FAs. We did not include the TOIs with PC disposition because this disposition is not conclusive.

\begin{table}[htb!]
 \centering
\caption{Confusion matrix of TESS classification using a basic version of \ExoMiner}
\label{table:TESS-results}
\begin{threeparttable}
\begin{tabularx}{\linewidth}{@{}Y@{}}
\begin{tabular}{cccc}
\toprule
Class & PC & Non-PC & Total \\
\midrule
 KPs  & 185 & 21 & 206 \\

 CPs  & 32 & 57 & 89 \\

 FPs  & 27 & 67 & 94 \\

 FAs  & 2 & 16 & 18 \\
 Total  & 246 & 161 & 407 \\
 \bottomrule
\end{tabular}
\end{tabularx}
\end{threeparttable}
\end{table}

\begin{figure}[htb!]
	\centering
	\subfigure{\includegraphics[width=\columnwidth]{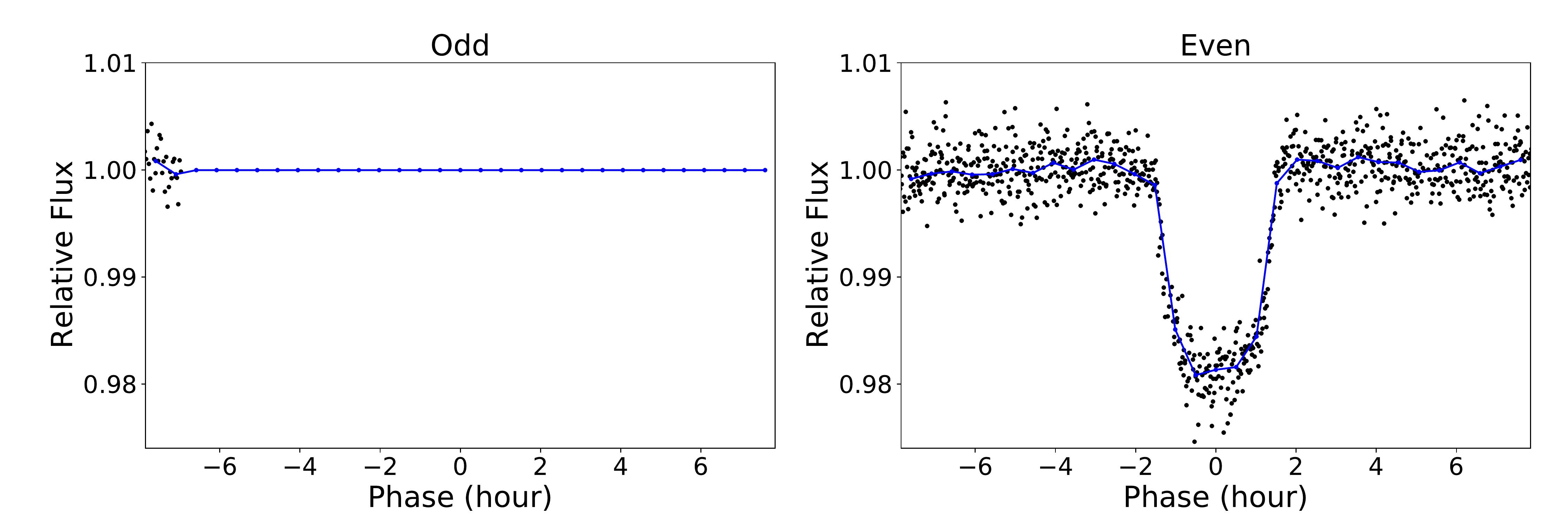}}
	\caption{Odd \& even transit-view flux phase folded time series generated for TOI 822.01. The blue dots show the average bin value, and the black dots show the cadences in the phase folded time series.}	
\label{fig:tess-no_oddeven}
\end{figure}

\begin{figure*}[htb!]
	\centering
	\subfigure{
\includegraphics[width=\textwidth]{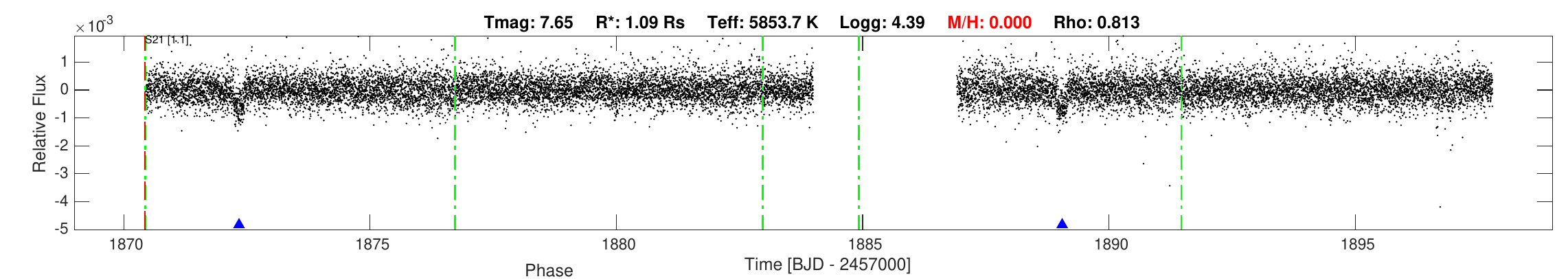}}
		\caption{Full-orbit flux time series for TCE TIC 4897275.1 (associated with TOI 1774.01) obtained from the DV summary generated by the SPOC pipeline for Sector 21.} 
\label{fig:tess-low_num_tr}
\end{figure*}

\begin{figure}[htb!]
	\centering
	\subfigure{\includegraphics[width=\columnwidth]{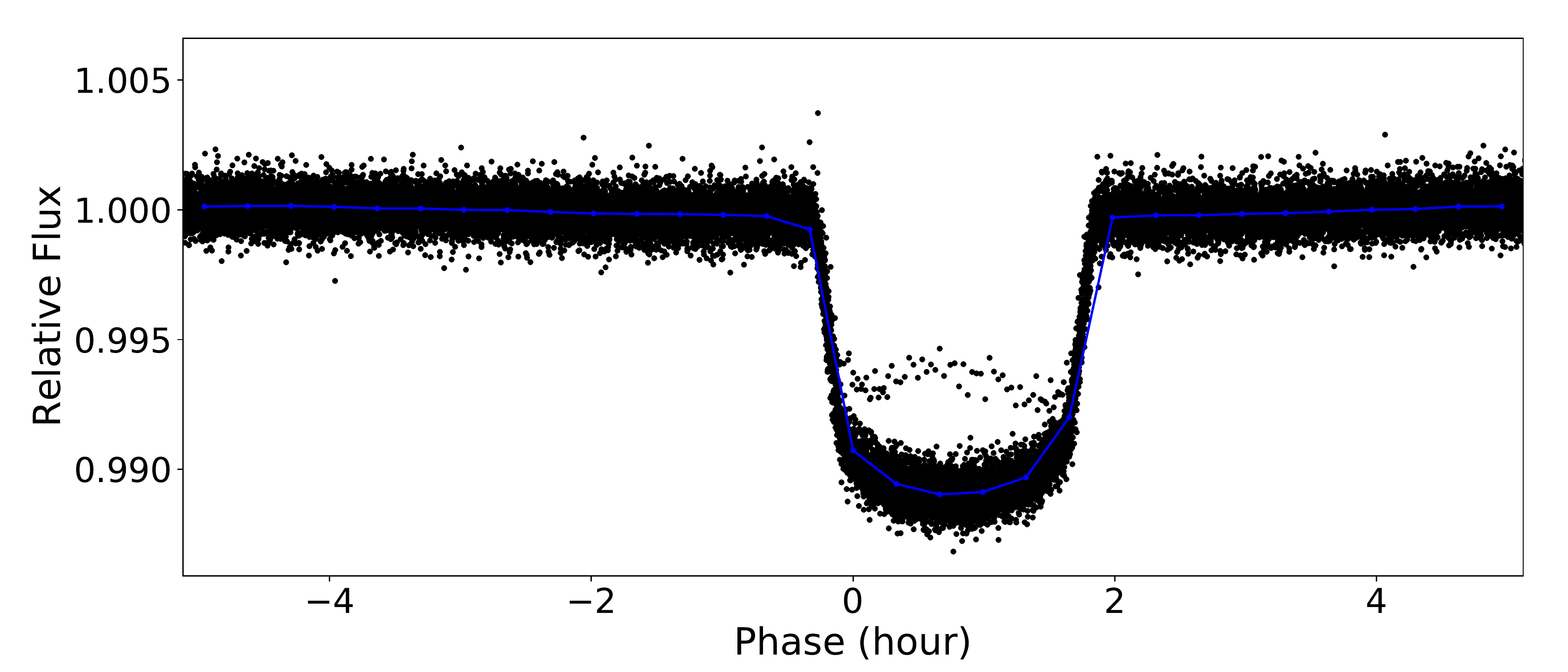}}
		\caption{Inaccurate epoch information leads to off-centered transit-views.}
\label{fig:tess-transit_shift}
\end{figure}

Table~\ref{table:TESS-results} shows the confusion matrix of applying \ExoMiner-Basic to TESS data for the standard classification threshold of 0.5. \ExoMiner-Basic performs reasonably well on the TESS data, resulting in a precision value of 0.88 and a recall value of 0.73.  
We expect the performance of the model to increase if the model has access to the DV diagnostics that the full \ExoMiner\ model uses for classifying \kepler\ data. However, an analysis of the misclassified KP and CP TOIs gave insight into some issues that must be addressed in the future, as shown in Figures \ref{fig:tess-no_oddeven},~\ref{fig:tess-low_num_tr}, and~\ref{fig:tess-transit_shift}.

There are cases of TOIs with transits located inside the downlink window for TESS. Data are downlinked to Earth at the perigee, which happens approximately every 13.7 days and lasts about a day. During that time, new data are not collected, which prevents the observation of some transits. For example, TOI 822.01 (WASP-132b) was observed in Sector 11 and has an orbital period of approximately 7.13 days, but the corresponding TCE (TIC~127530399.1) is reported as having double the period. Given that one transit was located in the downlink window, TPS did not fold the missing transit onto the observed transits, incorrectly determining the period of the TOI, which is known from previous observations. Since we are using the true orbital period from the TOI catalog, the preprocessing pipeline generates one of the odd \& even transit-views to be flat (Figure \ref{fig:tess-no_oddeven}), rendering this diagnostic test unreliable and leading the model to incorrectly classify this TOI as a false positive, even though it is a KP. As we discussed earlier, this problem can be addressed by using additional input features that take into account the number of observed transits and their variability for the odd and even views.

Another issue with TOIs that only have a small number of observed transits is that averaging the phase folded time series is not enough to filter the stochastic noise in the time series; this becomes even more critical when S/N is low. We expect to see more of these cases in TESS data considering that for significant areas of the sky, the total observation time for many TCEs is much shorter than for \kepler, which had a fixed field of view that was observed for four years. For example, TOI 1774.01 was observed in Sector 21 and only has two observed transits, as shown in Figure \ref{fig:tess-low_num_tr}, which creates flux views with significant levels of noise. 

A further issue arises when incorrect epoch values are reported in the TOI catalog because it causes the transit to be off-centered in the view. Since \ExoMiner\ expects the transit to be located in the center of the view, the model classifies such TOIs as false positives. For example, the transit for TOI 185.01 is shifted to the right in our view, as shown in Figure \ref{fig:tess-transit_shift}. 
A possible approach that can be used to make the model less sensitive to the location of the transit source in the views is to provide the model examples in which the transit is slightly shifted from the center of the view during training. Using such a data augmentation technique may make the model more robust to inaccuracies in the orbital period and epoch estimates.

\section{Discussion}
\label{sec:discussions}
We introduced \ExoMiner, a new deep learning classifier that outperforms existing machine classifiers in terms of various classification and ranking metrics. \ExoMiner\ is a robust and explainable deep learning model that allows us to reliably and accurately build new catalogs of PCs. The key to \ExoMiner's performance is its architecture that mimics the process by which domain experts vet transit signals by examining multiple types of diagnostic tests in forms of scalar values and time series data. This architecture also enabled us to design a preliminary branch-occlusion explainability framework that provides interpretability in terms of which diagnostic tests \ExoMiner\ utilizes in order to vet a signal. Moreover, since the general concept behind vetting transit signals is the same for both \kepler\ and TESS data, and \ExoMiner\ utilizes the same diagnostic metrics as expert vetters do, we expect an adapted version of this model to perform well on TESS data. Our preliminary results on TESS data verify this hypothesis. Using \ExoMiner, we also demonstrate that there are hundreds of new exoplanets hidden in the 1922 KOIs that require further follow-up analysis. Out of these, 301 new exoplanets are validated with confidence using \ExoMiner. 

\subsection{Caveats}
There are caveats related to the ML approach taken in this work, which include:
\begin{itemize}
\item As we reported in Section~\ref{sec:misclassified-cases}, \ExoMiner\ sometimes fails to adequately utilize diagnostic tests, e.g., centroid test, or does not have the data required for some FP scenarios, e.g., flux contamination. This leads to misclassification of transit signals. To provide the extent to which this causes problems, some statistics are in order. Out of a total of 8053 KOIs, there are 1779 KOIs that have centroid test flag on. \ExoMiner\ misclassifies 23 (0.29\%) of these of which only one had a score $>0.99$ (0.01\%). After applying the priors, the number of misclassified cases reduces to thirteen (0.16\%) out of which two (0.025\%) have scores $>0.99$. Out of 1087 KOIs that have flux contamination, \ExoMiner\ misclassifies sixteen (0.2\%) and give a score $>0.99$ to only one (0.01\%). After applying the priors, only one KOI with flux contamination is misclassified by \ExoMiner\ with a score of 0.84. Overall, applying the scenario priors alleviates this problem; however as we have suggested in the validation process discussed in Section~\ref{sec:validation-setup}, the classification performed by \ExoMiner\ should be vetoed using FP flags provided in the Cumulative KOI catalog in order to reduce the risk of FPs being validated as exoplanets. 

\item The assumption made in Equation~\ref{eq:prior_information} that $x$ and $I$ are independent might not be valid. This is because the information used in $I$, e.g., crowded field, might be present in the flux or centroid data ($x$). When this assumption does not hold, we are counting evidences for or against supporting an exoplanet classification twice (one from $I$ and another from $x$). Thus, the the application of the priors could lead to underestimating or overestimating the true probability. However, as we reported in Table~\ref{table:classification_results_prior}, \ExoMiner\ and other existing classifiers benefit from the scenario priors overall in terms of precision and recall. Therefore, the application of priors is beneficial. 

\item The performance comparison between different classifiers should also be taken with caution due to the existence of label noise. However, the relative performance of different classifiers is reasonably sound, as we showed in Section~\ref{sec:radial-velocity} and~\ref{sec:santerne} using additional independent datasets. 

\item In section~\ref{sec:training_set_sensitivity}, we only studied the effect of random label noise and underrepresented scenarios in the training set. However, the true nature of label noise or underrepresented scenarios could be systematic, limiting the reliability of our analysis. The presence of systematic label noise or underrepresented data scenarios could potentially lead to misclassification of FP scenarios for which there is significant label noise or lack of enough representative samples. 
\end{itemize}

There are also astrophysical caveats, which include:
\begin{itemize}
\item Given that there is no viable way to distinguish between BD FPs and real exoplanets using only \kepler\ photometry, \ExoMiner\ and all similar photometry based data-driven models are not able to correctly classify BD FPs. However, BDs are very rare in \kepler\ data: \citet{Santerne_independent_2016} found only three BDs in a follow-up study of 129 giant planet candidates from \kepler. The authors also calculated a BD occurrence rate of 0.29\% for periods less than 400 days, 15 times lower than what they find for giant planets. Considering the geometric probability of alignment, which we can assume is $\sim$5\% (conservatively), the chance of seeing a transiting BD is $\sim0.05\times0.29 \approx 0.01\%$. Moreover, note that there is only a handful of validated large planets in our list, i.e.,  17/301 with $4\,R_\oplus<R_p<10\,R_\oplus$ and only 7/301 with $5\,R_\oplus<R_p<10\,R_\oplus$. Thus, given that transiting BDs can often be identified via RV measurements, and the low occurrence rate of such systems, the BD FPs should be of little concern. 
\item There could exist blended EBs that do not exhibit secondary eclipses but might be exposed by high resolution imaging and photometric observations with better spatial resolution. There could still exist relevant patterns in the photometry data for such objects that machine can utilize to correctly classify them, even though those are not easy for human vetters to find and use. However, we did not provide any specific studies of the performance of deep learning on such objects in this work. Thus, we assume \ExoMiner\ is not able to detect such FPs.   
\item Due to the different stellar samples, pixel scales and blending issues, TESS will have a different distribution of false positives, and this will affect classifier results. Thus, special care needs to be taken transferring to TESS where validation is concerned.
\end{itemize}

\subsection{Future direction}
The analysis performed in this paper will allow us to pinpoint a number of future directions for improving and building upon this work. First, we will change the \ExoMiner\ architecture to include new important features, including 1) the number of transits and the uncertainty about the depth of the odd and even views to ensure that the model's use of the odd \& even branch is more effective and 2) the \kepler\ magnitude to allow the model to recognize saturated stars for which the centroid test is invalid. Second, we will perform a full explainability study by utilizing more advanced explainability techniques and design a framework that can provide experts with insight about \ExoMiner's labels and scores. Third, we will be doing more depth analysis of the effect of label noise and underrepresented scenariors and provide a list of possible inaccurate labels in the Q1-Q17 DR25 dataset. Finally, we will soon publish the results of utilizing an adapted version of \ExoMiner\ to classify TESS transit signals by extending the set of features in a similar way to the one used for \kepler\ data, and addressing some of the issues mentioned previously that affect TESS data. 

\section*{Acknowledgements}
Hamed Valizadegan, Miguel Martinho, Laurent Wilkens, and Nikash Walia are suported through NASA NAMS contract NNA16BD14C. Jeffrey Smith, Douglas Caldwell, and Joseph Twicken are supported through NASA Cooperative Agreement 80NSSC21M0079. We would like to thank multiple people who directly or indirectly contributed to this work. We are very grateful to Porsche M. Parker and Haley E. Feck from USRA and Krisstina Wilmoth from NASA ARC who have helped us tirelessly to recruit interns through NASA ARC $I^2$ intern program and NASA ARC Office of STEM Engagement. Without these amazing interns, this work would not have been possible. We are grateful to Patrick Maynard, who has recently joined our team as an intern and generated new insights into TESS data. We also appreciate David J. Armstrong in providing us with the detailed results of their ML models~\citep{armstrong-2020-exoplanet}, and Megan Ansdell, who answered our questions regarding \ExoNet\ code~\citep{Ansdell_2018}. Our discussion with David J. Armstrong on the caveats of using ML for this problem helped us tremendously in improving this work. This paper includes data collected by the \kepler\ and TESS missions and obtained from the MAST data archive at the Space Telescope Science Institute (STScI). Funding for the Kepler mission is provided by the NASA Science Mission Directorate. Funding for the TESS mission is provided by the NASA Explorer Program. STScI is operated by the Association of Universities for Research in Astronomy, Inc., under NASA contract NAS 5–26555. We acknowledge the use of public TESS data from pipelines at the TESS Science Office and at the TESS Science Processing Operations Center. Resources supporting this work were provided by the NASA High-End Computing (HEC) Program through the NASA Advanced Supercomputing (NAS) Division at Ames Research Center for the production of the SOC Kepler and the SPOC TESS data products.

\bibliography{ExoPlanetV2}{}
\bibliographystyle{aasjournal}



\end{document}